\documentclass{siamonline1116}
\pdfoutput=1
\usepackage[T1]{fontenc}
\usepackage{graphicx}
\usepackage{wrapfig}
\usepackage{amsmath,amssymb,amsfonts}
\usepackage{amsopn}
\numberwithin{theorem}{section}
\usepackage{multirow,multirow}
\usepackage{makecell}
\usepackage[small, labelfont={bf, sf, color=header1}, textfont=it, tableposition=top]{caption}
\usepackage{xcolor}
\usepackage{url}
\usepackage{afterpage}
\usepackage{textcomp}
\usepackage{rotating}
\usepackage{siunitx}
\usepackage{paralist}
\usepackage{booktabs}
\usepackage{lscape}
\usepackage{bbold}
\usepackage{bm}
\usepackage{algorithm}
\usepackage{algpseudocode}
\floatname{algorithm}{Procedure}

\usepackage{subcaption}
\usepackage{longtable}
\usepackage[version=4]{mhchem}
\usepackage[toc,page]{appendix}
\newcounter{runidnum}

\newsiamremark{remark}{Remark}
\usepackage{tikz}
\usetikzlibrary{shapes,arrows,trees,snakes}
\usepackage{pgfplots}
\usepackage{pgfplotstable}
\usepackage{tkz-euclide}
\usepackage{pifont}
\usetikzlibrary{calc}
\usetikzlibrary{decorations.pathreplacing,arrows}
\usetikzlibrary{shapes,arrows,trees,snakes,positioning}
\usetikzlibrary{decorations.markings}

\newcolumntype{R}{>{\columncolor{gray!20}}r}
\newcolumntype{L}{>{\columncolor{gray!20}}l}
\newcolumntype{C}{>{\columncolor{gray!20}}c}
\DeclareGraphicsExtensions{.pdf,.png}
\usepackage{lineno}
\modulolinenumbers[5]
\allowdisplaybreaks

\newcommand{\vect}[1]{\boldsymbol{#1}} 

\DeclareMathOperator*{\argmin}{arg\,min}
\DeclareMathOperator*{\argmax}{arg\,max}

\newcommand{\alert}[1]{\textcolor{red}{#1}} \renewcommand{\alert}[1]{}

\definecolor{light-gray}{gray}{0.80}

\newcommand{\rbr}[1]{\left(#1\right)}

\algnewcommand{\algorithmicand}{\textbf{ and }}
\algnewcommand{\algorithmicor}{\textbf{ or }}
\algnewcommand{\OR}{\algorithmicor}
\algnewcommand{\AND}{\algorithmicand}

\newcommand{\opA}{\ensuremath{f}}
\newcommand{\uq}{\ensuremath{\vect{x}}}
\newcommand{\uqmap}{\ensuremath{\vect{x}_{\mathrm{MAP}}}}

\newcommand{\DT}{\ensuremath{\mathbb{Y}}}
\newcommand{\ind}{\ensuremath{\boldsymbol{\mathrm{1}}}}

\graphicspath{figs}

\newcommand{\TheTitle}{BIMC: The Bayesian Inverse Monte Carlo method
for goal-oriented uncertainty quantification. Part II.}
\newcommand{\TheAuthors}{Siddhant Wahal and George Biros}

\headers{\TheTitle}{\TheAuthors}

\title{{\TheTitle}}

\author{
  Siddhant Wahal\thanks{
    Oden Institute for Computational Engineering and Sciences, 
    The University of Texas at Austin, Austin, TX, 78712,
(\email{siddhant@oden.utexas.edu}, \email{biros@oden.utexas.edu}). }
  \and
  George Biros\footnotemark[1]}

\ifpdf
\hypersetup{
  pdftitle={\TheTitle},
  pdfauthor={\TheAuthors}
}
\fi

\begin{document}
\maketitle

\begin{abstract}
In Part I of this article, we proposed an importance sampling
algorithm to compute rare-event probabilities in forward uncertainty
quantification problems.  The algorithm, which we termed the
\emph{``Bayesian Inverse Monte Carlo (BIMC) method''}, was shown to be
optimal for problems in which the input-output operator is nearly
linear. But applying the original BIMC to highly nonlinear systems can
lead to several different failure modes. In this paper, we modify the
BIMC method to extend its applicability to a wider class of
systems. The modified algorithm, which we call \emph{``Adaptive-BIMC
  (A-BIMC)''}, has two stages. In the first stage, we solve a sequence of
optimization problems to roughly identify those regions of parameter
space which trigger the rare-event. In the second stage, we use the
stage one results to construct a mixture of Gaussians that can be then
used in an importance sampling algorithm to estimate rare event
probability. We propose using a local surrogate that minimizes costly
forward solves. The effectiveness of A-BIMC is demonstrated via
several synthetic examples. Yet again, the modified algorithm is prone
to failure. We systematically identify conditions under which it fails
to lead to an effective importance sampling distribution.
\end{abstract}



\section{Introduction}

Part I of this paper presented an importance sampling algorithm to address the
following goal-oriented uncertainty quantification (UQ) question. Given
\begin{itemize}
    \item a smooth nonlinear function $f(\uq):\mathbb{R}^m\rightarrow\mathbb{R}$,  
    \item a probability distribution for its inputs $p(\uq)$, and,
    \item a target interval $\mathbb{Y} \subset \mathbb{R}$, 
\end{itemize}
what is the probability of the event $f(\uq) \in \mathbb{Y}$? Our interest was
in computing this probability efficiently, i.e., with as few evaluations of
$f$ as possible, especially when the event  $f(\uq) \in \mathbb{Y}$ is
rare. 

The algorithm presented in Part I, called the Bayesian Inverse Monte Carlo
(BIMC) method, employed a fictitious Bayesian inverse problem to identify
regions of parameter space that evaluate inside $\mathbb{Y}$. BIMC was proven
to lead to an optimal IS density for affine $f$ and Gaussian $p$. As such,
when applied to maps $f$ that appear nearly affine at the scale of the
covariance of $p$, BIMC outperformed a simple Monte Carlo method by several
orders of magnitude.

We also showed that when this is not the case, that is, when $f$ is
significantly nonlinear, BIMC leads to a poor-quality IS distribution. This in
turn can lead to inaccurate estimates of the rare-event probability. In Part II
of this two-part article, we propose modifications to BIMC in order to address
this major limitation. 

\paragraph{Summary of the methodology}
The modifications result in a two-stage algorithm, which we christen
Adaptive-BIMC (A-BIMC). Stage-1 of A-BIMC solves a sequence of optimization
problems in order to adaptively explore the input parameter space where $f(\uq)
\in \mathbb{Y}$ (more precisely, this region is the pre-image of
$\mathbb{Y}$, defined as the set $\{\uq \in \mathbb{R}^m: f(\uq) \in
\mathbb{Y}\}$  and denoted  $f^{-1}(\mathbb{Y})$). While BIMC also relies on
the solution of an auxiliary, \emph{``fictitious''}, inverse problem, the formulation and interpretation of the optimization problems in
Stage-1 of A-BIMC is quite different. 

BIMC solves a single optimization problem to arrive at a pseudo-MAP point and
then samples around this point to explore the pre-image $f^{-1}(\mathbb{Y})$. 
By sampling around a single point, BIMC only achieves local exploration of 
$f^{-1}(\mathbb{Y})$. While this may work for nearly affine maps, such limited
exploration is insufficient for more nonlinear problems.

On the other hand, Stage-1 of A-BIMC solves optimization problems in an
iterative fashion; the optimization problem in some iteration, aided by
specially designed algorithmic components,  explores the pre-image
$f^{-1}(\mathbb{Y})$ away from regions explored in the preceding iterations.
This allows global exploration of $f^{-1}(\mathbb{Y})$. Stage-1 keeps
iterating till a termination condition based on user-specified tolerances is
met. Then, the local minima so obtained are collected into a Gaussian mixture
that roughly approximates the theoretically ideal (zero-error) importance
sampling density $q^* \propto \ind_{\DT}(f(\uq))p(\uq)$.

The Gaussian mixture which crudely approximates $q^*$ is  refined using the
Mixture Population Monte Carlo Algorithm~\cite{cappe2008adaptive}.  The MPMC
algorithm modifies the mixture weights, means, and covariances of this Gaussian
mixture so that it closely approximates the ideal IS density $q^*$. This,
however, requires further evaluations of $f$, raising the computational cost of
the algorithm. In order to circumvent this problem, we replace evaluations of
$f$ in $q^*$ with that of a heuristically constructed surrogate of $f$. Next, we
list the contributions and limitations of our approach.

\paragraph{Contributions}

\begin{itemize}
    \item We extend our work in Part I and propose a novel scheme which adaptively
          explores the pre-image $f^{-1}(\mathbb{Y})$ on a global scale. In particular,
          we describe algorithmic strategies, such as parameter continuation, and a
          modified penalty algorithm required to achieve this exploration. 
    \item Our scheme employs the derivatives of $f$ to accelerate exploration of
          the pre-image $f^{-1}(\DT)$, as opposed to only pointwise evaluations
          of $f$. 
    \item We have attempted to keep tunable algorithmic parameters to a minimum, and as
          a result, A-BIMC possesses just one user-defined parameter.
    \item A-BIMC's performance is studied on synthetic problems. Experiments demonstrate
          that the performance of our method doesn't depend significantly on how small the target probability is. Rather it depends on the nonlinearity of the input-output map.
\end{itemize}

\paragraph{Limitations}

\begin{itemize}
    \item A-BIMC is a purely heuristic algorithm. In particular, the algorithmic
          components chosen preclude its theoretical analysis. Hence, unlike 
          BIMC, it is difficult to a-prior predict its performance. 
    \item A-BIMC is not without its own failure mechanisms. These are described in
          \Cref{section:failure}, along with strategies for diagnosing and 
          mitigating them.
    \item While our algorithm possesses only one tunable parameter, we 
          don't have an \emph{a priori} prescription for choosing it. 
          We recommend a default value in \Cref{section:experiments} 
          but cannot provide guarantees on whether this value will work or not. 
    \item A-BIMC relies crucially on the MPMC method, which doesn't scale to
          high-dimensional problems (say, for instance, greater than 64). As a
          result, A-BIMC is not suitable for problems with high intrinsic dimension. 
\end{itemize}

\paragraph{Notation}
\label{section:notation}
Key notation used in this paper is summarized in \Cref{table:notation}.
\begin{table}[h]
\footnotesize
\centering
\begin{tabular}[c]{l l}
\toprule
Symbols/Acronyms   & Meaning\\
\midrule
MC                 & Monte Carlo \\
IS                 & Importance Sampling\\
BIMC               & Bayesian Inverse Monte Carlo\\
A-BIMC             & Adaptive Bayesian Inverse Monte Carlo\\
ESS                & Effective Sample Size\\
MPMC               & Mixture Population Monte Carlo\\
RMS                & Root Mean Square\\
MAP                & Maximum \emph{A Posteriori}\\
\midrule
$f$                & input-output, or the forward, map\\
$\uq$              & vector of input parameters to $f$\\
$p(\uq)$           & input or nominal probability density for $\uq$\\
$\DT$              & target interval for $f(\uq)$\\
$\ind_{\DT}$       & indicator function, $\ind_{\DT}(z)$ is 1 if $z \in \DT$,  0 otherwise \\
$f^{-1}(\DT)$      & pre-image of the interval $\DT$, $\{\uq \in \mathbb{R}^m: f(\uq) \in \DT\}$\\
$m$                & dimension of $\uq$\\
$\mathcal{N} \rbr{\uq_0, \bm{\Sigma}_0}$ 
                   & normal distribution with mean $\uq_0$
                     and covariance $\bm{\Sigma}_0$\\
$\mathbb{P}$       & probability of an event \\
$\mathbb{E}_p$     & expectation of a random variable with 
                     respect to some density $p$ \\
$\mathbb{V}_p$     & variance of a random variable with respect 
                     to some density $p$\\
$\mu$              & $\mathbb{P}\rbr{f(\uq) \in \DT}$, equivalently, 
                     $\mathbb{E}_p\rbr{\ind_{\DT}\rbr{f(\uq)}}$.\\
$q^*$              & ideal importance sampling density\\
$Q_k$              & importance sampling mixture density at the $k$-th 
                     iteration of Stage-1\\
$Q$                & final importance sampling mixture density\\
$N$                & number of samples\\
$\hat{\mu}^N$      & MC estimate for $\mu$ computed using $N$ samples\\
$\tilde{\mu}^N$    & IS estimate for $\mu$ computed using $N$ samples\\
$\hat{e}_{\mathrm{RMS}}^N$ 
                   & RMS error in $\hat{\mu}^N$\\
$\tilde{e}_{\mathrm{RMS}}^N$ 
                   & RMS error in $\tilde{\mu}^N$\\
$y$                & pseudo-data \\
$p(y | \uq)$       & pseudo-likelihood density\\
$p(\uq | y)$       & pseudo-posterior density\\
$\uqmap$           & MAP point of $p(\uq|y)$\\
$\mathbf{H}_{\mathrm{MAP}}$ 
                   & Hessian of $-\log p(\uq | y)$ at $\uqmap$\\
$\mathbf{H}_{\mathrm{GN}}$ 
                   & Gauss-Newton Hessian of $-\log p(\uq | y)$ at some $\uq$\\
$\epsilon_{\mathrm{rel}}$ 
                   & relative tolerance for perplexity change across A-BIMC iterations\\
$\epsilon_{\mathrm{abs}}$
                   & absolute tolerance for perplexity change across A-BIMC iterations\\
$f_{\mathrm{surrogate}}$    
                   & cheap-to-evaluate surrogate for $f$ \\
$N_{\mathrm{MPMC}}$ 
                   & number of samples used by MPMC\\
$m_{\mathrm{int}}$ & intrinsic dimension of the rare-event problem \\
$D_{\mathrm{KL}}(p||q)$ 
                   & Kullback-Leibler (KL) divergence between 
                     densities $p$ and $q$\\
\bottomrule
\end{tabular}
\caption{Summary of key notation used in this paper.}
\label{table:notation}
\end{table}

\paragraph{Outline of the paper} In \Cref{section:BIMC}, we briefly review the
BIMC method, drawing attention to the core theoretical ideas behind it, as well
as its failure mechanisms. We describe A-BIMC in detail in
\Cref{section:methodology}. A-BIMC's performance is studied via several
numerical experiments in \Cref{section:experiments}. In \Cref{section:failure},
we discuss how A-BIMC can fail, and provide strategies for diagnosing and
mitigating failure. We conclude in \Cref{section:conclusion} and discuss several
avenues for future research. Supporting theory and additional results
accompanying our computational experiments are provided in the
supplement~\cite{wahal2019bimc_supplement_part_2}.

\section{BIMC}
\label{section:BIMC}

To provide context and motivation for developing A-BIMC, this section provides a
brief review of the BIMC algorithm. BIMC identifies regions of parameter space
that evaluate inside $\DT$ by setting up a fictitious inverse problem. This
fictitious inverse problem is defined by re-interpreting the ingredients of the
posed rare-event probability estimation problem (namely, $f(\uq)$, $p(\uq)$, 
and $\DT$) in the following manner:

\begin{enumerate}
    \item some $y \in \DT$ is selected to represent the observations,
    \item $p(\uq)$ is interpreted as the prior, and,
    \item a likelihood model, assuming additive Gaussian error of some magnitude
          $\sigma$ is constructed.
\end{enumerate}

These ingredients lead to a fictitious posterior distribution, $p(\uq | y)
\propto p(y | \uq) p(\uq)$. BIMC obtains an IS density $q(\uq)$ by
approximating the fictitious posterior $p(\uq | y)$ with a Gaussian
distribution. The Gaussian approximation invokes Laplace's
method~\cite{laplace1986memoir, wong1989asymptotic}.
to yield $q(\uq) =
\mathcal{N}\left(\uq_{\mathrm{MAP}}, \mathbf{H}_{\mathrm{MAP}}^{-1}\right)$,
where $\uq_{\mathrm{MAP}}$ is the maximum \emph{a posteriori} point of $p(\uq |
y)$, and
$\mathbf{H}_{\mathrm{MAP}}$ is the Hessian matrix of $-\log p(\uq | y)$ at
$\uq_{\mathrm{MAP}}$.  In this procedure, the inference problem is termed
fictitious because both the observation $y$ and the likelihood
variance $\sigma^2$ are auxiliary variables that are selected by the algorithm. This inverse problem is solved for the sole purpose of constructing an effective importance sampling
density.%
\footnote{Subsequently, the prefix \emph{pseudo} will be appended to the
components of the inference problem. That is, $y$ will be referred to as the
pseudo-data, $p(y | \uq)$ as the pseudo-likelihood, $p(\uq)$ as the
pseudo-prior, and $p(\uq | y)$ as the pseudo-posterior.}

Being contrived quantities, appropriate values for the pseudo-data $y$ and
pseudo-likelihood variance $\sigma^2$ are not known \emph{a priori}.  However,
they can have a profound impact on the performance of the importance sampling
distribution and must be tuned with care. The tuning procedure that BIMC adopts
is reproduced in the next subsection.

\subsection{Tuning parameters}
In BIMC, the strategy for tuning $\sigma^2$ and $y$ is based on
theoretical analysis of the case when $f(\uq)$ when affine and $p(\uq)$ is
Gaussian. Suppose that this is indeed so, that is,
$\opA(\uq) = \vect{v}^T\uq +\beta$, $p(\uq) = \mathcal{N}(\uq_0, 
\mathbf{\Sigma}_0)$ for some $\vect{v}$,
$\uq_0 \in \mathbb{R}^m$, $\beta \in \mathbb{R}$ and $\mathbf{\Sigma}_0 \in
\mathbb{R}^{m\times m}$. 
Then, for some pseudo-data  $y \in \mathbb{R}$ and pseudo-likelihood variance
$\sigma^2 \in \mathbb{R}^+$, the expression for the IS density is: 
$q(\uq) = \mathcal{N}(\uqmap, \mathbf{H}_{\mathrm{MAP}}^{-1})$, where,

\begin{align}
    \uqmap = \uq_0 + \frac{y - f(\uq_0)}{\sigma^2 + \vect{v}^T
    \mathbf{\Sigma}_0\vect{v}} \mathbf{\Sigma}_0 \vect{v},\quad
    \mathbf{H}_{\mathrm{MAP}}^{-1} = \mathbf{\Sigma}_0 - \frac{1}{\sigma^2 + 
    \vect{v}^T \mathbf{\Sigma}_0 \vect{v}}\big(\mathbf{\Sigma}_0\vect{v}\big)
    {\big(\mathbf{\Sigma}_0 \vect{v}\big)}^{T}.
\label{eq:affine_expressions}
\end{align}
 
Recall that an efficient IS density is one that is similar to $q^*$. Hence, it
follows that suitable values for $\sigma^2$ and $y$ are those that make
$q(\uq)$, as parameterized by $\sigma^2$ and $y$, resemble $q^*$ the most. These
optimal values for $\sigma^2$ and $y$ can be found by minimizing the
Kullback-Leibler (KL) divergence between $q^*$ and $q$:

\begin{align}
    \begin{pmatrix}
        \sigma^*\\
        y^*
    \end{pmatrix}    &= \argmin_{\sigma, y} D_{\mathrm{KL}}(q^* || q;
    \sigma, y)
    \label{eq:param_selection}
\end{align}

In the affine-Gaussian case, $D_{\mathrm{KL}}(q^* || q)$,
$\sigma^*$, and $y^*$ have analytical expressions. Part 1 of this paper derived
these expressions and proved 
that in the affine-Gaussian case, $q(\uq; \sigma^*, y^*)$  is the Gaussian distribution
closest in KL divergence to $q^*$. For situations where the input-output
map is not affine, closed form expressions for $q(\uq; \sigma,
y)$ and $D_{\mathrm{KL}}(q^* || q; \sigma, y)$ can rarely be written down. BIMC
tunes the parameters $\sigma^2$ and $y$ in such cases by first linearizing
$f(\uq)$ around a suitable point. Linearization yields an affine approximation of $f$
which is plugged into the procedure described above. Although this 
yields parameters that are optimal for the affine approximation and not for the
full nonlinear problem, it precludes additional calls to $f(\uq)$, thus keeping
computational costs low.

This methodology to devise an efficient IS density can fail in several cases.
These failure cases are detailed next. 

\begin{figure}[H]
    \centering
    \includegraphics[width=0.5\textwidth]{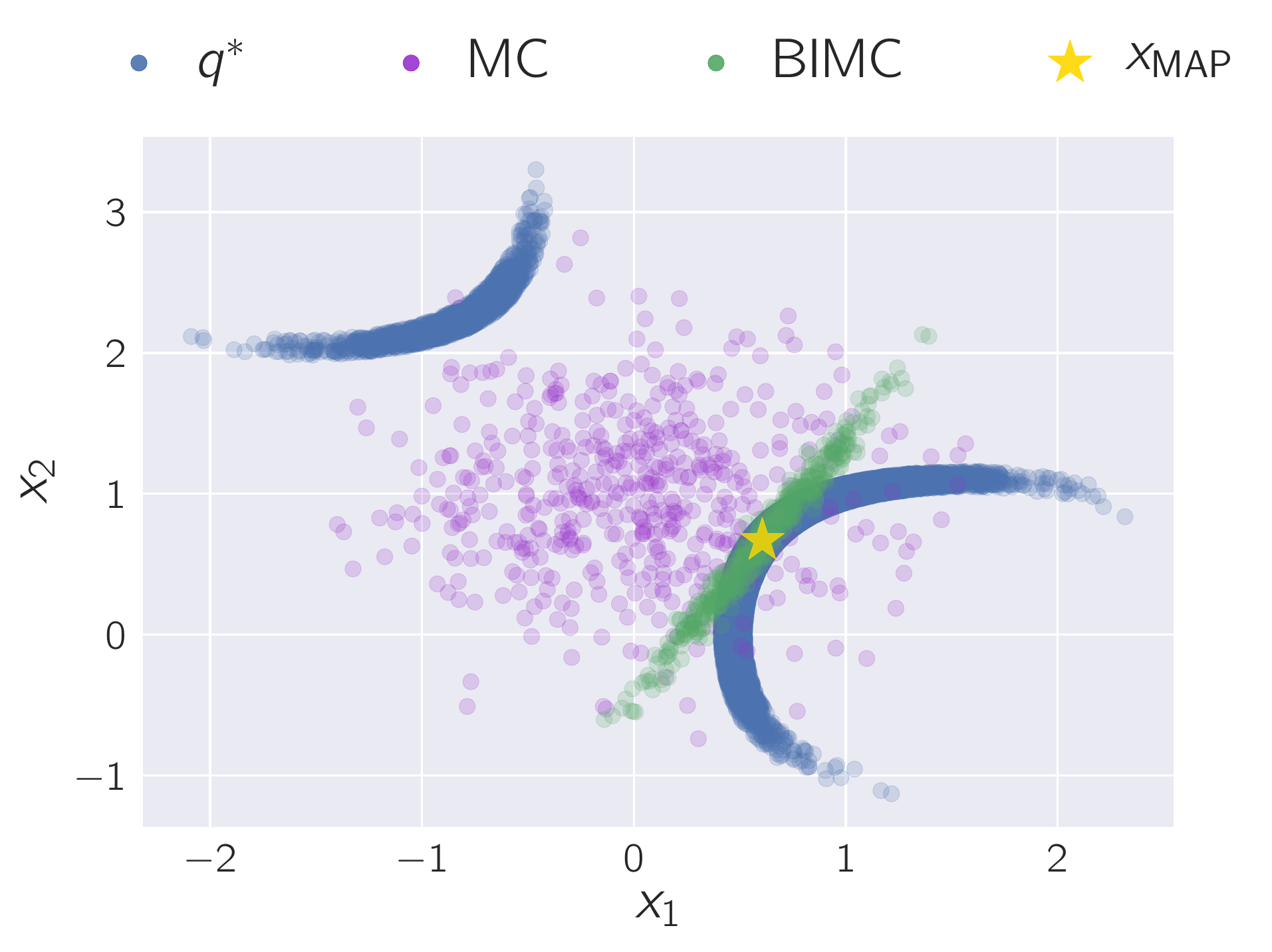} 
    \caption{A single Gaussian distribution fails to cover curved and/or
             disjoint $f^{-1}(\DT)$ when $f(\uq) = \sin(x_1)\cos(x_2)$. 
             Shown are 5000 samples from $q^*$ and 500 MC and BIMC samples.}
\label{fig:failTaylorGreen}
\end{figure}

\subsection{Failure modes of BIMC}
\label{subsection:bimc_failure}

The failure modes of BIMC can be categorized broadly into four cases:

\begin{itemize}
\item \emph{Curvature of $f^{-1}(\DT)$}:
Since BIMC results in an optimal approximation of $q^*(\uq)$ in the
affine-Gaussian case, it is expected to perform well for input-output maps that
are nearly affine. Using BIMC when this is not the case is susceptible to
failure, an unsurprising fact; using a single Gaussian distribution,
with convex, elliptical contours may prove to be insufficient to cover
arbitrary $f^{-1}(\DT)$. This phenomenon is captured in
\Cref{fig:failTaylorGreen} where $f: \mathbb{R}^2 \rightarrow \mathbb{R}$ is
defined as $f(\uq) = \sin(x_1)\cos(x_2)$ and the pre-image $f^{-1}(\DT)$ 
is a circular region in space.  
In conclusion, this mode of failure occurs when $f(\uq)$ is strongly nonlinear at 
the scale of the covariance of $p(\uq)$. 

\item \emph{Multiply-connected $f^{-1}(\DT)$}:
BIMC can also fail if $f^{-1}(\DT)$ is the union of disjoint regions in
space, which is also evident in \Cref{fig:failTaylorGreen}. 
In this case, $q^*$ has multiple modes whereas a single Gaussian can only
capture one. Because
the mean of $q(\uq)$ is found using numerical optimization, which mode is found
depends on the initial guess provided to the numerical optimization routine.

\item \emph{Poor selection of tunable parameters}: 
Yet another failure mode occurs when the optimal tunable
parameters that result from linearizing $f$ aren't appropriate for the full nonlinear
problem. As \Cref{eq:affine_expressions} reveals, the covariance of $q(\uq)$ is a rank-1
non-positive definite update of the covariance of $p(\uq)$. 
The degree of this update is inversely
proportional to $\sigma^2$. If the optimal likelihood variance, $\sigma^2$, for the
linearized problem is much smaller than the
optimal likelihood variance for the full nonlinear problem, the covariance
update will be needlessly large, and consequently, $q(\uq)$ will not have
enough support to cover $f^{-1}(\DT)$. This can cause large IS ratios, and at small
to moderate sample sizes, biased estimates for the rare event probability.
\item \emph{Intractable inverse problem}: It may be the case that the
optimization problem formulated to compute $\uq_{\mathrm{MAP}}$ is highly
ill-conditioned and non-convex, making $\uq_{\mathrm{MAP}}$ inaccessible via standard numerical
optimization routines. Since the BIMC methodology is crucially dependent on
successfully computing $\uq_{\mathrm{MAP}}$, we declare failure if the
numerical optimization routine employed fails at its task.
\end{itemize}

This concludes our recap of the BIMC method and its limitations. The next
section develops the Adaptive-BIMC (A-BIMC) methodology in detail.

\section{Methodology} \label{section:methodology}

A-BIMC is a heuristic algorithm designed to address the drawbacks in BIMC.
Perhaps the most crippling drawback of BIMC is its inability to conform to
disconnected or curved $f^{-1}(\DT)$. A-BIMC attempts to improve the 
performance of BIMC in this regime  by constructing a
mixture of Gaussians for importance sampling. Gaussian mixtures are a class of
probability distributions of the form $Q(\uq) = \sum_{k=1}^{K}\alpha_k
\mathcal{N}(\uq_k, \boldsymbol{\Sigma}_k)$, $\sum_k \alpha_k = 1$, and are
sufficiently rich to simulate complicated phenomena.  Constructing a suitable IS
density in the Gaussian mixture context amounts to identifying the mean $\uq_k$,
covariance $\boldsymbol{\Sigma}_k$, and mixture weight $\alpha_k$ of each
component. The algorithm does so in two stages, both of which proceed
iteratively. Stage-1 enriches the IS mixture with a new component at every
iteration, till the algorithm is satisfied that no more components are required
in the mixture.  The aim of this stage is to arrive at a mixture distribution
that roughly approximates $q^*(\uq)$. Stage-2 tunes the mixture that results at
the end of Stage-1 using the Mixture Population Monte Carlo
algorithm~\cite{cappe2008adaptive} so that it
forms a better approximation of $q^*$. Tuning the IS mixture in this manner also
removes the effect of any poorly selected parameters, such as $\sigma^2$ or $y$.
The following subsections describe each stage in more detail.

\subsection{Stage-1} \label{subsection:stage_1}

As mentioned previously, this stage constructs a crude approximation of $q^*$ by
adding a new component to the IS mixture at every iteration. To establish
notation, the IS mixture at the $k$-th iteration of this stage is denoted by
$Q_k$. Since each iteration in Stage-1 adds a new component to the IS mixture, 
$Q_k$ always has $k$ components. The $i$-th component of $Q_k$ is denoted by
$q_i$.  Further, the concept of a fictitious inverse problem (including its
components: the pseudo-prior, pseudo-likelihood, and pseudo-posterior)
introduced in \Cref{section:BIMC} is re-used here. 

First, the mixture is initialized using the BIMC procedure to ${Q}_{1} =
\mathcal{N}(\uq_{\mathrm{MAP}}, \boldsymbol{\Sigma}_{\mathrm{MAP}})$. For $k =
2, 3, \ldots$, the algorithm adds a new component $\mathcal{N}(\uq_k,
\boldsymbol{\Sigma}_k)$ to ${Q}_{k - 1}$ with some weight $\alpha_k$. The
center $\uq_k$ of every new component is found by solving an optimization
problem. The covariance $\boldsymbol{\Sigma}_k$ and weight $\alpha_k$ are
evaluated heuristically, but still depend on $\uq_k$.  The number of components
in the mixture grows till a termination condition based on user-specified
tolerances is satisfied.
\Cref{subsubsection:tuning_covariance,subsubsection:identify_wts} respectively
describe how $\boldsymbol{\Sigma}_k$ and $\alpha_k$ are chosen once $\uq_k$ is
known.  \Cref{subsubsection:identify_centers} describes how the center $\uq_k$
of a new component is chosen in the first place, and the condition that
terminates this stage of A-BIMC.

\subsubsection{Tuning covariance}
\label{subsubsection:tuning_covariance}
Suppose a new component of the Gaussian mixture, denoted $q_k$, is to be placed
at $\uq_k$. Then, for its covariance $\boldsymbol{\Sigma}_k$, we choose the
inverse of the Gauss-Newton Hessian matrix of the pseudo-posterior $-\log p(\uq
| y)$ at $\uq_k$. This way $\mathcal{N}(\uq_k, \boldsymbol{\Sigma}_k)$ crudely
approximates $q^*$ around $\uq_k$. The inverse of the Gauss-Newton Hessian, and
hence $\boldsymbol{\Sigma}_k$ is given by:

\begin{align}
\begin{split}
    \mathbf{H}_{\mathrm{GN}}^{-1} &= 
            \left(\frac{\nabla f(\uq_k) \nabla f(\uq_k)^T}{\sigma^2} 
            + \boldsymbol{\Sigma}_0^{-1}\right)^{-1}\\
         &= \boldsymbol{\Sigma}_0 - \frac{1}{\sigma^2 
            + \nabla f(\uq_k)^T\boldsymbol{\Sigma}_0\nabla f(\uq_k)}    
              \left(\boldsymbol{\Sigma}_0\nabla f(\uq_k)\right)
              \left(\boldsymbol{\Sigma}_0\nabla f(\uq_k)\right)^T,
\end{split}
\label{gaussNewtonHessInv}
\end{align}

and again an appropriate value for $\sigma$ remains is unknown. 
As in BIMC, this covariance matrix is parameterized by $\sigma^2$, the
variance of the pseudo-likelihood density. Note that unlike BIMC,
$\mathcal{N}(\uq_k, \boldsymbol{\Sigma}_k)$ isn't
dependent on $y$, the pseudo-data point: BIMC's dependence on $y$ arose out
of $\uq_{\mathrm{MAP}}$'s dependence on $y$. Now, however, the center of this
new component is fixed at $\uq_k$, severing its $y$-dependence. As a
result, the procedure for selecting $\sigma^2$ is slightly different in this
case. A-BIMC selects that value of $\sigma^2$ which minimizes the
Kullback-Leibler divergence between the linearized pushforwards of $q^*$ and
$q_k$. 
  
Suppose $f^{\mathrm{lin}}(\cdot) = \nabla f(\uq_k)^T (\cdot - \uq_k) + f(\uq_k)$
is the linearized approximation of $f$ at $\uq_k$
and$q^*_{\sharp}$ and $q_{k,\sharp}$ are the
pushforward densities  of $q^*$ and $q_k$ under $f^{\mathrm{lin}}$. Then,
the algorithm selects $\sigma_k^* = \argmin D_{\mathrm{KL}}(q^*_{\sharp} || q_{k,\sharp})$
for constructing $\mathbf{H}_{\mathrm{GN}}^{-1}$.
 The reason for resorting to the KL divergence between linearized
pushforwards is that an analytical expression for $D_{\mathrm{KL}}(q^* ||
q_k)$ is difficult to obtain even in the affine-Gaussian case,  let alone for
arbitrary $f$. On the other hand, $q^*_{\sharp}$ and
$q_{k,\sharp}$ being univariate densities, both
$D_{\mathrm{KL}}(q^*_{\sharp} || q_{k,\sharp})$ and
$\sigma^*_k$ have analytical expressions (provided in
\Cref{supplement_push_forward_kl_div,supplement_optimal_noise_var}).

%

\subsubsection{Identifying weights}
\label{subsubsection:identify_wts}

Whenever a new component $q_k = \mathcal{N}(\uq_k, \boldsymbol{\Sigma}_k)$ is
added to the IS mixture, A-BIMC readjusts the weight of all components in the
IS mixture so that they satisfy following rule:
  
\begin{align}
    \alpha_i \propto \int q_i(\uq) p(\uq) \mathrm{d}\uq,
    \label{eq:mix_weight_definition}
\end{align}

where $q_i$ is the $i$-th component of ${Q}_{k}$.
This heuristic for the mixture weights $\alpha_i$ is motivated by the
knowledge that components in regions where $p(\uq)$ is large should dominate in
the resultant IS mixture. If $p(\uq)$ is a Gaussian, then the integrals in
\Cref{eq:mix_weight_definition} can be evaluated analytically.

\subsubsection{Identifying centers}
\label{subsubsection:identify_centers}

The center of the first component in the mixture, $\uq_1$, is always fixed to be
the $\uqmap$, the MAP point of the pseudo-posterior $p(\uq | y)$. Now, suppose
that at the beginning of some
iteration $k$, the IS mixture has $k - 1$ components whose centers are at $\uq_1,
\uq_2, \ldots, \uq_{k - 1}$. Identifying the center of a new component, denoted
$\uq_{k}$, is a delicate balancing act. It must
\begin{enumerate}
    \item not be extremely close to an existing center $\uq_1, \ldots, \uq_{k -
1}$,
        as this makes $\uq_{k}$ redundant, wasting time and effort spent in
        discovering it,
    \item be in regions that correspond to high probability mass under $p(\uq)$
        so that the mixture resembles $q^*(\uq)$, but this
requirement can conflict with the previous one, and,
    \item must lie inside $f^{-1}(\DT)$ to maintain efficiency.
\end{enumerate}

A-BIMC constructs an optimization problem whose cost functional mathematically
captures these requirements. The solution of the optimization problem should
then be a suitable location to place $\uq_{k}$, as it represents the best
compromise between the potentially conflicting requirements. Requirement 1 is
represented mathematically by defining a fictitious repulsive force between
$\{\uq_1, \uq_2, \ldots, \uq_{k - 1}\}$ and $\uq_{k}$. This repulsive force
defines a fictitious potential energy, $\mathcal{U}(\uq_{k})$, that becomes the
first term in the cost functional of this optimization problem. 
Requirement 2 is met by leveraging $p(\uq)$ to
define an attractive potential, $\mathcal{P}(\uq_{k})$, which becomes the second
term in the cost functional. And the third requirement is enforced by adding the
constraint $f(\uq_{k}) \in \DT$.

The repulsive force is defined by imagining like-charged particles placed
at $\{\uq_1, \ldots, \uq_{k-1}\}$ and $\uq_{k}$. The charges at $\{\uq_1, \uq_2,
\ldots, \uq_{k-1}\}$ are considered stationary, while the charge at $\uq_{k}$ is
allowed to move. Then, the electrostatic repulsion between $\uq_{k}$ and
$\{\uq_1, \ldots, \uq_{k-1}\}$ forces $\uq_{k}$ away from the latter. Note that
it is possible to enrich the set $\{\uq_1, \ldots, \uq_{k-1}\}$ with more
(stationary) particles at locations where it is undesirable for a new component
to be placed. Henceforth, the set of undesirable locations for $\uq_{k}$ will be
denoted $\boldsymbol{\chi}^{(k)}_{\mathrm{fixed}} =
\{\uq_{\mathrm{fixed}}^{(i)}, i = 1, \ldots, n_{\mathrm{fixed}}\}$.  The set
$\boldsymbol{\chi}_{\mathrm{fixed}}^{(k)}$ always contains the existing particle
centers, $\uq_1, \ldots, \uq_k \subseteq
\boldsymbol{\chi}_{\mathrm{fixed}}^{(k)}$, but is allowed to contain any other
locations discovered by the algorithm (how
$\boldsymbol{\chi}^{(k)}_{\mathrm{fixed}}$ is modified is described in a later
section).

Since $\uq_{\mathrm{fixed}}^{(1)}, \ldots,
\uq_{\mathrm{fixed}}^{(n_{\mathrm{fixed}})}$
 are fixed, the electrostatic potential energy,
$\mathcal{U}$, only varies with $\uq_{k}$, and the contribution to potential 
energy due to the pairwise interactions of the members of
$\boldsymbol{\chi}_{\mathrm{fixed}}^{(k)}$
remains a constant. Subsequently, $\mathcal{U}$ is expressed only 
as a function of $\uq_{k}$ and the constant potential energy due to 
$\uq_{\mathrm{fixed}}^{(1)}, \ldots,
\uq_{\mathrm{fixed}}^{(n_{\mathrm{fixed}})}$ is omitted.

\begin{align*}
    \mathcal{U}(\uq_{k}) = \sum_{i = 1}^{n_{\mathrm{fixed}}}
    \frac{1}{\|\uq_{k} - \uq_{\mathrm{fixed}}^{(i)}\|}
\end{align*}

The attractive potential $\mathcal{P}(\uq)$ of a particle at some location $\uq$
is set to be $ -\beta \log p(\uq)$, where $\beta$ is a scale factor. Thus, the
total attractive potential energy due to this ``pseudo-prior'' potential energy
is $\mathcal{P}(\boldsymbol{\chi}_{\mathrm{fixed}}, \uq_{k}) = - \log p(\uq_k) - \beta \sum_{i =
1}^{n_{\mathrm{fixed}}} -\log p(\uq_{\mathrm{fixed}}^{(i)})$. 
Again, the constant contribution to
$\mathcal{P}(\boldsymbol{\chi}_{\mathrm{fixed}}, \uq_{k})$ by the
$n_{\mathrm{fixed}}$ fixed charges is ignored and the pseudo-prior potential energy
is expressed only as $\mathcal{P}(\uq_{k}) = - \beta \log(\uq_{k})$.  Hence, the
total potential energy of the system as a function of the center of the
prospective $k$-th component is:

\begin{align*}
    \mathcal{J}(\uq_{k}) &= \mathcal{U}(\uq_{k}) + \mathcal{P}(\uq_{k})
\end{align*}

Now, $\uq_{k}$ can be found by minimizing the potential energy of the system
provided the minima lies in the pre-image of the target interval $\DT$. This
third requirement is met by adding the constraint $y_k = f(\uq_{k})$, where $y_k
\sim \mathbb{U}(\DT)$ is randomly chosen from $\DT$ with uniform probability.
Thus, the following optimization problem is formulated:

\begin{align}
    \begin{split}
        \uq_{k} &= \argmin_{\uq \in \mathbb{R}^m}\, \sum_{i = 1}^{n_{\mathrm{fixed}}} \frac{1}{{\|\uq -
        \uq_{\mathrm{fixed}}^{(i)}\|}} - \beta \log p(\uq)\\
        \text{s.t.}\, y_k &= f(\uq_{k})
    \end{split}
    \label{minPotEnergy}
\end{align}

\begin{figure}[tbp]
    \centering
    \begin{subfigure}{0.5\textwidth}
        \includegraphics[width=0.9\textwidth]{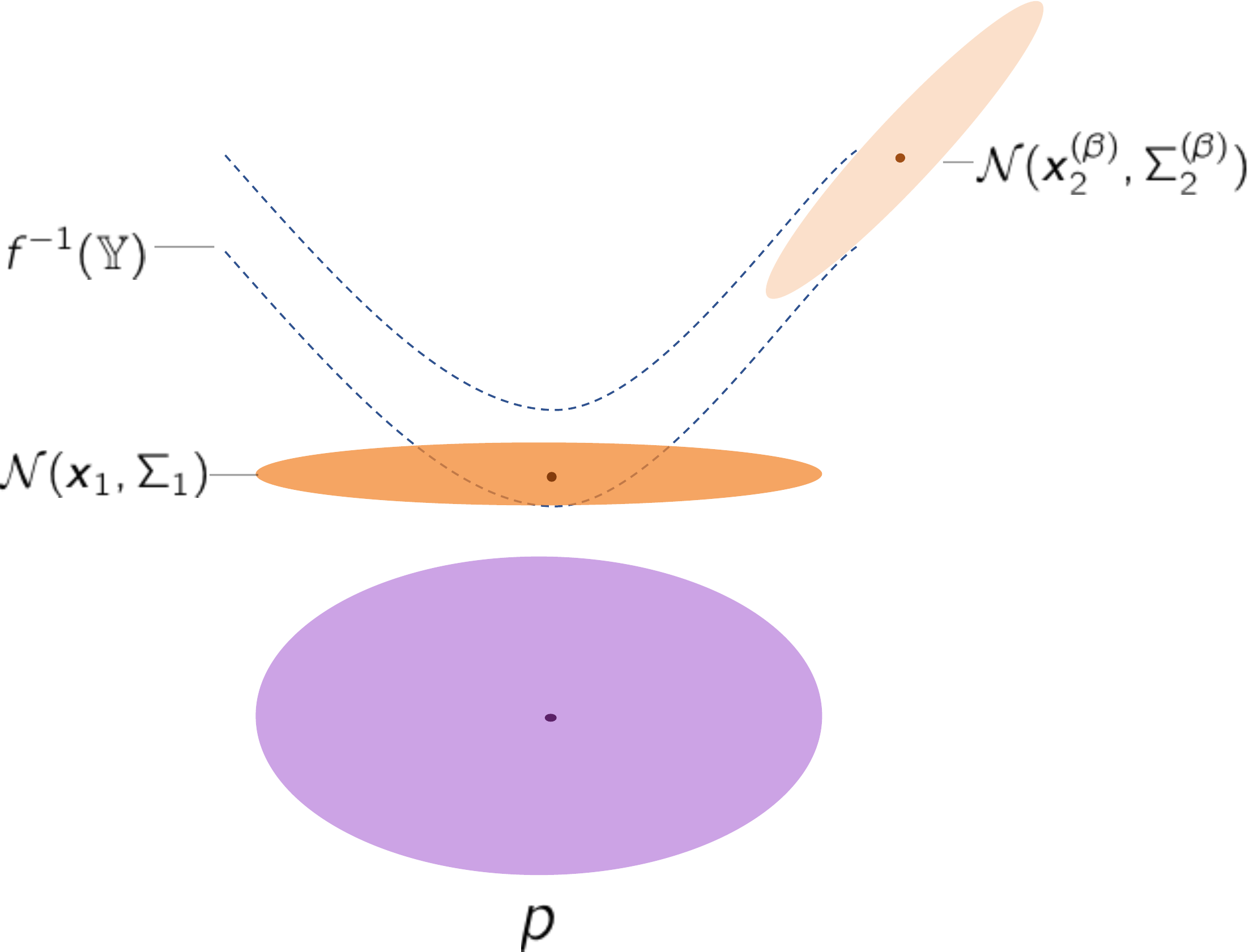}
        \caption{Small $\beta$}
    \end{subfigure}
    \begin{subfigure}{0.45\textwidth}
        \includegraphics[width=0.9\textwidth]{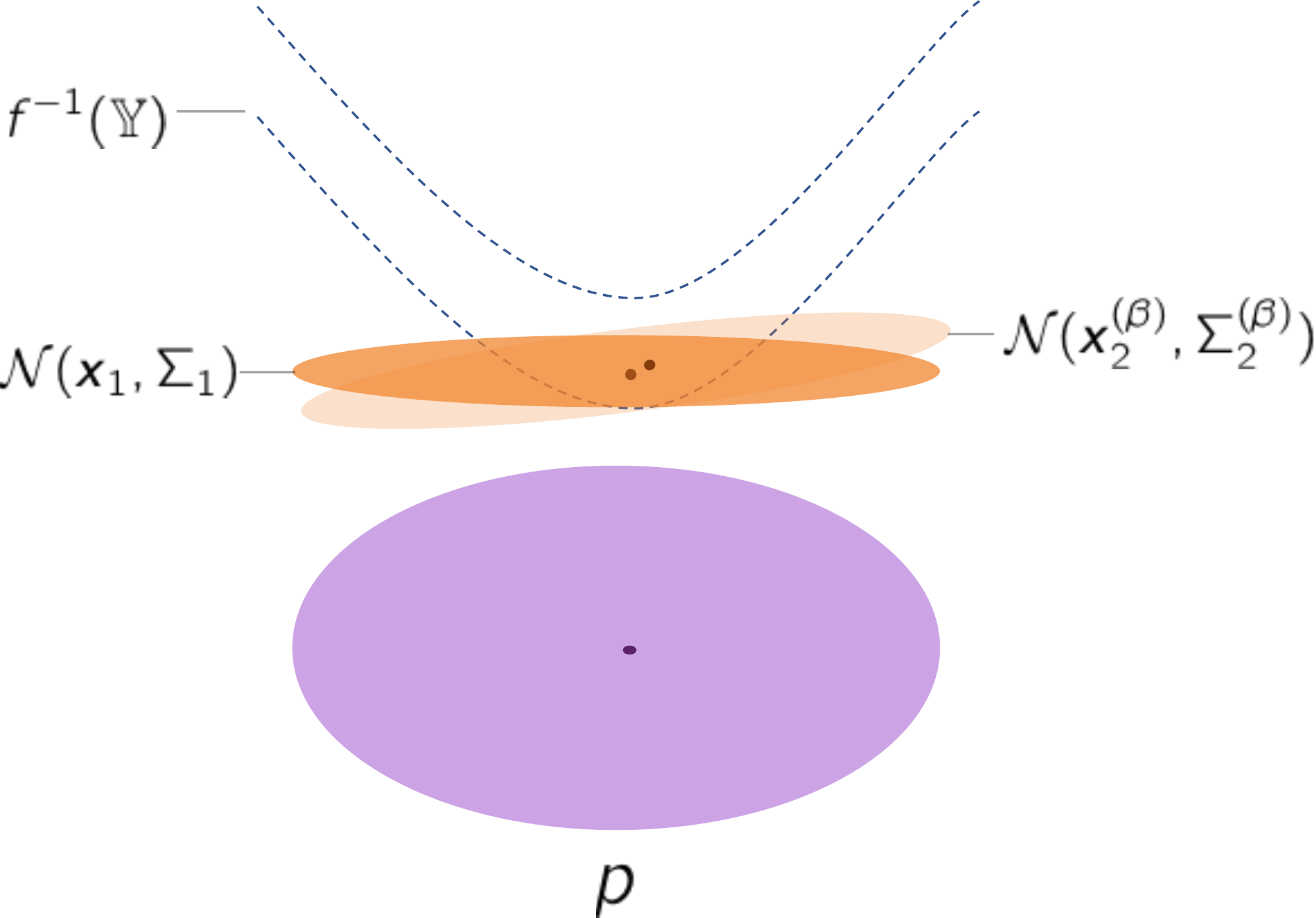}
        \caption{Large $\beta$}
    \end{subfigure}
    \caption{Illustration of the effect of the scale parameter $\beta$. A small
value of $\beta$ leads to the new component with center $\uq_2$ being placed far
away from $p(\uq)$. On the other hand, if $\beta$ is too large, then the new
component center will be undesirably close to $\uq_1$. The continuation
procedure described in \Cref{algo:continuation} is designed to yield an
intermediate value of $\beta$ that avoids both extremes.}
    \label{fig:continuation_illustration}
\end{figure}

The scale parameter $\beta$ controls the relative tradeoff between the repulsive
and the attractive potentials, and consequently, the spacing between $\uq_{k}$ 
and the fixed centers $\uq_{\mathrm{fixed}}^{(1)}, \ldots,
\uq_{\mathrm{fixed}}^{(n_{\mathrm{fixed}})}$.  To see this, imagine that the
system contains a single fixed charge at $\uq_1$,
and the optimization problem in \Cref{minPotEnergy} has been set up to find an
appropriate location for a new Gaussian component, whose center will be at
$\uq_2$ (see \Cref{fig:continuation_illustration} for an illustration).  
If $\beta$ is very small, then the total potential energy will be dominated by
electrostatic repulsion and the minimizer of $\mathcal{J}(\uq_2)$ will be far
away from $\uq_{1}$, possibly where $p(\uq_2)$ is small. On
the other hand, if $\beta$ is large, then the pseudo-prior potential energy will
dominate and the minimizer will be close to the nearest local minimum of
$p(\uq)$, preventing exploration of $f^{-1}(\DT)$ away from $\uq_1$.
Both these situations are undesirable, and hence, $\beta$ must be chosen
suitably. However, a suitable value of $\beta$ isn't known \emph{a priori}.
To ensure a spacing that's appropriate for the purposes of importance
sampling, A-BIMC employs a continuation scheme to implicitly fix $\beta$, which
is described next.

\paragraph{Parameter continuation}
The basic idea behind the continuation scheme in A-BIMC is the following. The
scheme tries several values of $\beta$ to minimize \Cref{minPotEnergy}, creates prospective IS
mixtures corresponding to each minimizer, and then selects one that is most
appropriate (in a sense that will be made precise shortly).

At the beginning of Stage-1, the minimum and maximum values for $\beta$,
$\beta_{\min}$ and $\beta_{\max}$, are computed by studying the relative
magnitudes of $\mathcal{U}(\uq)$ and $\mathcal{P}(\uq)$. Then, the scheme tries
five different values of $\beta$, logarithmically spaced between $\beta_{\min}$
and $\beta_{\max}$, starting at $\beta_{\min}$. Each value of $\beta$ leads to a
solution of the optimization problem in \Cref{minPotEnergy}, say
$\uq_k^{(\beta)}$. Through \Cref{gaussNewtonHessInv,eq:mix_weight_definition} 
each minimizer $\uq_k^{(\beta)}$
is in turn associated with a new prospective component
$\mathcal{N}(\uq_k^{(\beta)}, \boldsymbol{\Sigma}_k^{(\beta)})$, which can be
added to the IS mixture with weight $\alpha_k^{(\beta)}$ leading to a
prospective IS mixture, say $Q_k^{(\beta)}$. To decide whether $Q^{(\beta)}_k$ is
suitable or not, the scheme computes a sample estimate of the KL divergence
$D_{\mathrm{KL}}(Q_{k -1}|| Q_k^{(\beta)})$
between the current IS mixture $Q_{k - 1}$ and $Q_k^{(\beta)}$. 
The scheme tries larger and larger
values of $\beta$ as long as $D_{\mathrm{KL}}(Q_{k -1}|| Q_k^{(\beta)})$
increases with $\beta$. If from one iteration to the next, $D_{\mathrm{KL}}(Q_{k
-1}|| Q_k^{(\beta)})$ decreases, the scheme stops and appends the prospective
mixture $Q_k^{(\beta)}$ from the previous iteration.  Hence, implicitly, the
scheme selects that value of $\beta$ which makes the resulting IS mixture most
different from the current one. 

As there's no guarantee that the minimizer at a given $\beta$, $\uq_k^{(\beta)}$,
will get accepted, \Cref{minPotEnergy} is solved inexactly for each value of
$\beta$.  Further, the minimizer at some $\beta$, $\uq_k^{(\beta)}$, is supplied
as the initial guess for solving \Cref{minPotEnergy} at the next value of
$\beta$. Each $\uq_k^{(\beta)}$ is also added to the set of fixed charges,
$\boldsymbol{\chi}_{\mathrm{fixed}}$, ensuring that any future center is placed
away from all $\uq_k^{(\beta)}$, in addition to the center that was finally
appended to the IS mixture.  This prevents A-BIMC from exploring regions that it
has previously explored and forces it to venture into previously unexplored
regions. 

The next paragraph describes how \Cref{minPotEnergy} is solved at a given value
of $\beta$.

\paragraph{Solving the constrained optimization problem}
For a given value of $\beta$, the nonlinearly constrained 
optimization is solved as follows. First, \Cref{minPotEnergy} is transformed to an
unconstrained optimization problem via the Augmented
Lagrangian (AL) method. In the AL method, the equality constraint is enforced by
augmenting the Lagrangian with a term that penalizes violation of the
constraint. For \Cref{minPotEnergy}, choosing a quadratic penalty term yields the following
objective function:

\begin{align}
    \mathcal{L}(\uq) = \mathcal{U}(\uq) + \mathcal{P}(\uq)
                       + \frac{1}{2\delta}\left(y - f(\uq)\right)^2
                       - \lambda \left(y - f(\uq)\right)
    \label{modifiedAugmentedLagrangian}
\end{align}

In a traditional implementation of the AL algorithm, the constraint is enforced
by solving a sequence of optimization problems where the coefficient of the
penalty term, $1 / \delta$ and the Lagrange multiplier term, $\lambda$ are modified
simultaneously. Eventually, the minimizer of the unconstrained Augmented
Lagrangian converges to the minimizer of the constrained optimization
problem (see~\cite{nocedal2006numerical} for a basic implementation of
this algorithm). Although A-BIMC adopts the AL approach, for the purposes of IS, 
enforcing $y = f(\uq)$ exactly is unnecessary. It suffices if $f(\uq) \in \DT$. 
Starting from some initial guess for $\delta$
and $\lambda$, $\delta_0, \lambda_0$, the algorithm adopts the 
iterative approach only if the
corresponding minimizer, $\uq^{*}$, does not evaluate inside $\DT$. The
moment $f(\uq^*) \in \DT$, the AL iterations are terminated. 

\begin{algorithm}[tbp] \caption{{\textsc{Continuation}}}
    \begin{algorithmic}[1]
        \footnotesize

        \Require current IS mixture $Q_{k-1}$,
                 set of fixed charges
                 $\boldsymbol{\chi}_{\mathrm{fixed}}$,
                 set of values for the continuation parameter
                 $B = \{\beta_1, \beta_2, \ldots, \beta_{|B|}\}$,
                 initial guess for the quadratic penalty parameter
                 $\delta_{\mathrm{start}}$,
                 initial guess for the Lagrange multiplier
                 $\lambda_{\mathrm{start}}$,
                 constraint value $y$,
        \Ensure IS mixture enriched with a new component 
                $Q_{k}$,
                new fixed charges discovered
                $\boldsymbol{\chi}_{\mathrm{new}}$,
                sample estimate of ${D}_{\mathrm{KL}}(Q_{k-1} || Q_{k})$
        \State $\boldsymbol{\chi}_{\mathrm{new}} \gets \varnothing$
        \State Choose $\uq_{\mathrm{start}} \in \mathbb{R}^m$
        \For{$1 \le j \le |B|$}

            \State Using $\delta_{\mathrm{start}}, \lambda_{\mathrm{start}}, \uq_{\mathrm{start}}$ 
                   as starting guesses for $\delta, \lambda, \uq$, $\beta =
                   \beta_j$ and $\boldsymbol{\chi}_{\mathrm{fixed}}$ as the set of fixed charges, 
            \Statex \hspace{1.5em}follow \cref{algo:modifiedAugmentedLagrangian}
                    to minimize \Cref{minPotEnergy}
            \Statex \hspace{1.5em}$\uq_{\mathrm{new}} 
                                   \gets \textsc{ModifiedAugmentedLagrangian}(y, \DT, 
                                                                              \lambda_{\mathrm{start}}, 
                                                                              \delta_{\mathrm{start}}, \uq_{\mathrm{start}})$ 

            \State Using $\uq = \uq_{\mathrm{new}}$ in \Cref{supplement_optimal_noise_var} 
                   compute a suitable value of the pseudo-likelihood variance $\sigma_{\mathrm{new}}$

            \State Using $\uq = \uq_{\mathrm{new}}, \sigma = \sigma_{\mathrm{new}}$ 
                   in \Cref{gaussNewtonHessInv}, compute $\mathbf{H}_{\mathrm{GN}}^{-1}$ and assign it to 
                   $\boldsymbol{\Sigma}_{\mathrm{new}}$

            \State Obtain a prospective IS distribution
                   $Q_{k}^{(\beta_j)}$ by adding 
                   $\mathcal{N}(\uq_{\mathrm{new}}, \boldsymbol{\Sigma}_{\mathrm{new}})$ 
                   to $Q_{k - 1}$             

            \State $\uq_{\mathrm{start}} \gets \uq_{\mathrm{new}}$

            \State $\boldsymbol{\chi}_{\mathrm{new}} 
                     \gets \boldsymbol{\chi}_{\mathrm{new}} \cup \{\uq_{\mathrm{new}}\}$

            \State Compute a sample estimate of
                   $D_{\mathrm{KL}}(Q_{k-1} || Q_{k}^{(\beta_j)})$ 
                   and assign it to $\hat{D}_{\mathrm{KL}}[j]$

            \If{$\hat{D}_{\mathrm{KL}}[j] > \hat{D}_{\mathrm{KL}}[j - 1]$}
                \Statex \hspace{3em}\emph{Try a larger value of $\beta$}
                \State \textbf{continue}
            \Else 
                \Statex \hspace{3em}\emph{$Q_{k}^{(\beta_{j-1})}$ was more different from $Q_{k-1}$ than $Q_k^{(\beta_j)}$}
                \State $Q_k \gets Q_{k}^{(\beta_{j-1})}$ 
                \State \Return $(Q_k, \boldsymbol{\chi}_{\mathrm{new}}, \hat{D}_{\mathrm{KL}}[j - 1])$
            \EndIf
        \EndFor
        \Statex \emph{$\hat{D}_{\mathrm{KL}}(Q_{k-1} || Q_k^{(\beta_j)})$ 
                increases with $j$, hence $Q_k^{(\beta_{|B|})}$ is most different from $Q_{k-1}$}
        \State $Q_k \gets Q_{k}^{(\beta_{|B|})}$ 
        \State \Return $(Q_k, \boldsymbol{\chi}_{\mathrm{new}}, \hat{D}_{\mathrm{KL}}[|B|])$
    \end{algorithmic}
\label{algo:continuation}
\end{algorithm}

This concludes the description of how A-BIMC finds centers of new components.
The continuation and modified Augmented Lagrangian
algorithms described above are reproduced in pseudo-code in 
\Cref{algo:continuation,algo:modifiedAugmentedLagrangian}. 
The next subsection describes the termination criterion employed in Stage-1.

\begin{algorithm}[t] \caption{{\textsc{ModifiedAugmentedLagrangian}}}
  \begin{algorithmic}[1]
    \footnotesize
      \Require 
                    desired constraint value $y$,
                    target interval $\DT$, 
                    starting guess for Lagrange multiplier $\lambda_{\mathrm{start}}$, 
                    starting guess for penalty coefficient $\delta_{\mathrm{start}}$, 
                    starting guess for the optimizer $\uq_{\mathrm{start}}$

      \Ensure Approximate minimizer of \Cref{minPotEnergy} $\uq^*$
    
      \State $\delta \gets \delta_{\mathrm{start}}$
      \State $\lambda \gets \lambda_{\mathrm{start}}$
      \State $\uq^* \gets \uq_{\mathrm{start}}$

      \While{$f(\uq^*) \not \in \DT$}
        \State Starting from $\uq_{\mathrm{start}}$, compute the approximate minimizer of
        $\mathcal{L}(\cdot; \delta, \lambda)$ and assign it to $\uq^*$        

        \State $\lambda \gets \lambda - \frac{1}{\delta} \left(y - f(\uq^*)\right)$

        \State $\delta \gets 0.5 \delta$

        \State $\uq_{\mathrm{start}} \gets \uq^*$

      \EndWhile

    \State \Return $\uq^*$
  \end{algorithmic}
\label{algo:modifiedAugmentedLagrangian}
\end{algorithm}

\subsubsection{Termination}
\label{subsubsection:termination}

The procedure described above adds new components to the IS mixture in every
iteration. Hence, the number of components in the mixture grows steadily. In
order to decide whether the IS mixture is sufficiently stable, or conversely, if more
components are required, the algorithm keeps track of how the IS mixture changes
across iterations. The change in the IS mixture is measured in terms of a
quantity that resembles perplexity, $\zeta = \exp(- D_{\mathrm{KL}}(Q_{k-1}
|| Q_k))$, where $Q_k$ is the IS mixture at the $k$-th iteration.
Since $D_{\mathrm{KL}}\left(Q_{k - 1} || Q_k\right)\ge 0$, $\zeta
\in (0, 1]$. 
The KL divergence here is once again a sample estimate, and is in fact the same
quantity computed in the continuation scheme.  Stage-1 is terminated if relative change in 
$\zeta$ falls below a user specified
relative change threshold, or its absolute value exceeds a user-specified
absolute change threshold. 

\subsubsection{Summary}
\label{subsubsection:stage_1_summary}
In summary, Stage-1 adaptively explores the pre-image of the interval
$f^{-1}(\DT)$, and yields an IS distribution with the following form:

\begin{align}
    Q(\uq) = \sum_{k = 1}^{K} \alpha_k \mathcal{N}\left(\uq_k,
    \boldsymbol{\Sigma}_k\right).
\end{align}

The centers of the Gaussian mixture components are obtained by solving a
sequence of optimization problems, the weights $\alpha_k$ are heuristically
evaluated, and the covariances $\boldsymbol{\Sigma}_k$  are local 
Gauss-Newton Hessians. This procedure is heuristic, with no guarantees on the
quality of the final Gaussian mixture for importance sampling. Hence, Stage-2 of
A-BIMC further modifies the Gaussian mixture that results at the end of Stage-1.
Details of Stage-2 are provided next.

\subsection{Stage-2}
\label{subsection:stage_2}

Stage-2 of A-BIMC modifies the Gaussian mixture obtained at the end of Stage-1
via the Mixture Population Monte Carlo algorithm. This subsection begins with a
brief introduction to the MPMC algorithm, before describing how it's employed
within A-BIMC. 

\subsubsection{Mixture Population Monte Carlo method}
\label{subsubsection:mpmc}

The MPMC algorithm generates an importance sampling
 distribution to approximate some target
distribution $\pi$ using the following approach. Among all
possible Gaussian mixtures with $K$ components, it seeks that 
mixture $Q^*$ which is closest in KL divergence to $\pi$:

\begin{align}
   Q^* = \argmin_{Q \in \mathcal{Q}_K} D_{KL}(\pi || Q),
\label{mpmc_q_star}
\end{align}

where $\mathcal{Q}_K$ denotes the family of $K$-component Gaussian mixtures.
Here, $K$ is assumed to be known and fixed. 
Importance sampling is then performed using $Q^*$. 

Seeking $Q^*$ is equivalent to seeking its $K$ mixture weights $(\omega_1^*,
\ldots, \omega_K^*)$ and the means $\boldsymbol{m}_k^*$ and covariances
$\mathbf{C}_k^*$ of its components. \Cref{mpmc_q_star} can be restated as:

\begin{align}
    (\omega_1^*, \ldots, \omega_K^*, 
     \boldsymbol{m}_1^*, \ldots, \boldsymbol{m}_K^*, 
     \mathbf{C}_1^*, \ldots, \mathbf{C}_K^*) 
                    = \argmin_{\substack{\sum_k \omega_k = 1\\
                               \mathbf{C}_k \succ \boldsymbol{0}\\
                               \boldsymbol{m}_k \in \mathbb{R}^m}} 
                      D_{\mathrm{KL}} \left(\pi 
                            || \sum_k \omega_k \mathcal{N}(\boldsymbol{m}_k, \mathbf{C}_k)\right),
\end{align}
%
%

which, from the definition of KL divergence, is equivalent to 

\begin{align}
    (\omega_1^*, \ldots, \omega_K^*, 
     \boldsymbol{m}_1^*, \ldots, \boldsymbol{m}_K^*, 
     \mathbf{C}_1^*, \ldots, \mathbf{C}_K^*) 
                    = \argmax_{\substack{\sum_k \omega_k = 1\\
                               \mathbf{C}_k \succ \boldsymbol{0}\\
                               \boldsymbol{m}_i \in \mathbb{R}^m}} 
                      \int \pi(\uq) \log\left( \sum_k \omega_k 
                           \mathcal{N}(\boldsymbol{m}_k, \mathbf{C}_k)\right)
                      \mathrm{d}\uq.
\label{mpmc_mle}
\end{align}

Here, the notation $\mathbf{C}_k \succ \boldsymbol{0}$ implies that
$\mathbf{C}_k$ is positive definite. The constraints $\sum_k \omega_k =
1$ and $\mathbf{C}_k \succ \boldsymbol{0}$ are necessary for $Q^*$ to
remain a valid Gaussian mixture.  

The optimization problem in \Cref{mpmc_mle}
strongly resembles maximum likelihood estimation of the  mixture parameters, for
which the Expectation Maximization (EM) algorithm~\cite{dempster1977maximum}
 is usually employed. As a
result, MPMC also closely follows the EM algorithm, except that the sum over
i.i.d. data is replaced with an integration over $\pi(\uq)$.  Like EM, MPMC is
an iterative algorithm. Starting from some initial mixture, the mixture
weights, and the means and covariances of its components are
updated in every iteration.  The update expressions involve
evaluating expectations with respect to $\pi$. These integrals are in turn
evaluated using autonormalized importance sampling, and hence, the algorithm
only requires the ability to evaluate $\pi(\uq)$ up to a constant.
 
The progress of the algorithm is tracked by measuring the normalized perplexity
of the IS weights used in computing the update integrals.  The algorithm is
terminated when, at some iteration, the normalized perplexity stagnates, or
becomes sufficiently close to 1. The Gaussian mixture obtained at the end of
this iterative procedure is then used for importance sampling of $\pi$. 
The next subsection describes how MPMC is used
within A-BIMC.

\subsubsection{MPMC and A-BIMC}
\label{subsubsection:mpmc_and_a_bimc}
As the mixture obtained at the end of Stage-1, $Q_K$, only roughly
approximates $q^*$, it is further refined using the MPMC algorithm. 
One way to do this could be to set 
the target distribution $\pi$ to $q^*$, and
supplying $Q_K$  as the initial guess to the algorithm. However, this would require additional evaluations
of $f$, driving up the computational cost of the method.  A-BIMC instead
constructs a cheap surrogate for $f$, denoted $f_{\mathrm{surrogate}}$, and
then sets $\pi(\uq) = 1_{\mathbb{Y}}(f_{\mathrm{surrogate}}(\uq))p(\uq)$ in order to
tune $Q_K$. The surrogate $f_{\mathrm{surrogate}}$ is constructed as
follows.

\begin{algorithm}[t]\caption{{\textsc{ForwardSurrogate}}}
    \begin{algorithmic}[1]
        \footnotesize

        \Require input-output map $f(\cdot)$,
                 set of fixed charges $\boldsymbol{\chi}_{\mathrm{fixed}}$,
                 query location $\uq$
        \Ensure A surrogate $f_{\mathrm{surrogate}}(\uq)$ for $f(\uq)$
        \State Find that member of $\boldsymbol{\chi}_{\mathrm{fixed}}$ to which
               $\uq$ is closest and assign it to $\uq_{{\min}}$ 
        \State $f_{\mathrm{surrogate}}(\uq) \gets 
                f(\uq_{{\min}}) + \nabla f(\uq_{{\min}})^T(\uq - \uq_{{\min}})
                + 0.5 (\uq - \uq_{{\min}})^T \nabla^2 f(\uq_{{\min}}) (\uq - \uq_{{\min}})$
        \State \Return $f_{\mathrm{surrogate}}(\uq)$
    \end{algorithmic}
    \label{algo:fsurrogate}
\end{algorithm}

During the continuation phase in Stage-1, A-BIMC saves
$(f(\uq_{\mathrm{fixed}}^i), \nabla f(\uq_{\mathrm{fixed}}^i), \nabla^2
f(\uq_{\mathrm{fixed}}^i))$ at each fixed charge $\uq_{\mathrm{fixed}}^i$. These
three quantities can be used to construct a second-order Taylor series expansion
around each fixed charge $\uq_{\mathrm{fixed}}^i$. In order to approximate
$f(\uq)$ at some $\uq$, the surrogate first finds the fixed charged in
$\boldsymbol{\chi}_{\mathrm{fixed}}$ which is closest in Euclidean distance to
$\uq$, let's
say $\uq_{\min}$. Then, $f(\uq)$ is approximated as:
$f(\uq_{\min}) + \nabla_{\uq} f(\uq_{{\min}}) ^T (\uq - \uq_{{\min}}) +
0.5 (\uq - \uq_{{\min}})^T \nabla_{\uq}^2 f(\uq_{{\min}}) (\uq -
\uq_{{\min}})$. Pseudo-code for this procedure is provided in
\Cref{algo:fsurrogate}.

The surrogate constructed here is merely a suggestion. If a better, more
principled surrogate is available, then that may be used in constructing the
target distribution $\pi$ in MPMC. Irrespective of how the surrogate is
constructed, it is only used while tuning $Q_K$ via MPMC and nowhere
else. The actual function $f$ is used for the final
importance sampling stage. This concludes the presentation of the 
A-BIMC methodology. The next subsection offers a summary of our algorithm.

\subsection{Summary}
\label{subsection:methodology_summary}

In summary, our algorithm involves the following steps:
\begin{itemize}
    \item \emph{Constructing the IS distribution}: First, the importance
           sampling distribution is constructed. This itself is a two stage
           process. In Stage-1, a sequence of optimization problems is solved to adaptively
           discover $f^{-1}(\DT)$. The resulting sequence of local minima are
           consolidated into a Gaussian mixture using heuristic estimates of the
           covariance and mixture weights. Stage-2 involves tuning the
           Gaussian mixture  that results
           from the first stage against $q^*(\uq)$ via MPMC. This involves 
           evaluations of $q^*(\uq)$, which in turn requires evaluations of
           $f(\uq)$. While tuning with MPMC, a cheap surrogate for $f$ is used, in
           order to keep function evaluations low. For simplicity, we will 
           denote the importance sampling mixture at the end of this two stage
           procedure using just $Q$.

    \item \emph{Sampling the IS distribution}: Finally, the mixture that is
          obtained at the end of the previous stage is used as an importance 
           sampling density to evaluate the rare-event
           probability. In this step, A-BIMC uses
           true evaluations of $f(\uq)$, instead of any surrogates.
\end{itemize}

Pseudo-code for the A-BIMC methodology is provided in \Cref{algo:adaptiveBIMC}.

\begin{algorithm}[h] \caption{{\textsc{AdaptiveBIMC}}}
    \begin{algorithmic}[1]
        \footnotesize
        \Require 
                    input-output map $f(\cdot)$, 
                    target interval $\DT$,
                    mean $\uq_0$ and covariance $\boldsymbol{\Sigma}_0$ of $p(\uq)$,
                    absolute and relative tolerances $\epsilon_{\mathrm{abs}}$ and $\epsilon_{\mathrm{rel}}$,
                    number of samples $N$
        \Ensure 
                        importance sampling estimate of the rare event probability $\tilde{\mu}^N$,
                        the associated relative root mean square error $\tilde{e}_{\mathrm{RMS}}^N$

        \Statex
        \Statex \emph{Construct the IS distribution}
        \Statex \emph{Stage-1}
        \State Find $\uq_{\mathrm{start}}$ such that $f(\uq_{\mathrm{start}})
\in \DT$

        \State Linearize $f(\uq)$ around $\uq_{\mathrm{start}}$, 
               and use this linearized approximation in Equation (SM7)  
               in~\cite{wahal2019bimc_supplement}  to compute the optimum 
               pseudo-likelihood variance ${\sigma^*}^2$ and optimum pseudo-data $y^*$

        \State Construct the pseudo-posterior $p(\uq | y)$ using $y = y^*$ and
               $\sigma = \sigma^*$ and compute its MAP point, assign it to $\uqmap$

        \State Using $\uq = \uqmap, \sigma = \sigma^*$ 
                   in \Cref{gaussNewtonHessInv}, compute $\mathbf{H}_{\mathrm{GN}}^{-1}$ and assign it to 
                   $\boldsymbol{\Sigma}_{\mathrm{MAP}}$

        \State $Q_1 \gets \mathcal{N}(\uqmap, \boldsymbol{\Sigma}_{\mathrm{MAP}})$

        \State $\boldsymbol{\chi}_{\mathrm{fixed}} \gets \{\uqmap\}$

        \State Choose initial guess for the Lagrange multiplier $\lambda_{\mathrm{start}} \in \mathbb{R}$

        \State Choose initial guess for the penalty $\delta_{\mathrm{start}} \in \mathbb{R}^+$

        \State Choose values for the scale parameter $B \gets \{\beta_1, \beta_2, \ldots, \beta_{|B|}\}$.

        \For{$k = 2, 3, \ldots$}
            \State Sample $y$ uniformly from $\DT$

            \State Add a new component to $Q_{k - 1}$ using \Cref{algo:continuation}.
                   As a side effect, obtain 
                   $\hat{D}_{\mathrm{KL}}(Q_{k-1} || Q_k)$ 
                   and $\boldsymbol{\chi}_{\mathrm{new}}$
            \Statex \hspace{1.5em}$ (Q_{k}, \boldsymbol{\chi}_{\mathrm{new}}, \hat{D}_{\mathrm{KL}}) 
                     \gets \textsc{Continuation}(Q_{k-1}, 
                                                 \delta_{\mathrm{start}}, 
                                                 \lambda_{\mathrm{start}}, 
                                                 y,
                                                \boldsymbol{\chi}_{\mathrm{fixed}}, 
                                                 B)$
            \State $\boldsymbol{\chi}_{\mathrm{fixed}} 
                    \gets \boldsymbol{\chi}_{\mathrm{fixed}} \cup \boldsymbol{\chi}_{\mathrm{new}}$
            \State $\zeta[k] \gets \exp(-\hat{D}_{\mathrm{KL}}(Q_{k-1} || Q_k))$
            
            \If{$ \zeta[k] > \epsilon_{\mathrm{abs}}$ \OR $|\zeta[k] - \zeta[k
- 1]|/ \zeta[k] < \epsilon_{\mathrm{rel}}$}
                break
            \EndIf
            
        \EndFor
        \Statex \emph{Assume Stage-1 required ${K}$ iterations. Denote the mixture obtained ${Q_K}$.}
        \Statex 
        \Statex \emph{Stage-2}
        \State Using $\boldsymbol{\chi} = \boldsymbol{\chi}_{\mathrm{fixed}}$, 
               employ \Cref{algo:fsurrogate} to construct a surrogate $f_{\mathrm{surrogate}}$ for $f$.
        \State Define $q^*_{\mathrm{surrogate}} \propto \ind_{\DT}(f_{\mathrm{surrogate}})p$.
        \State Tune the IS mixture $Q_{K}$ to match $q^*_{\mathrm{surrogate}}$ using the 
               MPMC algorithm, and assign the tuned mixture to $Q$
        \Statex
        \Statex \emph{Compute the rare event probability}
        \State Generate $N$ samples from the tuned mixture $\boldsymbol{X}_1,
\ldots,  \boldsymbol{X}_N \overset{\mathrm{i.i.d.}}\sim Q$
        \State  $w_i \gets \frac{\ind_{\DT}(f(\boldsymbol{X}_i))p(\boldsymbol{X}_i)}{Q(\boldsymbol{X}_i)}, 
                i = 1, \ldots, N$
        \State $\tilde{\mu}^N \gets \frac{1}{N}\sum_i w_i$
        \State $\tilde{e}^N_{\mathrm{RMS}} \gets \sqrt{\sum_i (w_i - \tilde{\mu}^N)^2} / N\tilde{\mu}^N$
        \State \Return $\tilde{\mu}^N, \tilde{e}_{\mathrm{RMS}}^N$
    \end{algorithmic}
    \label{algo:adaptiveBIMC}
\end{algorithm}

\section{Experiments}
\label{section:experiments}
This section presents results from a variety of numerical experiments. These
experiments have been designed to assess how A-BIMC performs, expose the
conditions that affect its performance, and in the process, unearth any
potential limitations. To provide a context in which the results can be
understood, the following remarks are in order.

\paragraph{Measuring performance}
Objectively assessing the performance of an IS scheme solely from samples is
still an unresolved question. Appealing to Chebyshev's
inequality~\cite{vershynin2018high} reveals a simple (but potentially
restrictive) criterion: a low relative RMSE is sufficient for $\tilde{\mu}^N$
to be close to $\mu$ with high probability.  As a diagnostic quantity, the
relative RMSE also appears in disguise elsewhere; for instance, as the
$\chi^2$-divergence between $q^*$ and $Q$ in
\cite{sanz-alonso2018importance,agapiou2017importance}, the Effective Sample
Size~\cite{mcbook,agapiou2017importance,sanz-alonso2018importance} or the
second moment of the IS weights \cite{agapiou2017importance}.  These studies
have established the central role that the relative RMSE and its variants play
in controlling the error between $\tilde{\mu}^N$ and $\mu$. For this reason,
A-BIMC's performance is primarily measured in terms of the relative RMSE. Since
the true relative RMSE is unavailable, it is approximated via samples. Let
$\boldsymbol{X}_1, \ldots, \boldsymbol{X}_N \overset{i.i.d.}{\sim} Q(\uq)$ be
$N$ samples from $Q$. The relative RMSE, defined in Equation (4) in Part I, can
be estimated via samples as follows:

\begin{align*}
    \tilde{e}_{\mathrm{RMS}}^N &= \sqrt{\frac{\sum_i w_i^2}{\left(\sum_i w_i\right)^2} 
                                        - \frac{1}{N}}
\end{align*}

where $w_i = \ind_{\DT}(f(\boldsymbol{X}_i))p(\boldsymbol{X}_i) /
Q(\boldsymbol{X}_i)$. The sample estimate of the relative RMSE has been shown to
be an inadequate indicator of performance in \cite{chatterjee2018sample}.
However, in the large $N$ asymptotic regime, $\tilde{e}_{\mathrm{RMS}}^N$
approximates $e_{\mathrm{RMS}}$ well, and therefore $\tilde{e}_{\mathrm{RMS}}$
and its variants are expected to perform adequately as diagnostics
\cite{agapiou2017importance}. In addition to the sample estimate of the relative
RMSE, the ESS is also reported. Taking inspiration from the the
function-specific ESS defined in~\cite{mcbook}, the following rare-event
specific ESS is employed:

\begin{align*}
    \mathrm{ESS} = \frac{1}{\sum_i \bar{w}_i^2},
\end{align*}

where $\bar{w}_i = \ind_{\DT}(f(\uq_i))p(\uq_i) / Q(\uq_i) / \sum_j
\ind_{\DT}(f(\uq_j))p(\uq_j) / Q(\uq_j)$. \Cref{supplement:relationship_erms_ess}
establishes that there exists a one-to-one correspondence between
$\tilde{e}_{\mathrm{RMS}}^N$ and the rare-event specific ESS. Hence, the ESS is only
reported for a few representative experiments.  

\paragraph{Measuring function evaluations}
Along with performance estimates, the number
of function evaluations required during Stage-1 of A-BIMC are also reported.
Recall that Stage-1 solves a sequence of optimization problems to iteratively
explore $f^{-1}(\mathbb{Y})$. Here, it is assumed that A-BIMC has access to
an oracle who, when queried, returns $f(\uq), \nabla f(\uq)$ and $\nabla^2
f(\uq)$ at some $\uq$. The function evaluations reported correspond to the
number of queries that Stage-1 made to this oracle. It is further assumed that
the cost of evaluating the surrogate $f_{\mathrm{surrogate}}$ during Stage-2
is negligible, and is not reported.

\paragraph{Forward maps}
As proof-of-concept, the toy periodic map presented in \Cref{section:BIMC} is revisited.
In addition, A-BIMC is tested on synthetic maps drawn from the following
classes of functions. Maps that belong to these classes are sufficiently rich
to expose both the advantages and drawbacks of A-BIMC. The manner in which
these maps are actually constructed is detailed while describing the setup for
each experiment. The function classes are:

\begin{itemize}
    \item Quadratic polynomials: The class of functions $f: \mathbb{R}^m \rightarrow
           \mathbb{R}$ of the form: $f(\uq) = \uq^T \mathbf{H} \uq
           + \vect{b}^T\uq + c$ for some $\mathbf{H} \in \mathbb{R}^{m \times m}$,
           $\vect{b} \in \mathbb{R}^m$ and $c \in \mathbb{R}$. In other words, 
           $f$ is nonlinear but with constant curvature. Note that for this
           class of functions, $f = f_{\mathrm{surrogate}}$.
    \item Cubic polynomials: The class of functions $f: \mathbb{R}^m \rightarrow
           \mathbb{R}$ of the form: $f(\uq) =
            \boldsymbol{\mathcal{S}}:\boldsymbol{\mathcal{X}} + \uq^T \mathbf{H} \uq
           + \vect{b}^T\uq + c$ for some $\boldsymbol{\mathcal{S}} 
            \in \mathbb{R}^{m\times m \times m}$, 
            $\mathbf{H} \in \mathbb{R}^{m \times m}$,
           $\vect{b} \in \mathbb{R}^m$ and $c \in \mathbb{R}$ and where the
           tensor contraction $\boldsymbol{\mathcal{S}} : \boldsymbol{\mathcal{X}}$ is defined as
           $\boldsymbol{\mathcal{S}}:\boldsymbol{\mathcal{X}} = \mathcal{S}_{ijk} x_i x_j x_k$. 
           Now, $f$ is nonlinear and again possesses varying curvature since
           $\nabla^2 f(\uq)$ varies with $\uq$.
\end{itemize}                    

\paragraph{Implementation}
A-BIMC has been implemented in MATLAB. The optimization problems posed in
\Cref{subsubsection:identify_centers} are solved using the Trust Region algorithm,
as implemented in MATLAB's in-built \texttt{fminunc} routine. In the
implementation, the gradients and the Hessian of the objective function are
passed to the optimizer. Because each individual optimization problem
the continuation phase of Stage-1 does not need to be solved exactly, 
the optimizer is terminated if the gradient is reduced by a factor of
$10^{-2}$ or if the number of iterations exceeds 5. 

\paragraph{Algorithmic parameters}
Unless otherwise noted, A-BIMC is run using the following
parameters: $\epsilon_{\mathrm{abs}} = 1 - 10^{-3}$ and
$\epsilon_{\mathrm{rel}} = 10^{-3}$. For Stage-2, the specific
variant of MPMC used is termed the Rao-Blackwellized version in
\cite{cappe2008adaptive}. By default, MPMC uses
$N_{\mathrm{MPMC}} = 10^6$ samples per iteration and a maximum of 50
iterations.

The following subsections present the results of numerical experiments. The
first experiment is a proof-of-concept that establishes whether all of BIMC's
drawbacks have been rectified. The next experiment subjects A-BIMC to problems
of increasing dimensionality. This is followed by a study of its performance as
the rarity of the problem is increased.

\begin{figure}[htbp]
    \centering
    \begin{subfigure}{0.4\textwidth}
        \includegraphics[width=\textwidth]{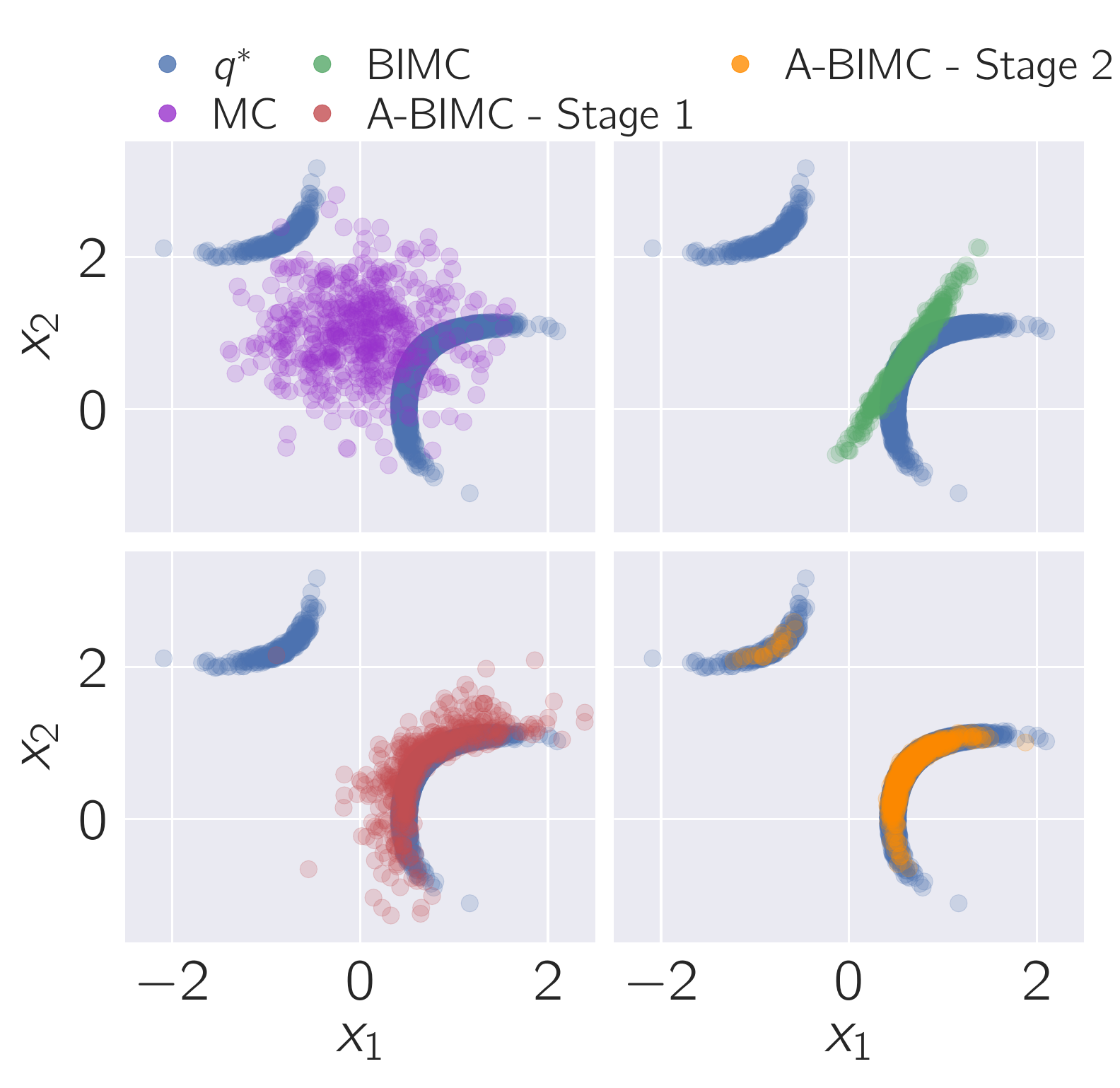}
        \caption{Samples}
        \label{subfig:toy_problem_compare_samples}
    \end{subfigure}
    \begin{subfigure}{0.42\textwidth}
        \includegraphics[width=\textwidth]{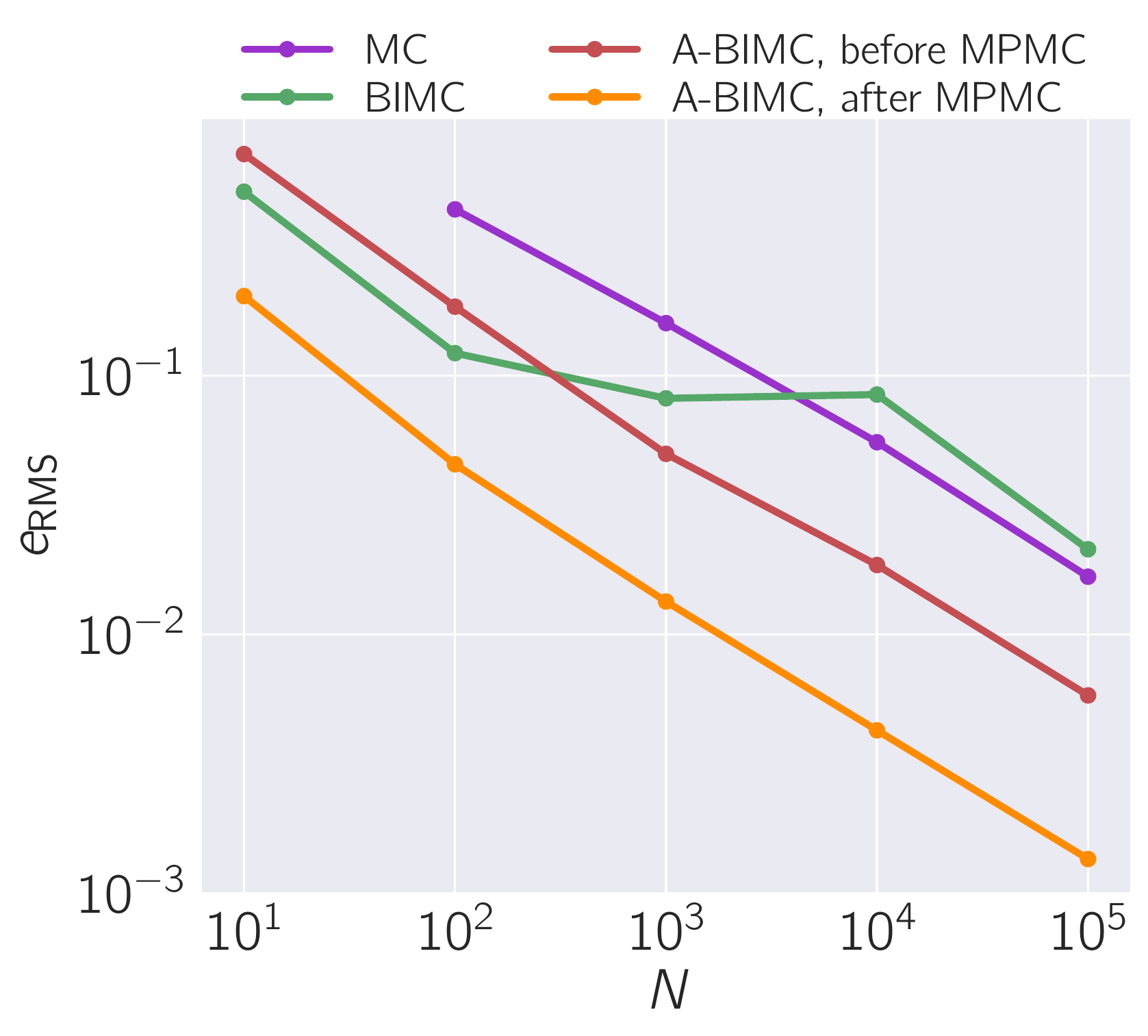}
        \caption{Relative RMSE versus number of samples $N$}
        \label{subfig:toy_problem_compare_err}
    \end{subfigure}
\caption{A comparison of simple Monte Carlo, BIMC, and the various stages of
A-BIMC using the toy problem 
- samples (\Cref{subfig:toy_problem_compare_samples}) and relative error
(\Cref{subfig:toy_problem_compare_err}).
\Cref{subfig:toy_problem_compare_samples} shows 5000 samples from $q^*$ and 500
from each estimator. While the IS mixture at the end of
Stage-1 of A-BIMC is an improvement over BIMC, tuning via MPMC drastically
improves its quality.}
\label{fig:toy_problem_compare}
\end{figure}

\subsection{A toy problem}
\paragraph{Purpose} This is a proof-of-concept experiment whose aim is to establish
that A-BIMC leads to a consistent importance sampling distribution and that all
of the drawbacks listed in \Cref{section:BIMC} have been fixed. 

\paragraph{Setup}
As noted earlier, for this toy problem, the forward map is defined as 
 $f(\uq) := \sin(x_1) \cos(x_2)$. The nominal distribution is 
$p(\uq) = \mathcal{N}(\uq_0, \boldsymbol{\Sigma}_0)$ where $\uq_0 = [0, 1]^T$ 
and $\boldsymbol{\Sigma}_0 = 0.3
\mathbf{I}_2$. A Monte Carlo simulation using $10^6$ samples results in the
following 99\% confidence interval: $3.413\times 10^{-2} \pm 4.68 \times
10^{-4}$.

\paragraph{Results and discussion}
\Cref{fig:toy_problem_compare} compares the quality of various estimators - MC,
BIMC, and A-BIMC at the end of Stages 1 and 2, in terms of the quality of
samples and the relative RMSE. As is evident, Stage-1 yields a mixture that
roughly approximates $q^*$. This mixture contains 46 components. 
MPMC takes this rough approximation of $q^*$ and and
tunes it so that it better resembles $q^*$. Also noteworthy is the fact that at
the end of Stage-1, A-BIMC finds the successfully finds the secondary mode of
$q^*$, albeit with an insufficient mixture weight, a discrepancy which is fixed
by MPMC. The total number of function evaluations required by A-BIMC was 3051.

\subsection{Effect of dimensionality}
\label{subsection:dimensionality_effect}

\paragraph{Purpose}
Importance sampling schemes are known to suffer from the curse of
dimensionality. This experiment investigates how increasing the dimensionality of the
rare-event probability estimation problem affects A-BIMC's performance.

\paragraph{Setup}
To obtain a complete understanding of A-BIMC's performance, two different
notions of dimensionality are varied. The first notion of dimensionality is the
ambient dimension of the problem, which is simply the dimension of the space in
which the uncertain parameters live. Therefore, if $\uq \in \mathbb{R}^m$, then
the ambient dimension of the problem is $m$. The second notion of dimension is
that of the intrinsic dimension $m_{\mathrm{int}}$ of the problem. 
The intrinsic dimension of the posed rare-event probability estimation problem
is defined to be the dimension of the subspace in which the ideal importance
sampling distribution $q^*$ differs from the nominal distribution $p$. Such
a situation can arise, for instance, if the forward map $f$ is sensitive only
to a few directions in the input parameter space. This definition of intrinsic
dimension for rare-events closely follows that for Bayesian inference, where it
is taken to be the dimension of the subspace where the posterior differs from
the prior~\cite{spantini2015optimal,cui2014likelihood}.

Given the ambient dimension  $m$, the intrinsic dimension of the
problem is specified as follows. First, $p$ is set to
 $p = \mathcal{N}(\boldsymbol{1}_m, \mathbf{I}_m)$, 
where $\boldsymbol{1}_m = [1, 1, \ldots 1]^T \in \mathbb{R}^m$ 
is the vector of all 1's and $\mathbf{I}_m$ is
the $m\times m$ identity matrix. Then forward maps $f(\uq)$ are constructed so
that they satisfy $\partial f / \partial x_i = 0$ for $i = m_{\mathrm{int}}
+ 1, \ldots, m$. This way, the maps are insensitive to any variation of the
input parameters in $\text{span}\{\vect{e}_{m_{\mathrm{int}} + 1}, \ldots,
\vect{e}_m\}$, where $\vect{e}_i$ is the $i$-th canonical basis vector. 
In addition, since $p$ is a Normal distribution with identity covariance,
$q^*$ differs from $p$ only in span$\{\vect{e}_1, \ldots,
\vect{e}_{m_{\mathrm{int}}}\}$. In span$\{\vect{e}_{m_{\mathrm{int} + 1}}, \ldots, e_m\}$,
$q^*$ is identical to $p$ by construction.

As for actually constructing $\boldsymbol{\mathcal{S}}, \mathbf{H}, \vect{b}$, 
the following procedure is adopted. Given $m_{\mathrm{int}}$ and $m$, we set (in
MATLAB notation)
$\boldsymbol{\mathcal{S}}[m_{\mathrm{int}} + 1 : m,
m_{\mathrm{int}} + 1 : m, m_{\mathrm{int}} + 1 : m] =
\mathbf{0}$, $\mathbf{H}[ m_{\mathrm{int}} + 1 : m,
m_{\mathrm{int}} + 1 : m] = \mathbf{0}$ and
$\vect{b}[m_{\mathrm{int}} + 1 : m] = \boldsymbol{0}$. 
Now, let $\bar{\boldsymbol{\mathcal{S}}}$, $\bar{\mathbf{H}}$ 
and $\bar{\vect{b}}$ denote the non-zero blocks of $\boldsymbol{\mathcal{S}}$,
$\mathbf{H}$, and $\vect{b}$, i.e., 
$\boldsymbol{\mathcal{S}}[1 : m_{\mathrm{int}}, 1 : m_{\mathrm{int}}, 1 :
m_{\mathrm{int}}], \mathbf{H}[1 : m_{\mathrm{int}}, 1 : m_{\mathrm{int}}],
\vect{b}[1 : m_{\mathrm{int}}]$. These are constructed as follows.

The tensor
$\bar{\boldsymbol{\mathcal{S}}}$ is constructed as:
$\bar{\boldsymbol{\mathcal{S}}} = 10 \boldsymbol{\mathcal{I}} +
\boldsymbol{\mathcal{G}}$. Here $\boldsymbol{\mathcal{I}}$ is defined as
$\boldsymbol{\mathcal{I}}_{ijk} := \boldsymbol{\delta}_{ij}\boldsymbol{1}_k$,
$\boldsymbol{\delta}_{ij}$ is the Kronecker delta, and $\boldsymbol{1}_k$
assumes the value 1 for all possible $k$. The tensor
 $\boldsymbol{\mathcal{G}}$ is a tensor of
i.i.d. standard Normal variables, $\boldsymbol{\mathcal{G}}_{ijk} \sim \mathcal{N}(0, 1)$.
The matrix $\bar{\mathbf{H}}$ is constructed as a sample from a
Wishart distribution with scale matrix $2 \mathbf{I}_{m_{\mathrm{int}}}$ and
$m_{\mathrm{int}} + 1$ degrees of freedom, 
$\bar{\mathbf{H}} \sim W(2 \mathbf{I}_{m_{\mathrm{int}}}, m_{\mathrm{int}} +
1)$. The vector $\bar{\vect{b}}$ is a vector of uniformly distributed random
numbers in $[0, 1]$, $\bar{\vect{b}}_i \sim \mathbb{U}({[0,1]})$.

The ambient dimension $m$ is set to $16, 32, 64$, and for $m$, the intrinsic
dimension $m_{\mathrm{int}}$ is  varied from a minimum of $m_{\mathrm{int}} = 4$
to a maximum of $m_{\mathrm{int}} = m$. 

\paragraph{Results and discussion}

\begin{figure}[H]
    \centering
        \begin{subfigure}{0.42\textwidth}
        \includegraphics[width=\textwidth]{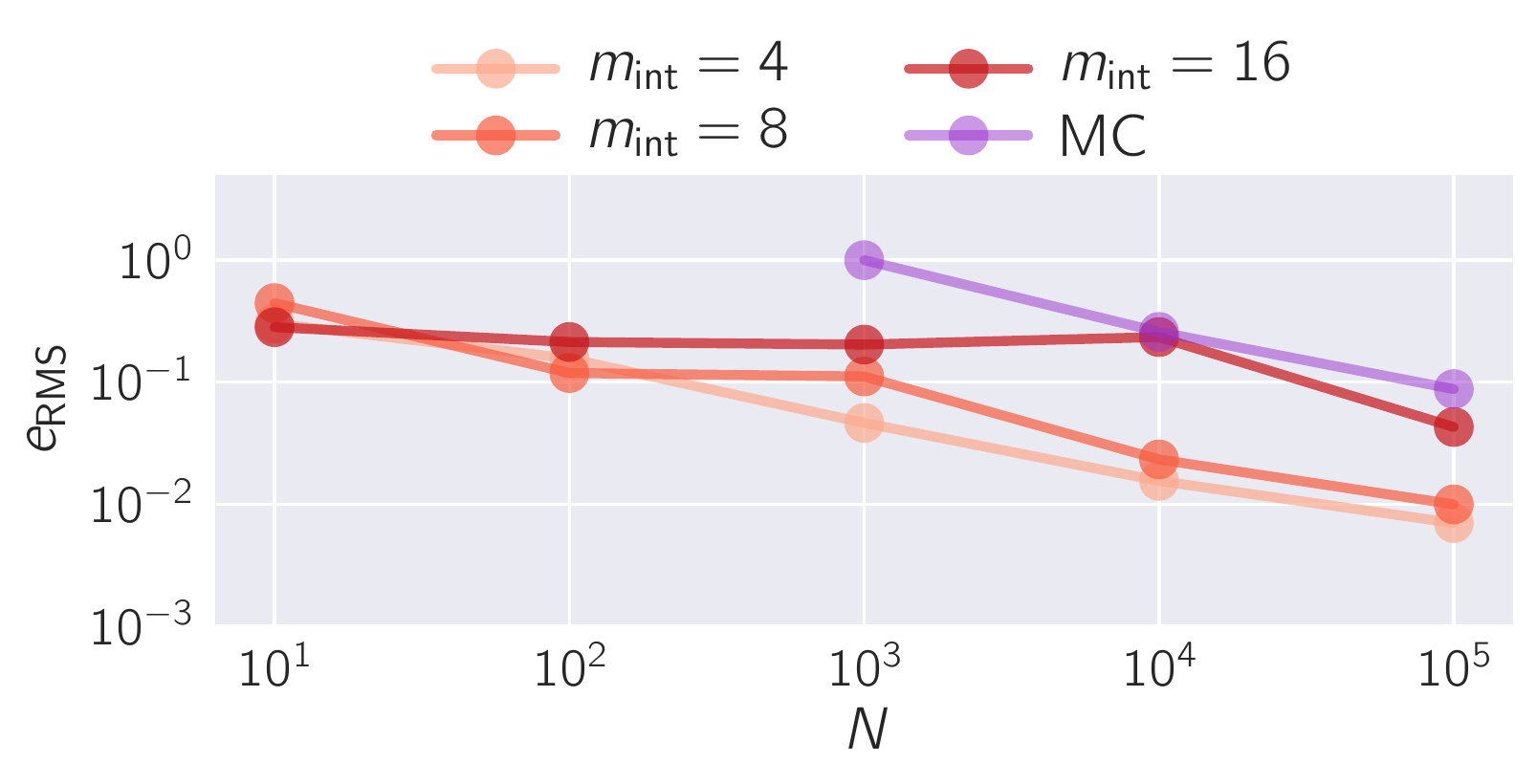}
        \caption{Stage-1, $m = 16$}
    \end{subfigure}
   \begin{subfigure}{0.42\textwidth}
        \includegraphics[width=\textwidth]{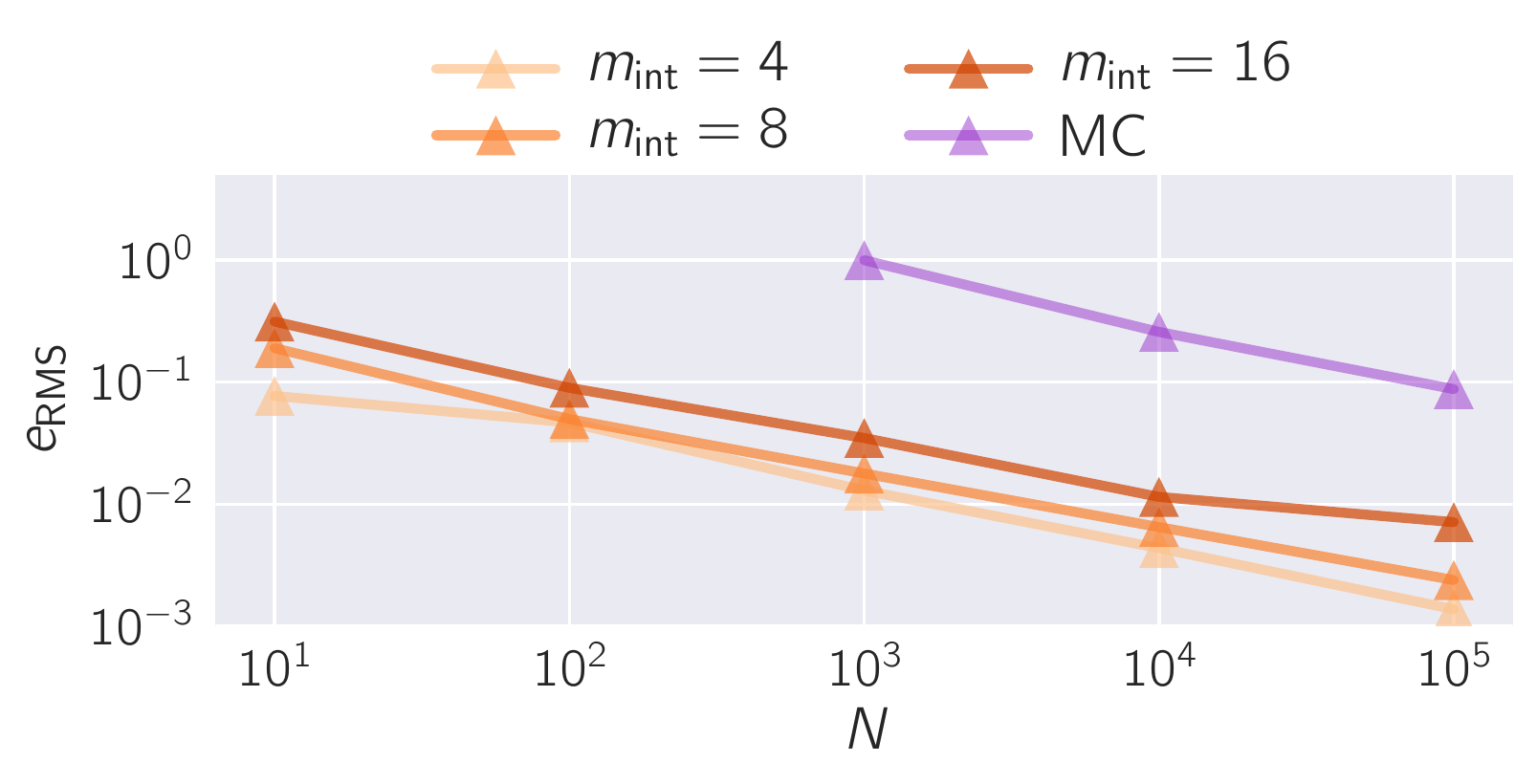}
        \caption{Stage-2, $m = 16$}
    \end{subfigure}
    \begin{subfigure}{0.42\textwidth}
        \includegraphics[width=\textwidth]{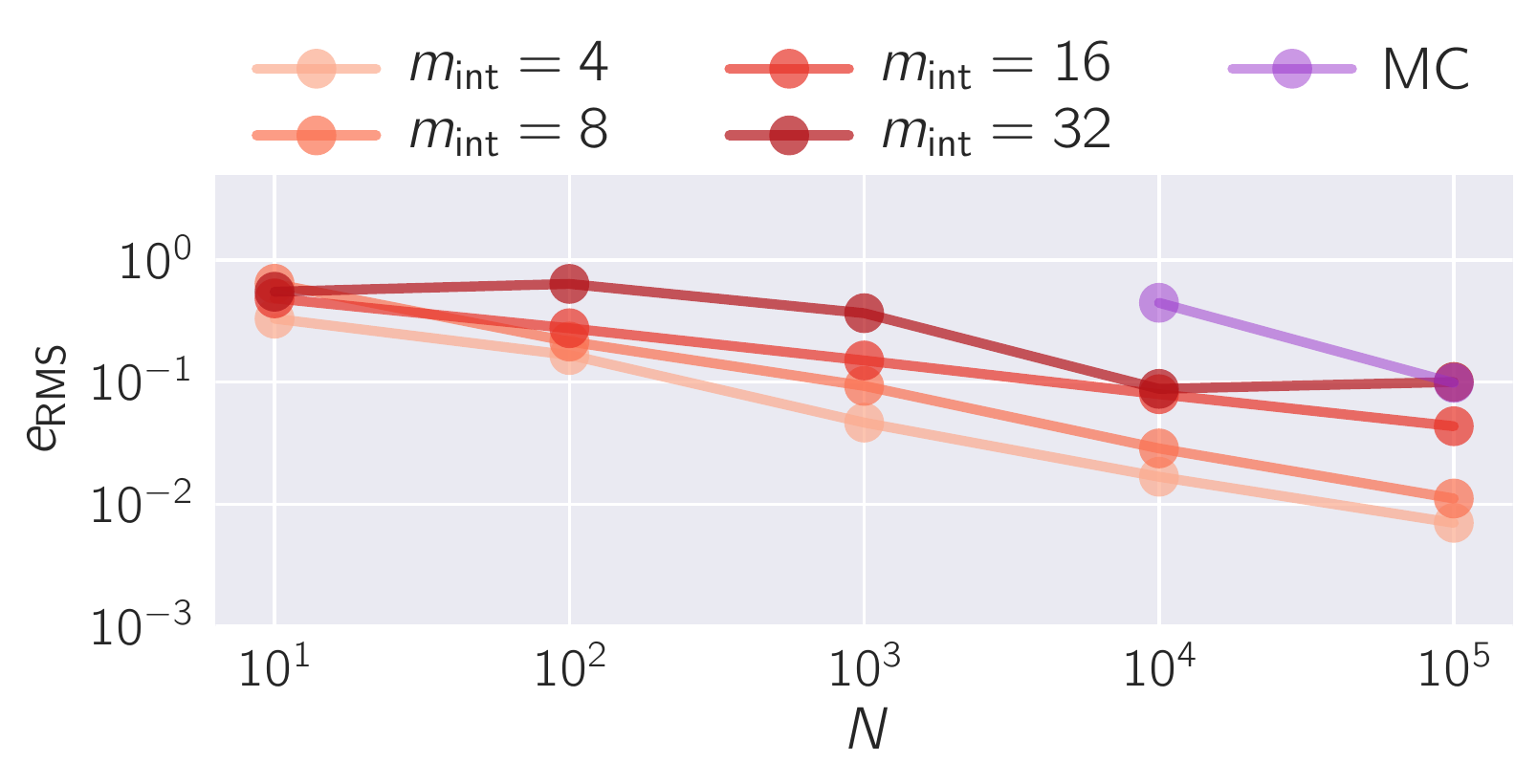}
        \caption{Stage-1, $m = 32$}
    \end{subfigure}
   \begin{subfigure}{0.42\textwidth}
        \includegraphics[width=\textwidth]{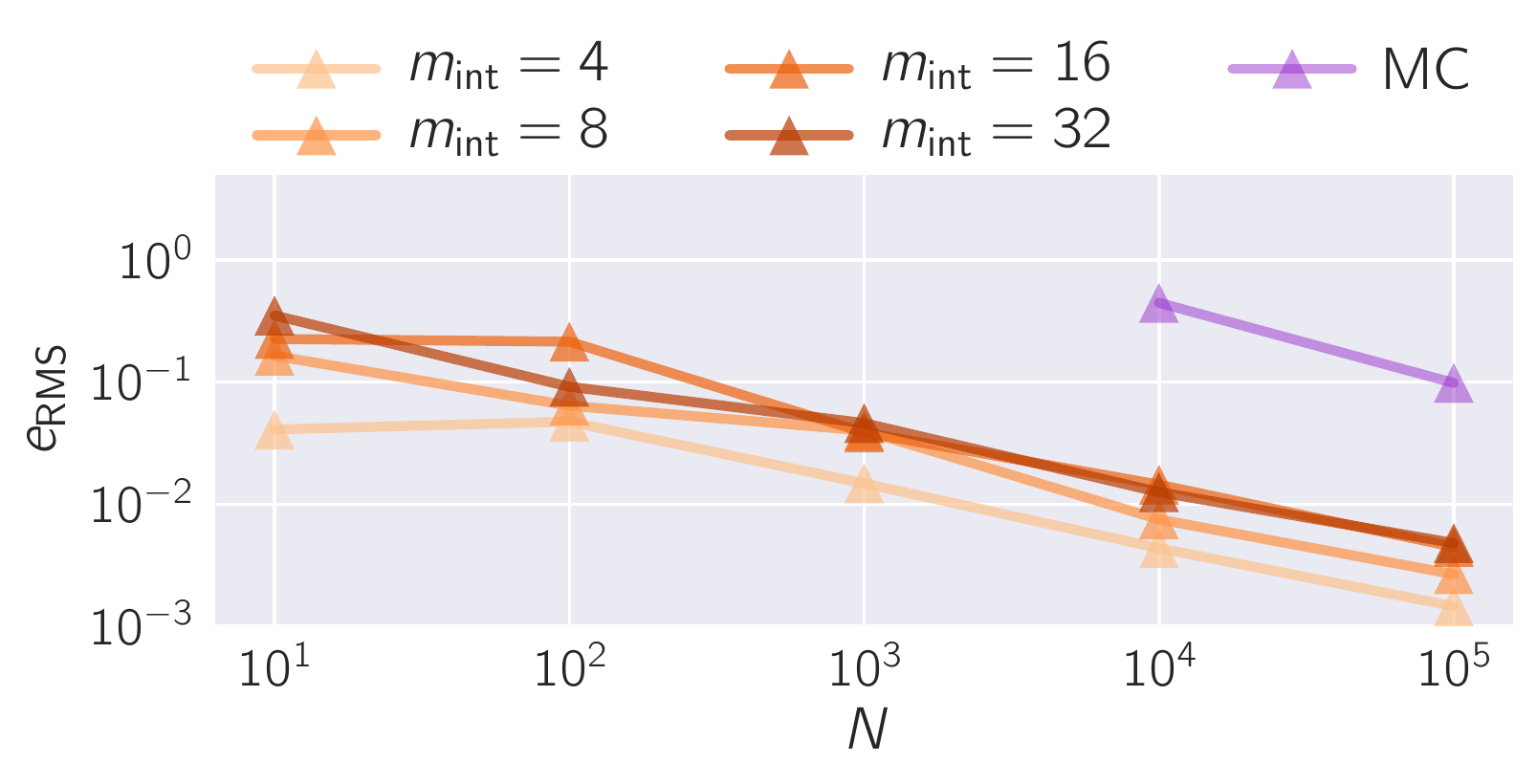}
        \caption{Stage-2, $m = 32$}
    \end{subfigure}
    \begin{subfigure}{0.42\textwidth}
        \includegraphics[width=\textwidth]{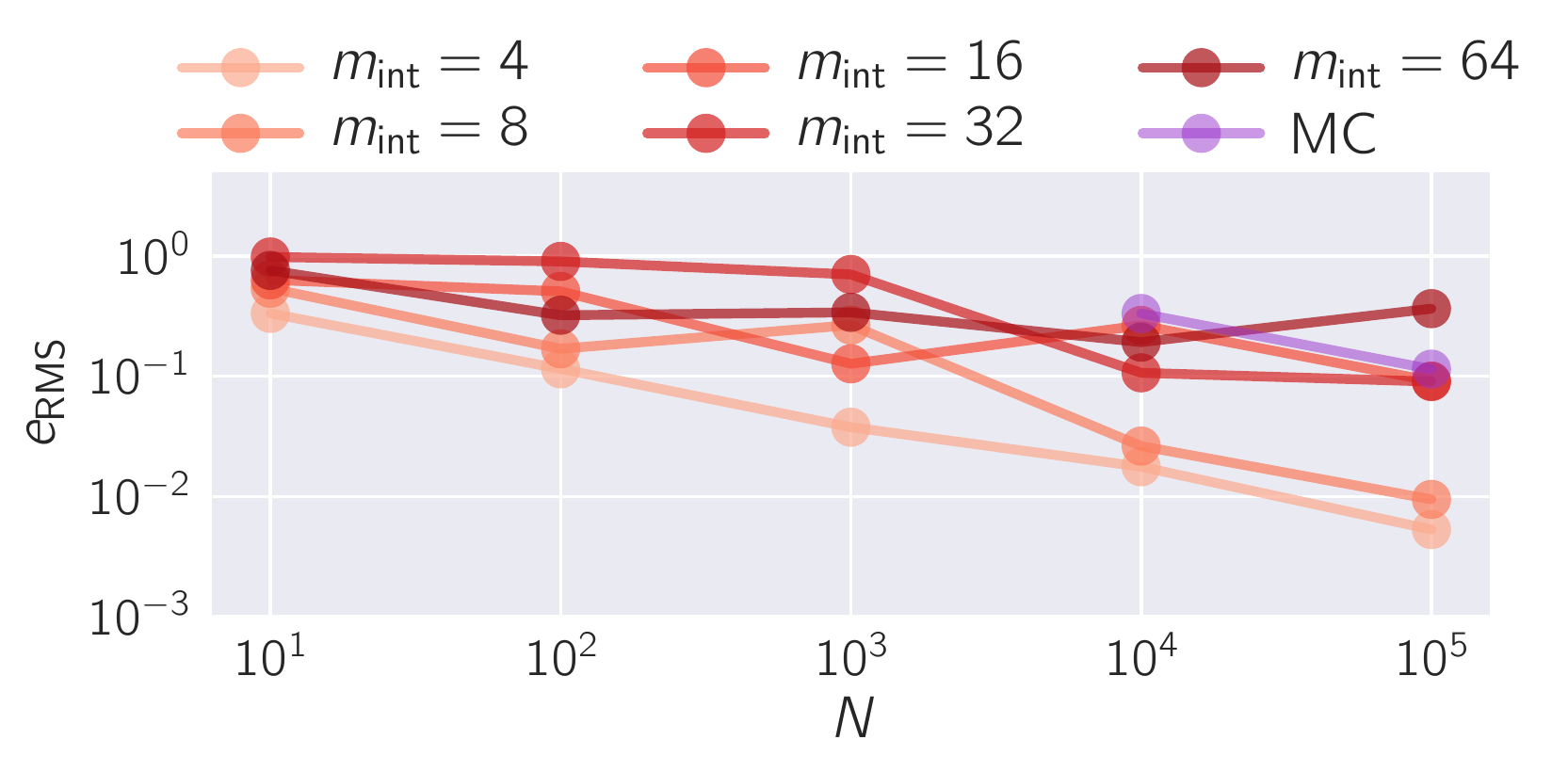}
        \caption{Stage-1, $m = 64$}
    \end{subfigure}
   \begin{subfigure}{0.42\textwidth}
        \includegraphics[width=\textwidth]{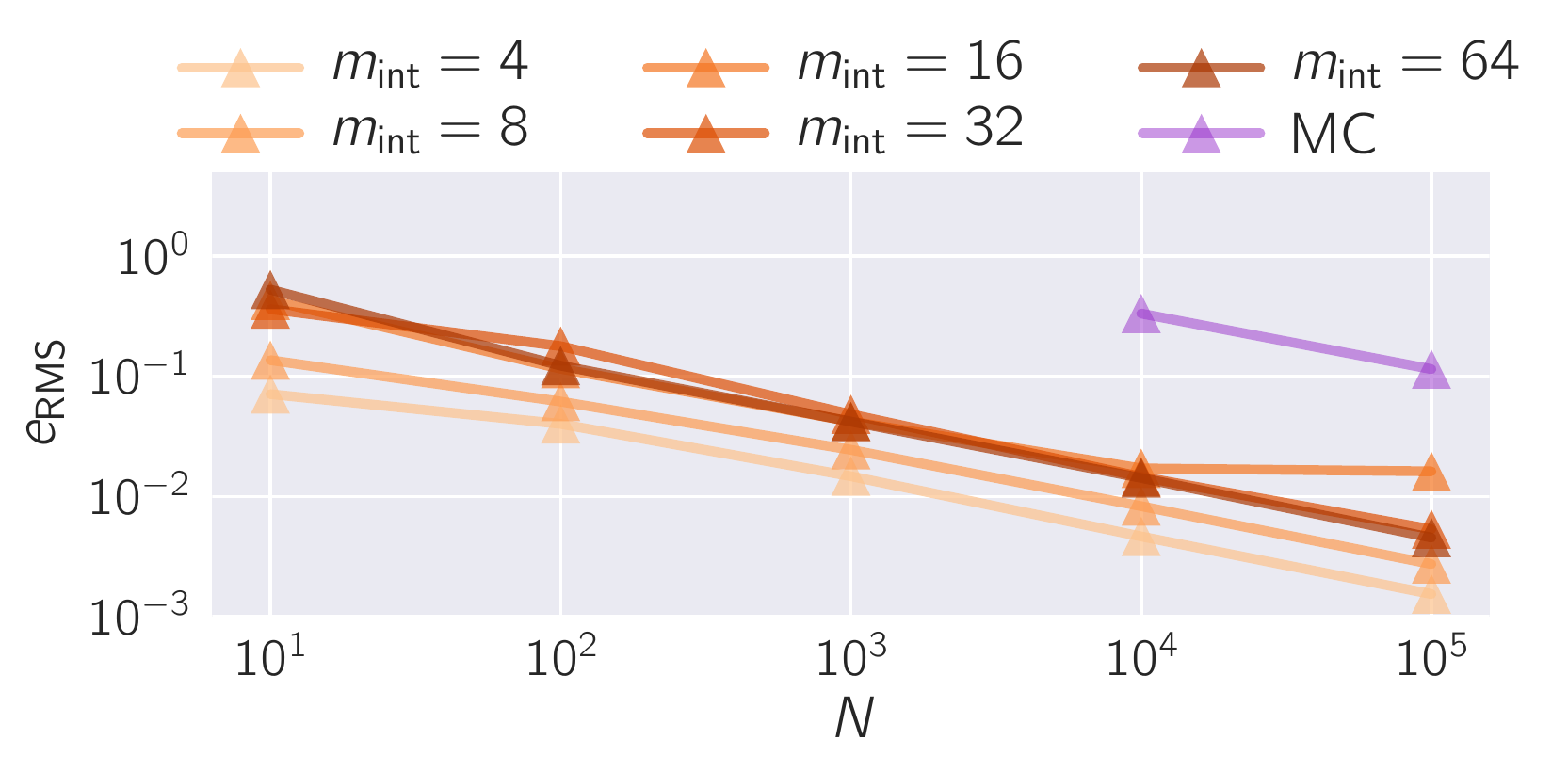}
        \caption{Stage-2, $m = 64$}
    \end{subfigure}
   \caption{Convergence of the relative RMSE, $e_{\mathrm{RMS}}$, with number of
samples $N$ at various combinations of $m$ and $m_{\mathrm{int}}$ for the
quadratic problem.}
   \label{fig:quadratic_dim_conv}
\end{figure}

\begin{figure}[htbp]
    \centering
        \begin{subfigure}{0.42\textwidth}
        \includegraphics[width=\textwidth]{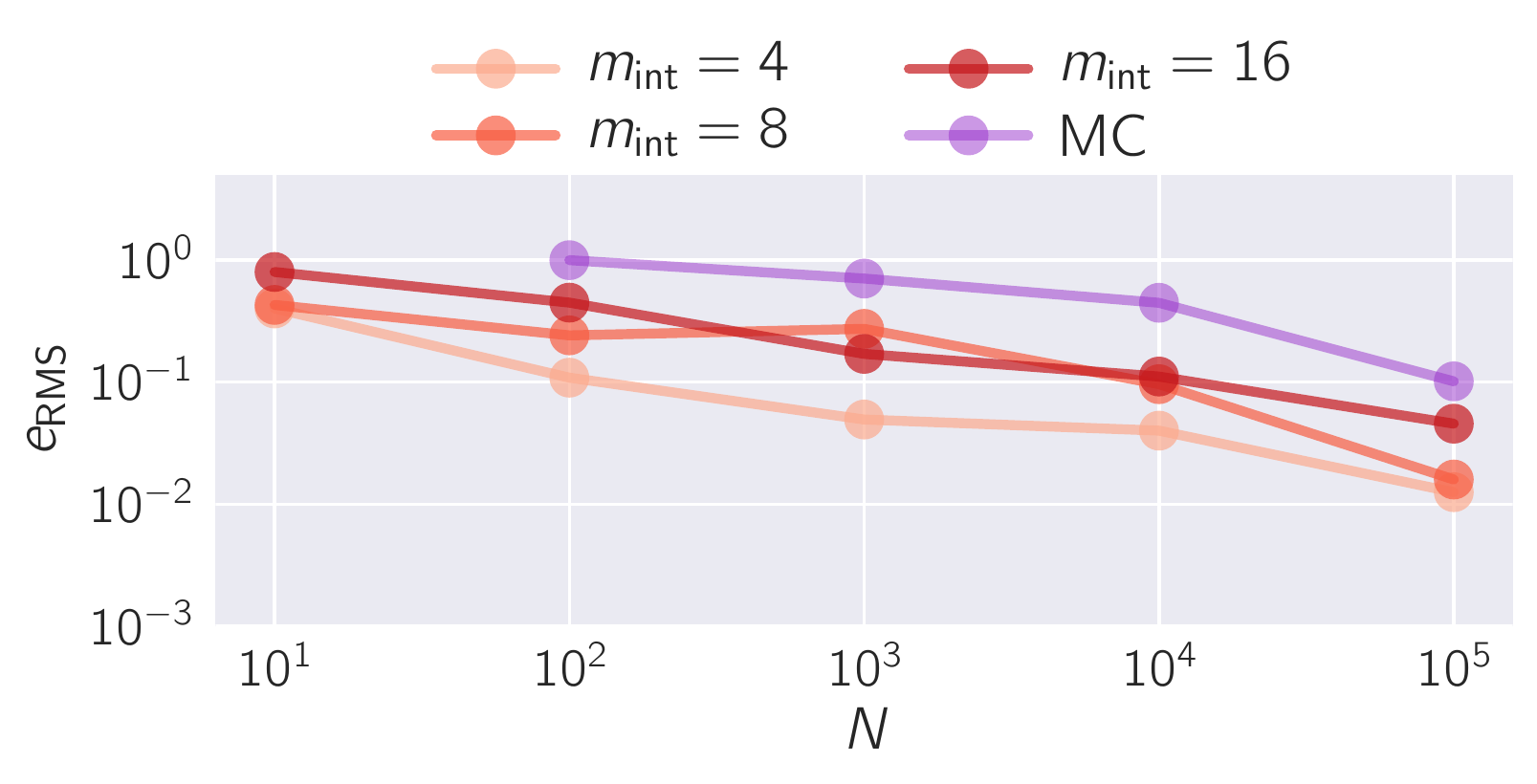}
        \caption{Stage-1, $m = 16$}
    \end{subfigure}
   \begin{subfigure}{0.42\textwidth}
        \includegraphics[width=\textwidth]{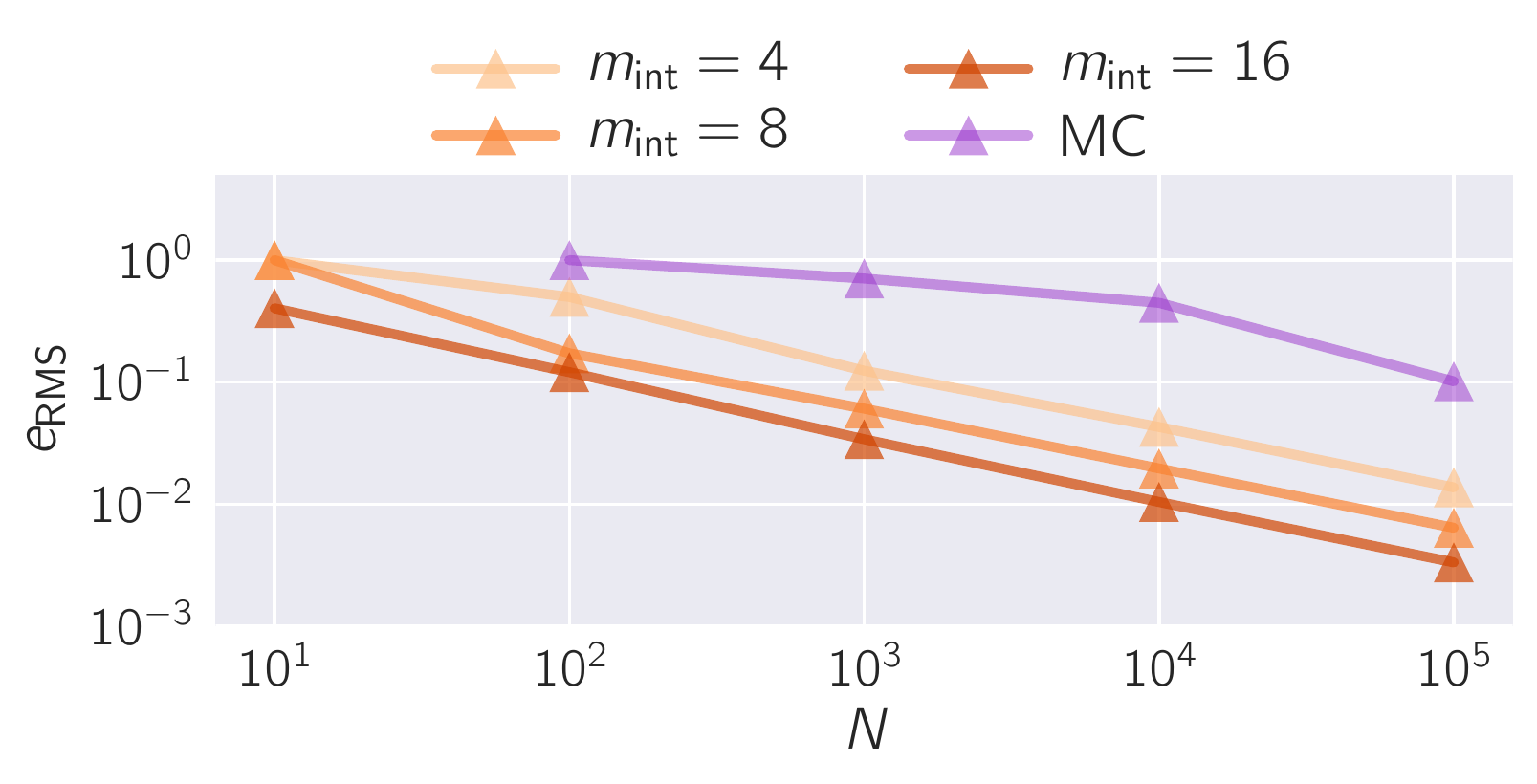}
        \caption{Stage-2, $m = 16$}
    \end{subfigure}
    \begin{subfigure}{0.42\textwidth}
        \includegraphics[width=\textwidth]{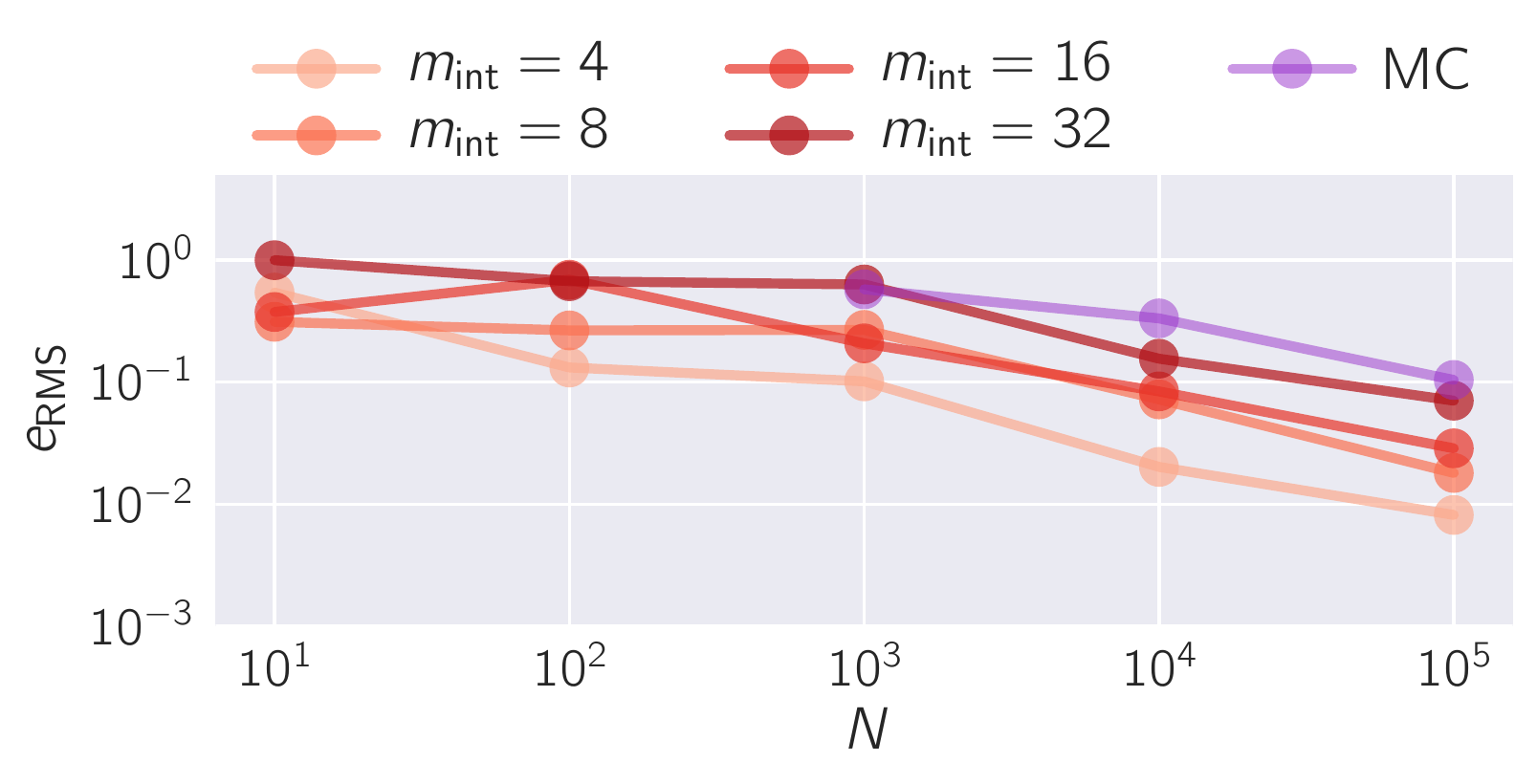}
        \caption{Stage-1, $m = 32$}
    \end{subfigure}
   \begin{subfigure}{0.42\textwidth}
        \includegraphics[width=\textwidth]{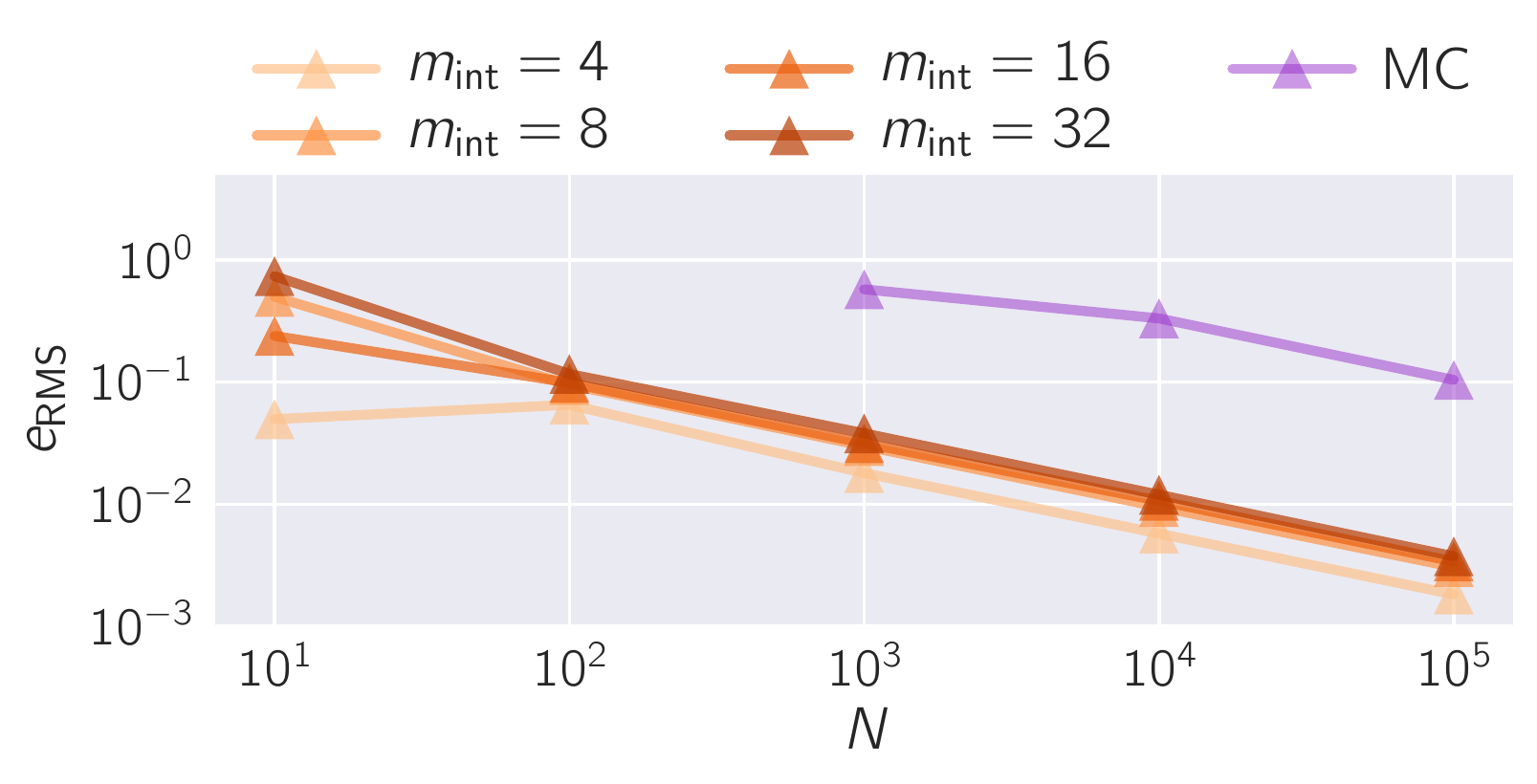}
        \caption{Stage-2, $m = 32$}
    \end{subfigure}
    \begin{subfigure}{0.42\textwidth}
        \includegraphics[width=\textwidth]{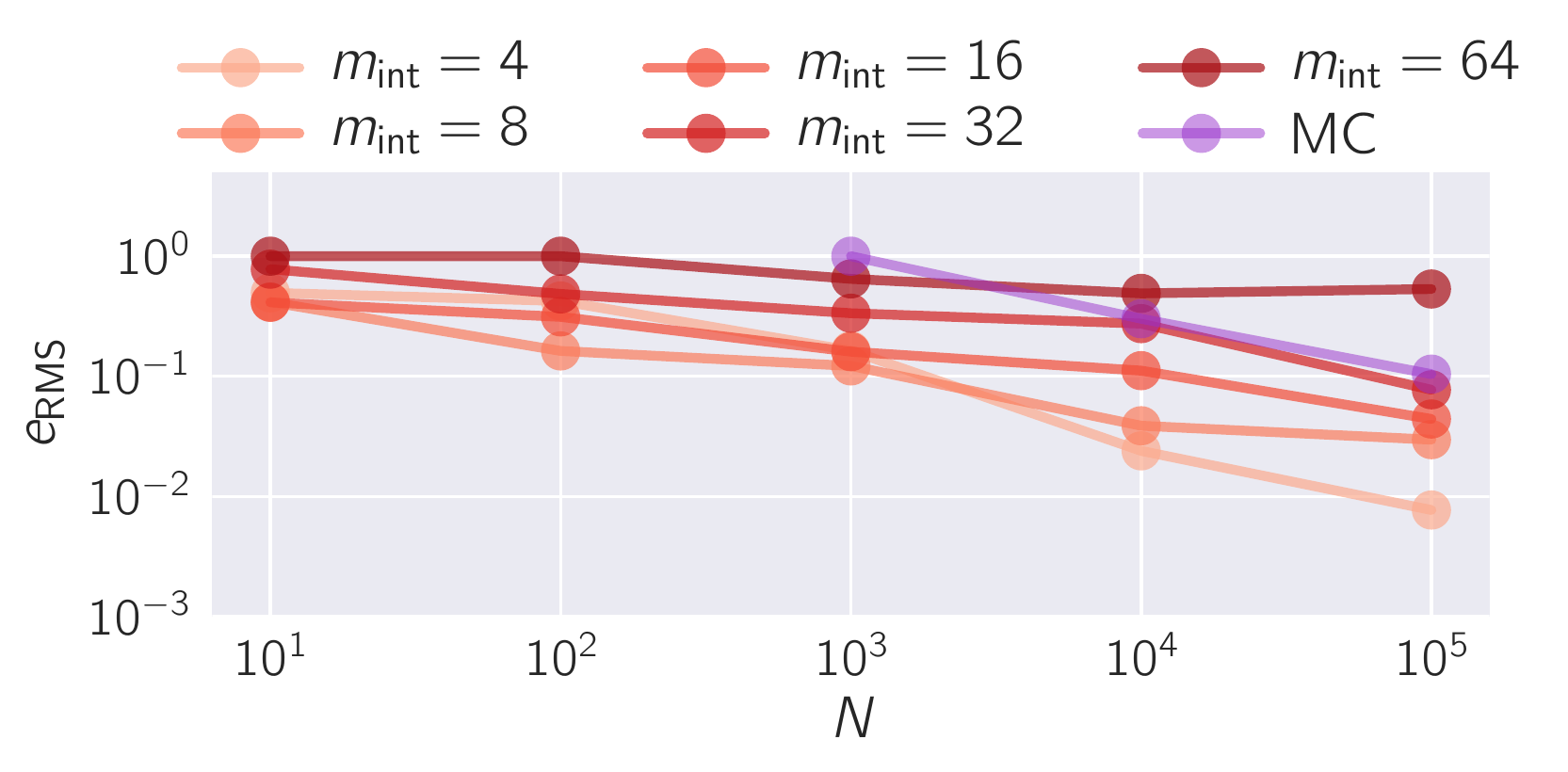}
        \caption{Stage-1, $m = 64$}
    \end{subfigure}
   \begin{subfigure}{0.42\textwidth}
        \includegraphics[width=\textwidth]{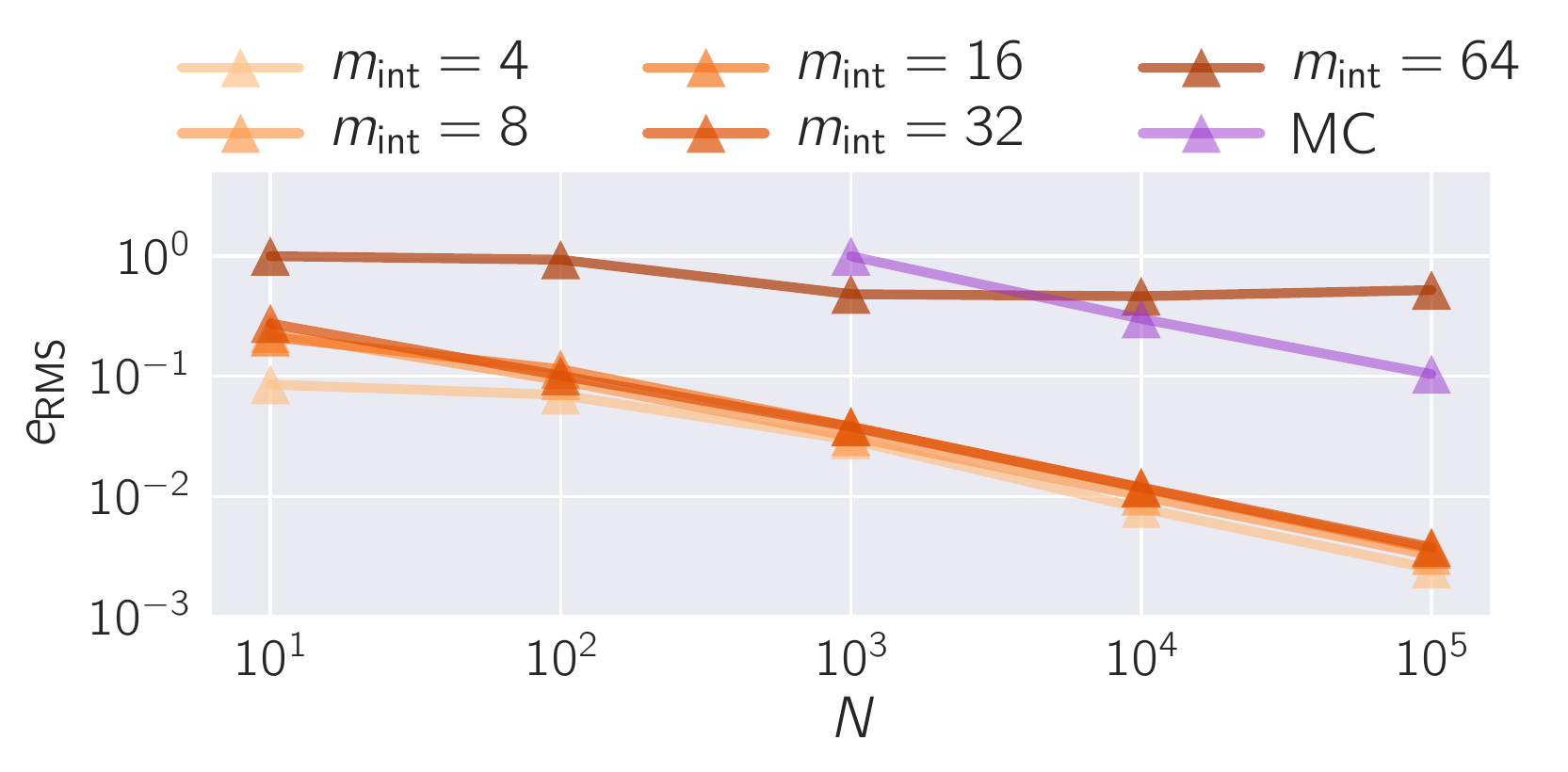}
        \caption{Stage-2, $m = 64$}
    \end{subfigure}
\caption{Convergence of the relative RMSE, $e_{\mathrm{RMS}}$, with number of
samples $N$ at various combinations of $m$ and $m_{\mathrm{int}}$ for the
cubic problem.}
   \label{fig:cubic_dim_conv}
\end{figure}

\Cref{fig:quadratic_dim_conv,fig:cubic_dim_conv} reports the relative RMSE
obtained. Stage-1 of A-BIMC only yields consistent estimates of the rare-event probability at low
$m_{\mathrm{int}}$. At large $m_{\mathrm{int}}$, the error after Stage-1
 fails to exhibit the expected $1/\sqrt{N}$ convergence.
Stage-2 of A-BIMC appears weakly dependent on the intrinsic
dimension of the problem. Notably, it consistently leads to low errors.
For instance, achieving a relative RMSE of around $10\%$ only requires $N = 100$
samples. Exceptions to this trend is the cubic case at  $m_{\mathrm{int}} = 4$ and $m =
16$ and $m_{\mathrm{int}} = 64, m = 64$. 

The poor performance at $m_{\mathrm{int}} = 4$ and $m = 64$ (this case is
referred to as F1 subsequently) can be attributed to
the surrogate $f_{\mathrm{surrogate}}$ not possessing sufficient accuracy. The
poor performance at $m_{\mathrm{int}} = m = 64$ (referred to as F2 subsequently)
is due to the relatively higher
ambient dimensionality of the problem, which causes MPMC to result in mixtures
whose components have rank-deficient covariance matrices. These limitations, as
well as a measures to diagnose and fix them, are discussed in
\Cref{section:failure}.

\begin{figure}[htbp]
\centering
    \begin{subfigure}{0.25\textwidth}
        \includegraphics[width=\textwidth]{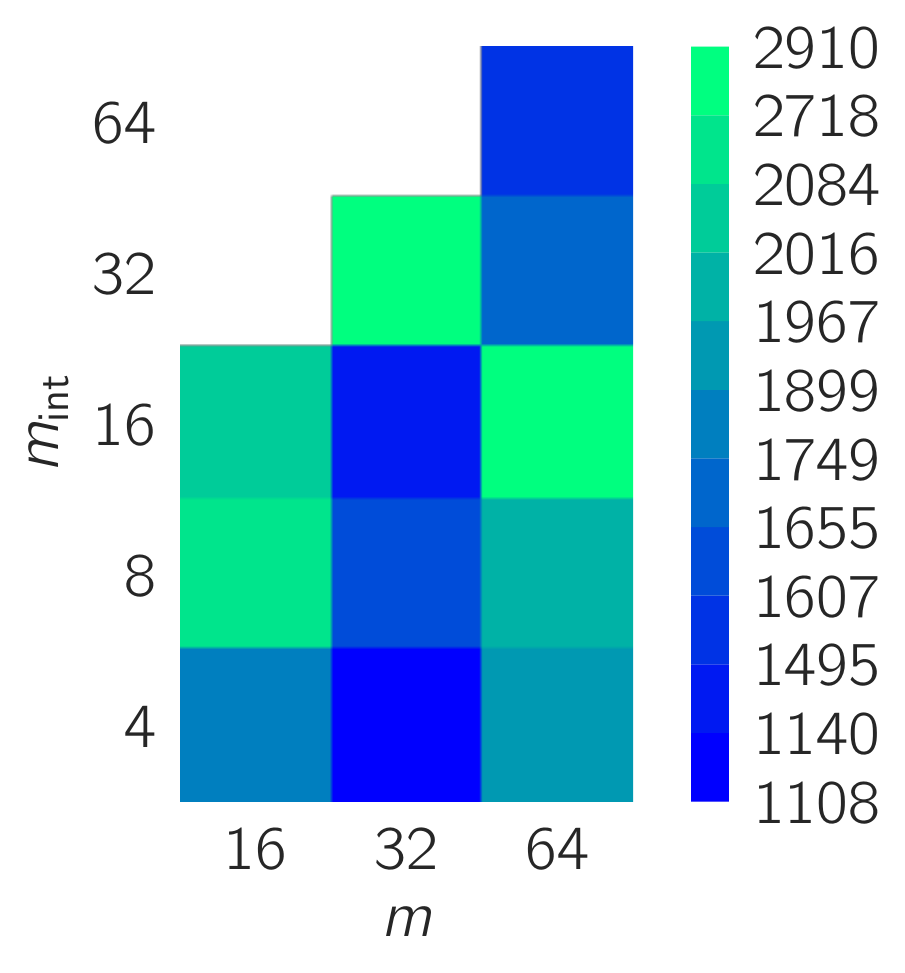}
        \caption{Quadratic problem}
    \end{subfigure}
    \begin{subfigure}{0.25\textwidth}
        \includegraphics[width=\textwidth]{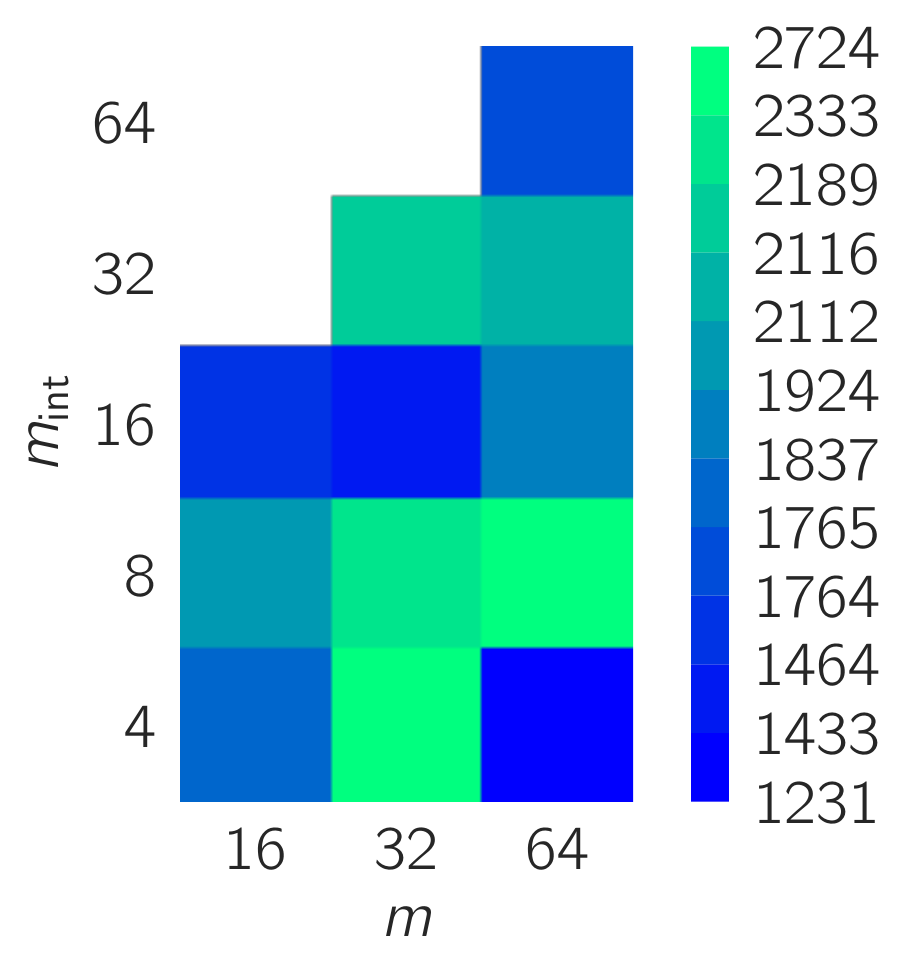}
        \caption{Cubic problem}
    \end{subfigure}
    \caption{Function evaluations required by Stage-1 of A-BIMC display no
significant trend with either $m$ or $m_{\mathrm{int}}$.}
    \label{fig:dim_conv_func_evals}
\end{figure}

\Cref{fig:dim_conv_func_evals} reports the number of function evaluations
required by Stage-1 of A-BIMC. The function evaluations display no significant
trend with either ambient, or the intrinsic dimension of the problem.
\Cref{table:quadratic_ESS,table:cubic_ESS} reports the best and worst observed
normalized-ESS, $\mathrm{ESS}/N$ for each ambient dimension at $N = 10^4$. The
normalized-ESS reflects the poorly performing cases noted earlier
($\{m_{\mathrm{int}} = 4, m = 16\}$ and $\{m_{\mathrm{int}} = 64, m = 64\}$.
The actual rare-event probabilities are reported in
\Cref{supplement:additional_results}.

\begin{table}[H]
\footnotesize
\centering
    \begin{tabular}{c c c}
    \toprule
    $m$ & Worst ESS (corresponding $m_{\mathrm{int}}$) & Best ESS (corresponding $m_{\mathrm{int}})$ \\
    \cmidrule(l{0.5em}r){1-3}
    16 & 0.43 (16) & 0.84 (4)\\
    32 & 0.32 (16) & 0.84 (4)\\
    64 & 0.25 (16) & 0.82 (4)\\
    \bottomrule
    \end{tabular}
    \caption{Worst and best observed normalized-ESS at various 
             ambient dimension $m$ for the quadratic problem. 
             The normalized-ESS is reported at $N = 10^4$.
             The values in parentheses indicate the
             intrinsic dimension $m_{\mathrm{int}}$ at which the normalized-ESS was observed.}
\label{table:quadratic_ESS}
\end{table}

\begin{table}[htbp]
\footnotesize
\centering
    \begin{tabular}{c c c}
    \toprule
    $m$ & Worst ESS (corresponding $m_{\mathrm{int}}$) & Best ESS (corresponding $m_{\mathrm{int}})$ \\
    \cmidrule(l{0.5em}r){1-3}
    16 & 0.051   (4)  & 0.48 (16)\\
    32 & 0.41    (32) & 0.75 (32)\\
    64 & 0.00046 (64) & 0.61 (4)\\
    \bottomrule
    \end{tabular}
    \caption{Worst and best observed normalized-ESS at various 
             ambient dimension $m$ for the cubic problem. 
             The normalized-ESS is reported at $N = 10^4$.
             The values in parentheses indicate the
             intrinsic dimension $m_{\mathrm{int}}$ at which the normalized-ESS was observed.}
\label{table:cubic_ESS}
\end{table}

\subsection{Effect of rarity}
\paragraph{Purpose} 
This experiment is designed to answer the following question: how does A-BIMC
perform as the magnitude of the rare-event probability is decreased? 

\paragraph{Setup}
Both the quadratic and cubic forward maps are constructed as in 
\Cref{subsection:dimensionality_effect}
but the dimensionality of the problem is fixed at $m = 16$ and
$m_{\mathrm{int}} = 8$ so that the effect of decreasing probability level can be
extracted. The nominal distribution for both forward maps 
 is also as in \Cref{subsection:dimensionality_effect}, $p(\uq) = \mathcal{N}(\mathbf{1},
\boldsymbol{I}_m)$. The rarity level is increased by choosing $\DT$ so that the
rare-event probability is approximately $\mathcal{O}(10^{-4}),
\mathcal{O}(10^{-5}), \mathcal{O}(10^{-6})$ respectively for each forward map. 

\begin{figure}[htbp]
\centering
    \begin{subfigure}{0.4\textwidth}
        \includegraphics[width=\textwidth]{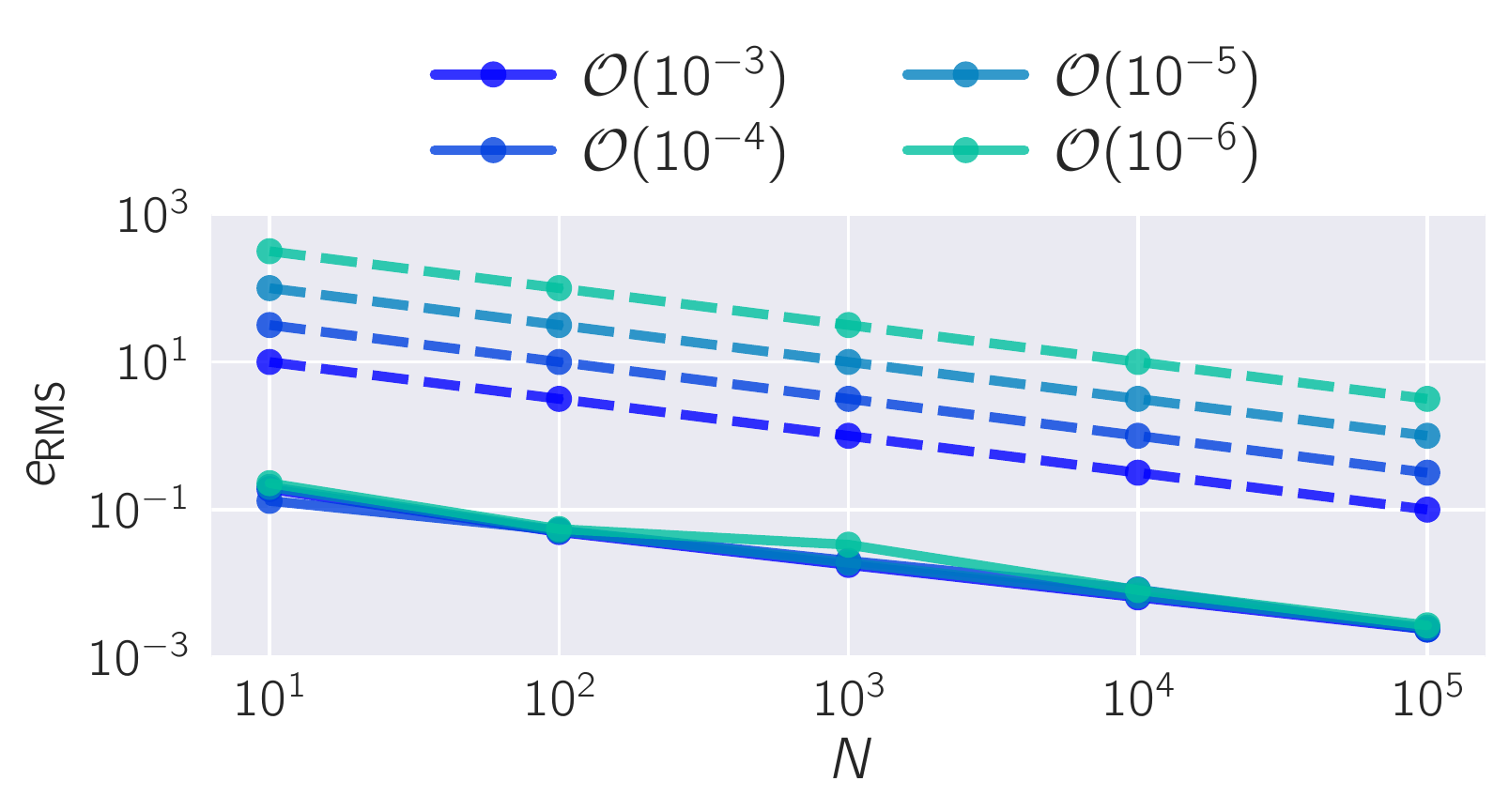}
        \caption{Quadratic problem}
    \end{subfigure}
    \begin{subfigure}{0.4\textwidth}
        \includegraphics[width=\textwidth]{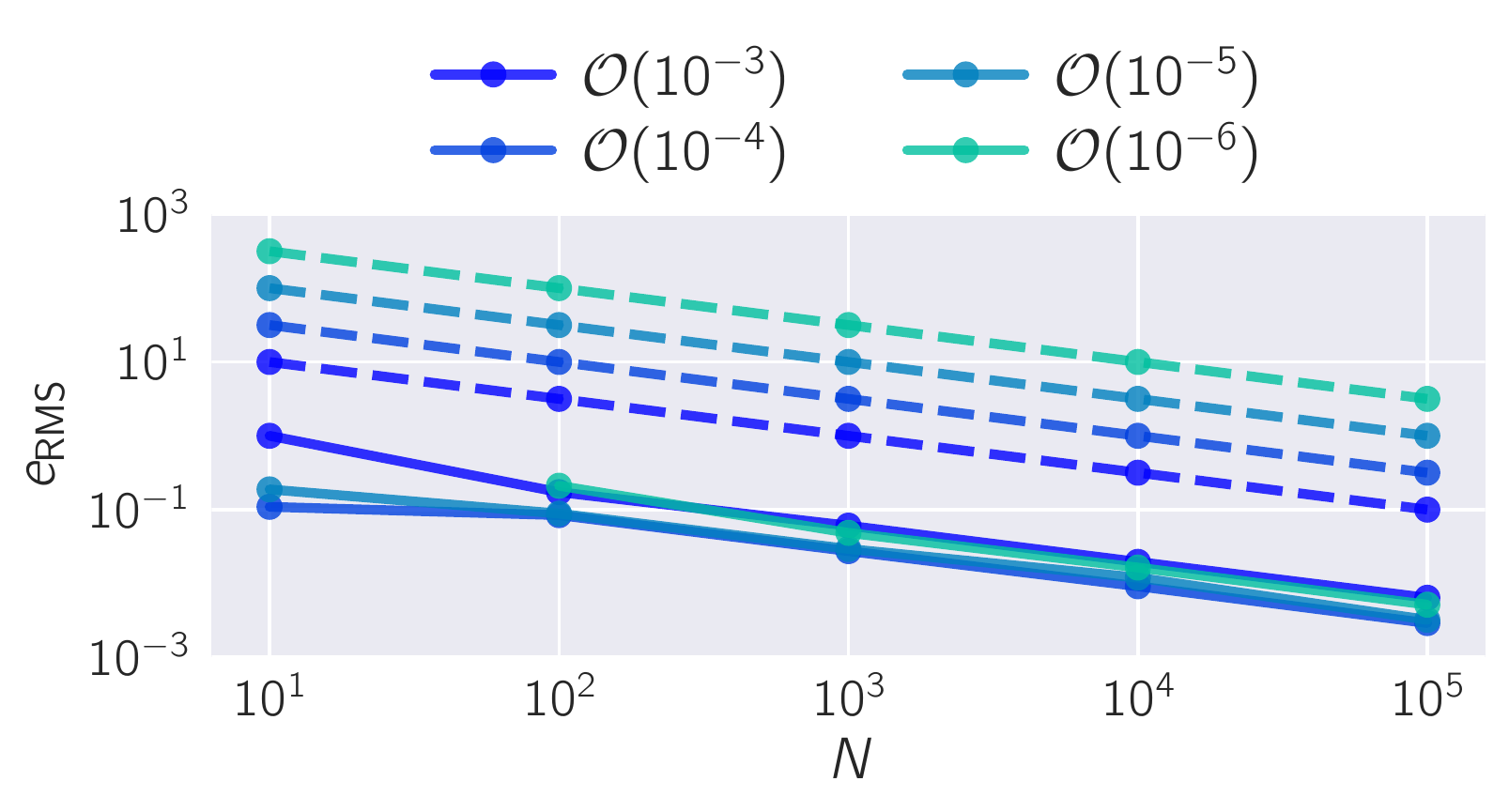}
        \caption{Cubic problem}
    \end{subfigure}
    \caption{This figure demonstrates that A-BIMC (solid lines) 
             performs consistently well even as the magnitude of 
             the rare-event probability is decreased from
             $\mathcal{O}(10^{-3})$ to $\mathcal{O}(10^{-6})$. The dashed lines
             show the expected theoretical performance of the 
             simple Monte Carlo estimator $\hat{\mu}$,
             computed using the expression $\sqrt{(1 - \hat{\mu}) / (\hat{\mu}N)}$ 
             (see Section 2.1 of Part I).}
    \label{fig:rarity_err_comparison}
\end{figure}

\begin{figure}[htbp]
\centering
    \begin{subfigure}{0.4\textwidth}
        \includegraphics[width=\textwidth]{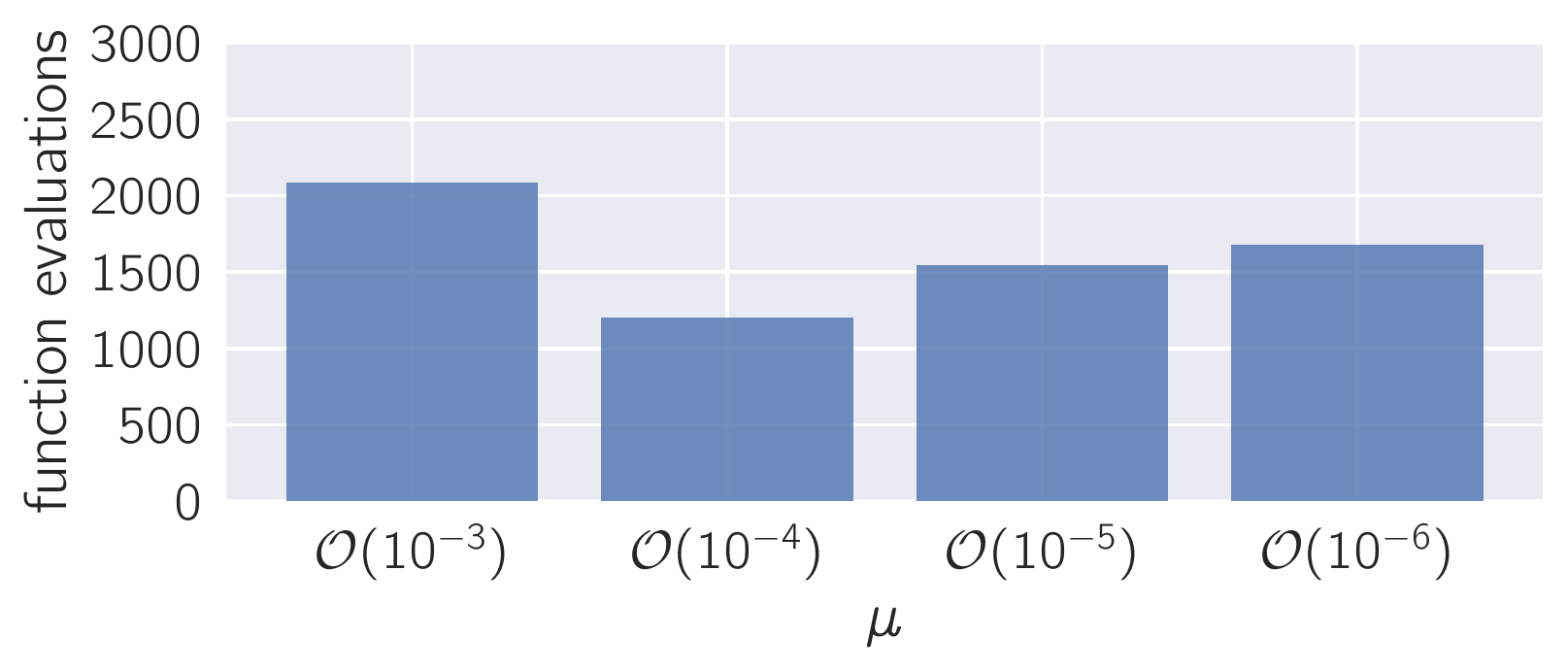}
        \caption{Quadratic problem}
    \end{subfigure}
    \begin{subfigure}{0.4\textwidth}
        \includegraphics[width=\textwidth]{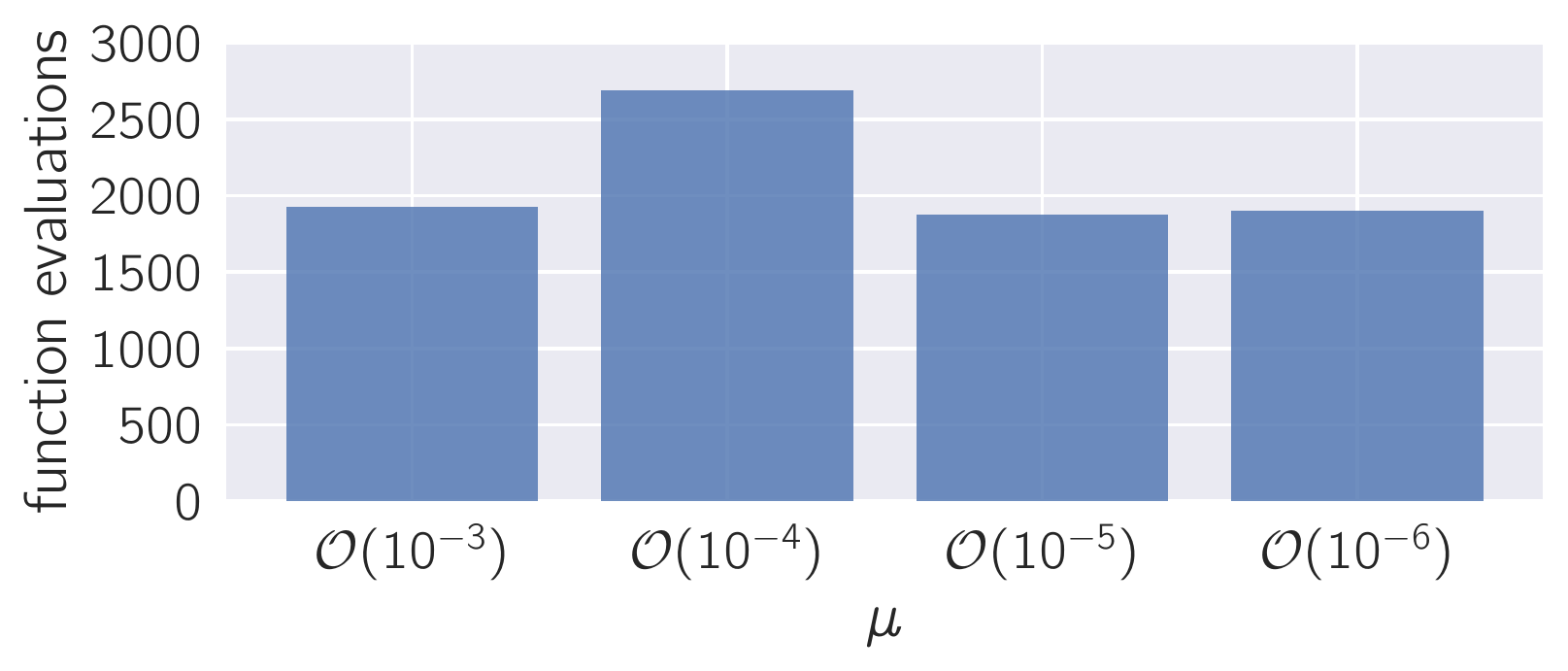}
        \caption{Cubic problem}
    \end{subfigure}
    \caption{Function evaluations required by A-BIMC remain independent of the
magnitude of the rare-event probability, as this probability is decreased from
$\mathcal{O}(10^{-3})$ down to $\mathcal{O}(10^{-6})$.}
    \label{fig:rarity_func_evals_comparison}
\end{figure}

\paragraph{Results and discussion}
As was demonstrated in Section 2.1 of part 1, the number of samples 
required by a simple Monte Carlo method to achieve a
specified accuracy in the rare-event probability increases as the probability
decreases. \Cref{fig:rarity_err_comparison} offers evidence that this is not the case for
A-BIMC. \Cref{fig:rarity_func_evals_comparison} demonstrates that the number of function
evaluations required to explore $f^{-1}(\DT)$ also remains approximately
constant as the probability magnitude decreases.


This concludes the presentation of results of numerical experiments. The next section explores
how A-BIMC can fail, and in particular, offers explanations for why A-BIMC
performed unsatisfactorily in experiments F1 and F2 in  \Cref{subsection:dimensionality_effect}.

\section{Failure}
\label{section:failure}
In this section, we explore how A-BIMC can fail. A-BIMC's failure is tied to
inappropriately choosing the tolerance ${\epsilon}_{\mathrm{abs}}$, or the
number of samples used by MPMC per iteration, $N_{\mathrm{MPMC}}$ (in order to
keep the number of tunable parameters to a minimum, we always choose
$\epsilon_{\mathrm{rel}} = 1 - \epsilon_{\mathrm{abs}}$). Hence,
this section doubles as a discussion on the consequences of choosing these
parameters inappropriately. In addition to exploring the mechanisms behind
failure, recommendations for diagnosing, as well as mitigating it, are offered.

\begin{figure}[H]
\centering
   \includegraphics[width=0.7\textwidth]{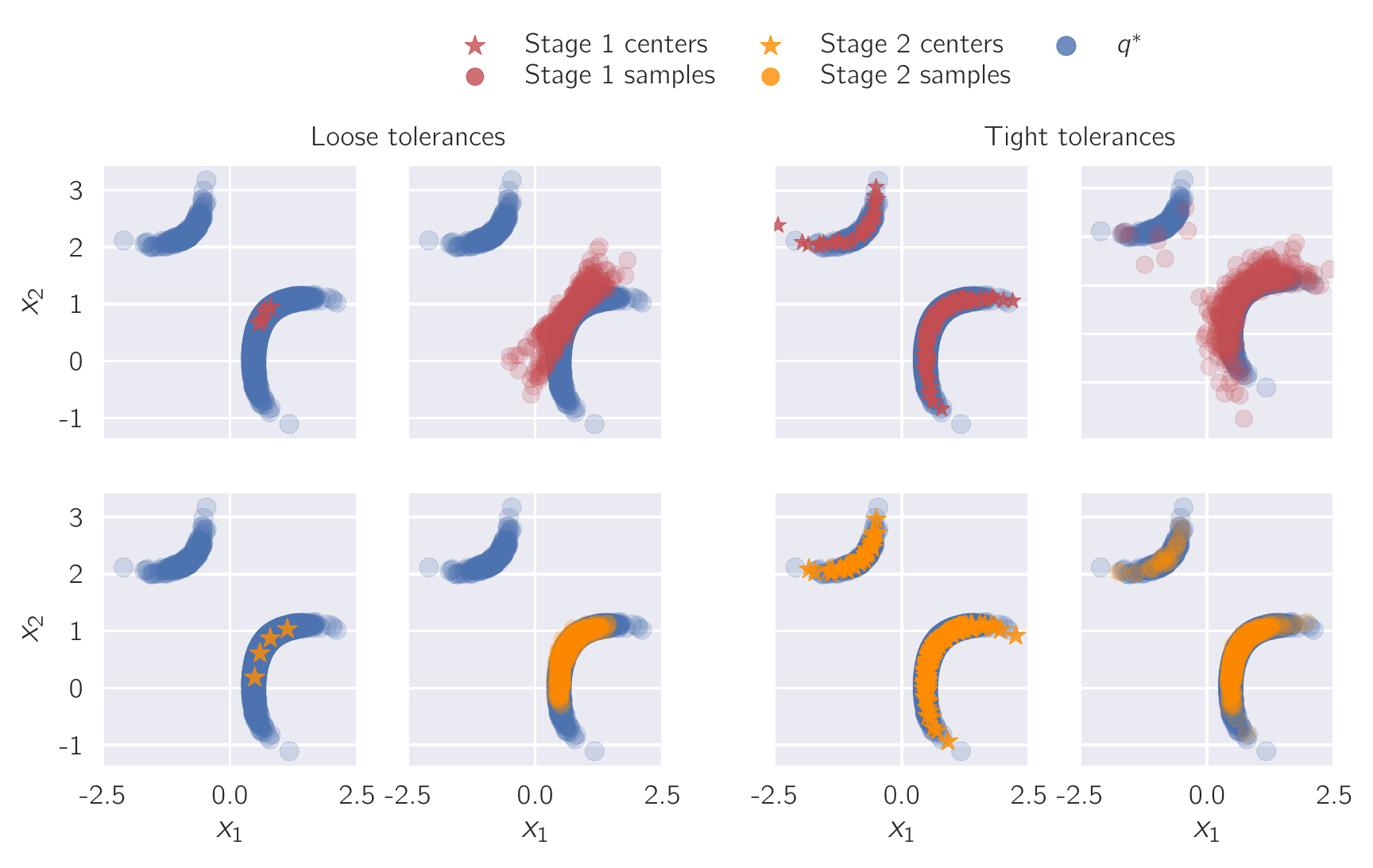}
   \caption{This figure illustrates the effect of varying the tolerances
$\epsilon_{\mathrm{abs}}$ and $\epsilon_{\mathrm{rel}}$. The left half of the
figure corresponds to a loose tolerance, $\epsilon_{\mathrm{abs}} = 1 - 10^{-1}$,
while the right half corresponds to a tighter tolerance,
$\epsilon_{\mathrm{abs}} = 1 - 10^{-5}$. In both cases, 
$\epsilon_{\mathrm{rel}} = 1 - \epsilon_{\mathrm{abs}}$ and $N_{\mathrm{MPMC}} =
10^6$. Tighter tolerances lead to a greater number of components in the mixture
and can aid the discovery of disjoint regions of $f^{-1}(\DT)$. However, they
increase the computational cost. For instance, $\epsilon_{\mathrm{abs}} = 1 -
10^{-1}$ required 266 function evaluations, whereas $\epsilon_{\mathrm{abs}} =
10^{-5}$ required 19749. All cases show 5000 samples from $q^*$ and 500 samples
from the IS mixture obtained after Stages 1 and 2. }
\label{fig:taylor_green_failure}
\end{figure}

\paragraph{Inappropriate $\epsilon_{\mathrm{abs}}$}
Recall that
$\epsilon_{\mathrm{abs}}$ is a tolerance on how similar mixtures are allowed to
be from one iteration to the next in Stage-1. A smaller value of
$\epsilon_{\mathrm{abs}}$ means mixtures are allowed to be similar, which in
turn implies more components will be added to the mixture in Stage-1. Adding more
components is advantageous for two reasons - it aids discovery of disjoint regions
of $f^{-1}(\mathbb{Y})$, and leads to a more accurate surrogate
$f_{\mathrm{surrogate}}$ (this fact illustrated in
\Cref{fig:taylor_green_failure} for the toy
problem). At the same time, adding more components requires solving more
optimization problems, driving up the computational cost of Stage-1.
Clearly, $\epsilon_{\mathrm{abs}}$ represents  trade-off between a smaller 
computational footprint and better discovery of $f^{-1}(\DT)$.
However, an $\emph{a priori}$ prescription for $\epsilon_{\mathrm{abs}}$ remains
elusive at this time. 

\begin{figure}[H]
\centering
\begin{subfigure}{0.45\textwidth}
    \includegraphics[width=\textwidth]{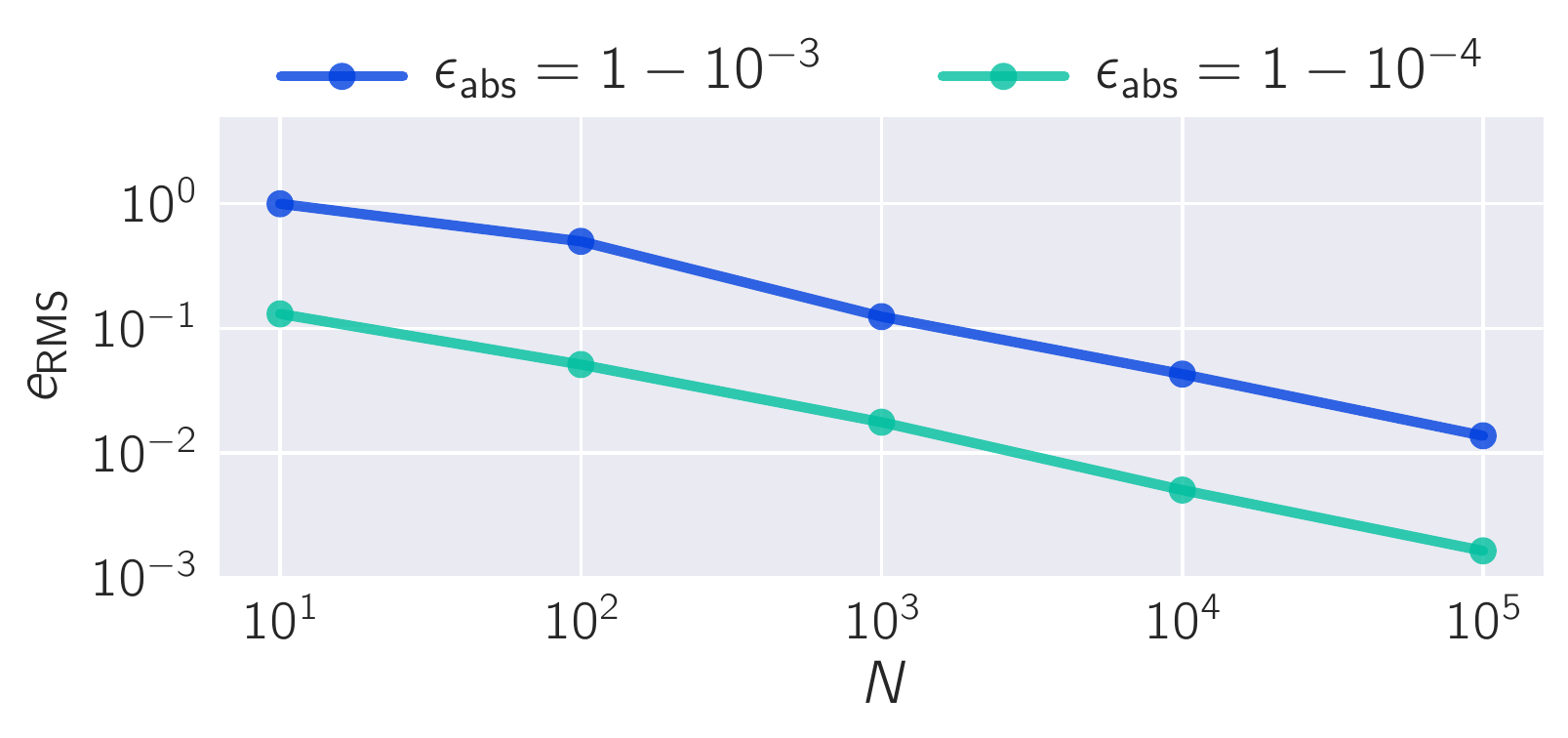}
    \caption{Failure case F1 is remedied by increasing $\epsilon_{\mathrm{abs}}$.}
    \label{fig:failure_f1}
\end{subfigure}
\begin{subfigure}{0.45\textwidth}
    \includegraphics[width=\textwidth]{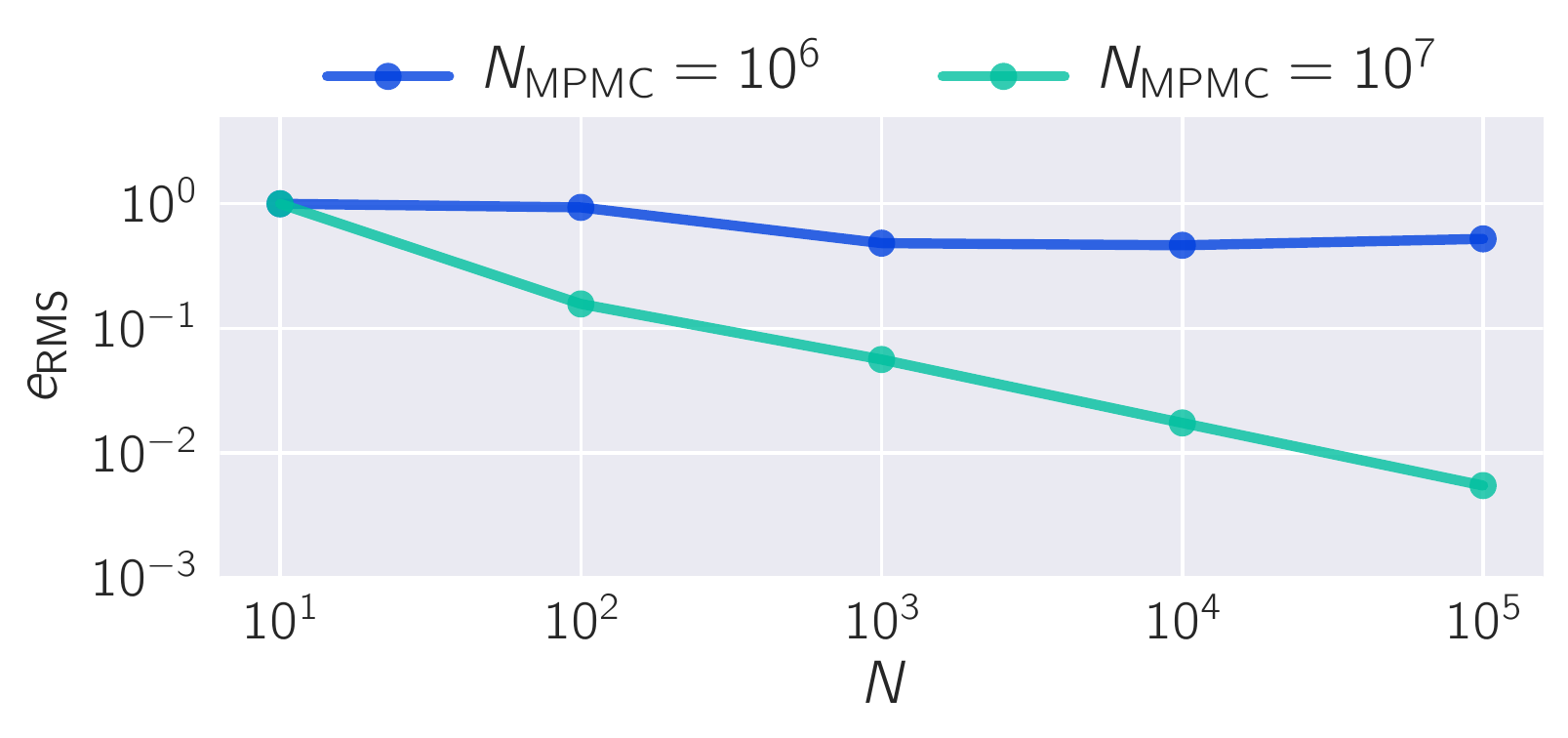}
    \caption{Failure case F2 is remedied by increasing $N_{\mathrm{MPMC}}$.}
    \label{fig:failure_f2}
\end{subfigure}
\caption{Fixing failure cases F1 and F2. Decreasing $\epsilon_{\mathrm{abs}}$ in
failure case F1 increased the number of function calls from 1765 to 3429.}
\end{figure}

A possible hypothesis for why A-BIMC performed poorly in experiment F1 is that
$\epsilon_{\mathrm{abs}} = 1 - 10^{-3}$ was too loose, leading to an inaccurate
surrogate. Running A-BIMC at $\epsilon_{\mathrm{abs}} = 1 - 10^{-4}$
immediately improves the surrogate, and consequently, A-BIMC's performance,
lending credibility to this hypothesis (see \Cref{fig:failure_f1}).  Therefore,
an inappropriate $\epsilon_{\mathrm{abs}}$ can be diagnosed by a low ESS.  Note
that A-BIMC need not be re-run from scratch to correct for loose
$\epsilon_{\mathrm{abs}}$. Let $Q_{1}^{\mathrm{loose}}$ and
$\boldsymbol{\chi}^{\mathrm{loose}}_{\mathrm{fixed}}$ denote the IS mixture and
set of fixed charges computed at the end of Stage-1 with a loose
$\epsilon_{\mathrm{abs}}$. Then, rectifying the effect of a loose
$\epsilon_{\mathrm{abs}}$ simply involves starting from
$Q_{1}^{\mathrm{loose}}$ and
$\boldsymbol{\chi}_{\mathrm{fixed}}^{\mathrm{loose}}$ and re-running lines
10-17 of \Cref{algo:adaptiveBIMC} using a tighter $\epsilon_{\mathrm{abs}}$.
This will have the same effect as a cold restart of A-BIMC using the tighter
tolerance. 

\paragraph{ High ambient dimensionality}
MPMC is particularly susceptible to
failure when the dimension of the parameter $\uq$ is high. This is because the
update formulae for $\alpha_k$'s, $\uq_k$'s and $\boldsymbol{\Sigma}_k$'s in
MPMC are
posed as high-dimensional integrals.  These integrals are approximated via IS,
and at a given sample size, become poorer and poorer approximations of the the
true updates as the dimension of the parameters increases. This failure
mechanism usually manifests as one of the $\boldsymbol{\Sigma}_k$'s becoming
rank-deficient during the tuning procedure.  This is the mechanism of failure
in experiment F2. In this case, it is truly the high ambient dimensionality of
the problem, and not other factors, such as an inaccurate surrogate, that cause
MPMC to fail. This is confirmed in \Cref{fig:failure_f2}, where MPMC is rerun using
$N_{\mathrm{MPMC}} = 10^7$, but with the same surrogate. Increasing
$N_{\mathrm{MPMC}}$ successfully improves performance. This failure mode 
usually manifests as a rank-deficient covariance matrix in one of the mixture 
components during MPMC, and it can be fixed by increasing $N_{\mathrm{MPMC}}$. 
Note that such a failure mode, compared to an inappropriate
$\epsilon_{\mathrm{abs}}$,  is relatively benign as it doesn't require additional
evaluations of $f$.

\paragraph{Poor surrogates}
The surrogate used to replace full evaluations of
$f(\uq)$ is a heuristic, and unfortunately, possesses no guarantees as
to its accuracy. If the surrogate poorly approximates $f(\uq)$, then the
resulting mixture will be a poor approximation of $\ind_{\DT}(f(\uq))$.
This failure is tied to item 1, but can also occur independently, 
for instance when no computationally tractable $\epsilon_{\mathrm{abs}}$
can deliver a surrogate that is accurate enough.

\section{Conclusion and future work}
\label{section:conclusion}

In Part II of this article, we extend the applicability of BIMC, an algorithm
for computing rare-event probabilities. The extended algorithm, called
Adaptive-BIMC, proceeds in two stages. In Stage-1, A-BIMC constructs a rough
approximation of the (theoretical) ideal importance sampling distribution by
exploring the regions that trigger the rare-event on a global scale. This is in
contrast to BIMC, which can only achieve local exploration of the region of
interest around the so-called pseudo-MAP point. Global exploration in A-BIMC is
achieved by solving a sequence of optimization problems to discover points along
the region of interest, and then accruing the discovered points into a Gaussian
mixture distribution. Both the optimization, and the covariances of the
components in the Gaussian mixture are derivative-aware. Stage 2 of A-BIMC
refines the rough approximation yielded by Stage 1 using the Mixture Population
Monte Carlo algorithm. While this would usually require further evaluations of
the forward map, we avoid doing so by using a heuristic surrogate which is
constructed on-the-fly.

Results from several numerical experiments allow us to make the following
conclusions:

\begin{itemize}
\item A-BIMC is independent of the rarity level of the problem, a trait desirable
      from any scheme that aims to efficiently compute rare-event probabilities
\item A-BIMC does break down as the ambient dimensionality of the problem
      increases. This breakdown is due to MPMC, the algorithm employed to refine the
      Gaussian mixture distribution yielded by Stage 1.
\end{itemize}

We suspect that if the intrinsic dimension of the problem is low, the breakdown
with increasing ambient dimension can be arrested. This can be done by
performing importance sampling only in the subspace in which the ideal
importance sampling density differs from the nominal distribution. The
challenge here will be to discover the subspace in question. In the future, we
aim to apply techniques developed for dimension reduction in Bayesian inverse
problems~\cite{zahm2018certified,constantine2015active} to reduce the dimension
of the rare-event probability estimation problem.

\bibliographystyle{siamplain}

\bibliography{refs}
\end{document}


\maketitle

\section{Kullback-Leibler divergence between linearized pushforwards}
\label{supplement:kl_div_push_forwards}

Recall that when a new component, say 
$q_k = \mathcal{N}(\uq_k, \mathbf{H}_{\mathrm{GN}}^{-1})$,
is added to the IS mixture in Stage-1, its covariance
$\mathbf{H}_{\mathrm{GN}}^{-1}$ is dependent on $\sigma^2$:

\begin{align}
    \mathbf{H}_{\mathrm{GN}}^{-1} = \boldsymbol{\Sigma}_0
                                    - \frac{1}{\sigma^2 
                                               + \nabla f(\uq_k)^T\boldsymbol{\Sigma}_0\nabla f(\uq_k)}
                                      \left(\boldsymbol{\Sigma}_0\nabla f(\uq_k)\right)
                                      \left(\boldsymbol{\Sigma}_0\nabla f(\uq_k)\right)^T.
\label{supplement_gaussNewtonHessInv}
\end{align}

For each $\uq_k$, A-BIMC finds a suitable value of $\sigma^2$ by

\begin{enumerate}
    \item linearizing $f(\uq)$ around $\uq_k$ to obtain 
          $f^{\mathrm{lin}}(\uq) = \nabla f(\uq_k)^T (\uq - \uq_k) + f(\uq_k)$
    \item computing the push-forward densities of $q_k$ and $q^*$ under
          $f^{\mathrm{lin}}$, denoted $q_{k,\natural}$ and
          $q^*_{\natural}$.
    \item setting $\sigma = \sigma^*$ in \Cref{supplement_gaussNewtonHessInv},
          where $\sigma^* = \argmin D_{\mathrm{KL}}(q^*_{\natural} || q_{k,\natural})$
\end{enumerate}

In this section, we derive expressions for
$D_{\mathrm{KL}}(q^*_{\natural} || q_{k, \natural})$ 
and $\sigma^*$. To begin with, notice that the univariate
densities $q^*_{\natural}$ and $q_{k,\natural}$ have the following functional form:

\begin{align*}
    q^*_{\natural}(z) &= \frac{\ind_{\DT}(z)}{\mu} 
                                 \mathcal{N}\left(z; f^{\mathrm{lin}}(\uq_0), 
                                             \nabla f(\uq_k)^T \boldsymbol{\Sigma}_0 \nabla f(\uq_k)\right)\\
    q_{k,\natural}(z) &= \mathcal{N}\left(z; f(\uq_k), 
                                \frac{\sigma^2\nabla f(\uq_k)^T \boldsymbol{\Sigma}_0 \nabla f(\uq_k) }
                                     {\sigma^2 + \nabla f(\uq_k)^T \boldsymbol{\Sigma}_0 \nabla f(\uq_k)} 
                           \right).
\end{align*}

Hence, the push-forward of $q^*$, $q^*_{\natural}$, is the Normal distribution
$\mathcal{N}\left(f^{\mathrm{lin}}(\uq_0), \nabla f(\uq_k)^T
\boldsymbol{\Sigma}_0\nabla f(\uq_k)\right)$ truncated over the interval $\DT$.
Further, $\mathcal{N}\left(f^{\mathrm{lin}}(\uq_0), \nabla
f(\uq_k)^T\boldsymbol{\Sigma}_0\nabla f(\uq_k)\right)$ is nothing but the
push-forward of $p$ under $f^{\mathrm{lin}}$. Denoting this push-forward by
$p_{\natural}$, $q^*_{\natural}$ can be expressed more succinctly as $q^*_{\natural}(z) =
\ind_{\DT}(z) p_{\natural}(z) / \mu$.

\begin{table}[htbp]
\small
\centering
\begin{tabular}[c]{l l}
\toprule
Symbol             & Meaning \\
\midrule 
$\vect{v}_k$         & $\nabla f(\uq_k)$\\
$\rho$               & ${\sigma /\sqrt{\sigma^2 
                            + \nabla f(\uq_k)^T\boldsymbol{\Sigma}_0\nabla f(\uq_k)}}$\\
$\nu_T$              & mean of $q^*_{\sharp}$\\
$\gamma_T^2$         & variance of $q^*_{\sharp}$\\
\bottomrule
\end{tabular}
\caption{Symbols used in \Cref{supplement:kl_div_push_forwards}}
\label{table:kl_div_push_forwards_notation}
\end{table}

For the remainder of this section, we shall use the symbols defined in
\Cref{table:kl_div_push_forwards_notation}. From the definition of K-L
divergence, we have,

\begin{align}
\begin{split}
    D_{\mathrm{KL}}(q_{\natural}^* || q_{k,\natural})
                    &= \int_{\mathbb{R}} \frac{\ind_{\DT}(z)}{\mu} 
                                        \log \frac{\ind_{\DT}(z)}{\mu}
                                             \frac{p_{\sharp}(z)} {q_{k,\sharp}(z)} 
                                        \mathrm{d}z\\
                    &= \frac{1}{\mu}\int_{\DT} p_{\natural}(z)
                       \log \frac{ p_{\natural}(z)}{q_{k, \natural}(z)}
                       \mathrm{d}z - \log \mu.
\end{split}
\label{supplement_kl_div_definition}
\end{align}

Now, it can be shown that 

\begin{align}
\begin{split}
    \log \frac{p_{\natural}(z)}{q_{k,\natural}(z)} = &\log \rho
                                         + \frac{1}{2\rho^2 \vect{v}_k^T\boldsymbol{\Sigma}_0\vect{v}_k}
                                           \left(z - f(\uq_k)\right)^2 
                                         -\frac{1}{2\vect{v}_k^T \boldsymbol{\Sigma}_0\vect{v}_k} 
                                          \left(z - f^{\mathrm{lin}}(\uq_0)\right)^2
\end{split}
\label{supplement_logp_by_logq}
\end{align}

Plugging \Cref{supplement_logp_by_logq} into
\Cref{supplement_kl_div_definition} and using the definition of $q_{\sharp}^*$, we
obtain


\begin{align}
\begin{split}
    D_{\mathrm{KL}}(q_{\natural}^* || q_{k,\natural})
                    = &\log \rho - \log \mu \\
                      &+ \frac{1}{2\rho^2\vect{v}_k^T\boldsymbol{\Sigma}_0\vect{v}_k}
                         \mathbb{E}_{q^*_{\natural}}\left[\left(z - f(\uq_k)\right)^2\right]\\
                      &- \frac{1}{2\vect{v}_k^T\boldsymbol{\Sigma}_0\vect{v}_k}
                         \mathbb{E}_{q^*_{\natural}}\left[\left(z - f^{\mathrm{lin}}(\uq_0)\right)^2\right]
\end{split}
\label{supplement_kl_div_intermediate}
\end{align}

The expectations with respect to $q^*_{\sharp}$ in
\Cref{supplement_kl_div_intermediate} can be related to the mean $\nu_T$ and
variance $\gamma_T^2$ of $q^*_{\sharp}$. This leads to the following
expression for the KL divergence between the push-forwards

\begin{align}
\begin{split}
    D_{\mathrm{KL}}(q^*_{\natural} || q_{k,\natural}) 
                    = &\log \rho - \log \mu + \frac{\gamma_T^2}{2\sigma^2}\\
                      &+ \frac{1}{2\rho^2\vect{v}_k^T\boldsymbol{\Sigma}_0\vect{v}_k}
                         \left[\nu_T - f(\uq_k)\right]^2\\
                      &- \frac{1}{2\vect{v}_k^T\boldsymbol{\Sigma}_0\vect{v}_k}
                         \left[\nu_T - f^{\mathrm{lin}}(\uq_0)\right]^2
\end{split}
\label{supplement_push_forward_kl_div}
\end{align}

The value of $\sigma$ that minimizes this equation is:

\begin{align}
    \sigma^* = \frac{\vect{v}_k^T\boldsymbol{\Sigma}_0\vect{v}_k
                       \left(\gamma_T^2 + (\nu_T - f(\uq_k))^2\right)}
                       {\vect{v}_k^T\boldsymbol{\Sigma}_0\vect{v}_k
                        - \gamma_T^2 - (\nu_T - f(\uq_k))^2}
\label{supplement_optimal_noise_var}
\end{align}

\section{Relationship between $e_{\mathsf{RMS}}$ and ESS}
\label{supplement:relationship_erms_ess}

This section establishes that the sample estimate of the relative RMSE,
$\tilde{e}_{\mathrm{RMS}}^N$, and the Effective Sample Size are closely related,
and in fact, are different ways of expressing the mismatch between the ideal
importance sampling density $q^*$ and the importance sampling density $Q$.

Let $\boldsymbol{X}_1, \ldots, \boldsymbol{X}_N \overset{\mathrm{i.i.d.}}\sim Q(\uq)$ be i.i.d.
samples from $Q$. Start by defining the normalized IS weights as follows:

\begin{align*}
    \bar{w}_i &= \frac{\ind_{\DT}(f(\boldsymbol{X}_i)) p(\boldsymbol{X}_i)/q(\boldsymbol{X}_i)}
                     {\sum_j \ind_{\DT}(f(\boldsymbol{X}_j)) p(\boldsymbol{X}_j)/q(\boldsymbol{X}_j)} 
\end{align*}

Noticing that the denominator $\sum_j \ind_{\DT}(f(\boldsymbol{X}_j)) p(\boldsymbol{X}_j)/q(\boldsymbol{X}_j)$
is nothing but $N \tilde{\mu}^N$, we have,

\begin{align*}
    \bar{w}_i = \frac{\ind_{\DT}(f(\boldsymbol{X}_i)) p(\boldsymbol{X}_i)/q(\boldsymbol{X}_i)}{N \tilde{\mu}^N}.
\end{align*}

We use the function-specific ESS introduced in~\cite{mcbook}:
\begin{align*}
    \mathrm{ESS} = \frac{1}{\sum_i \bar{w}_i^2} 
                 = \frac{(N\tilde{\mu}^N)^2}
                   {\sum_i \ind_{\DT}(f(\boldsymbol{X}_i))p^2(\boldsymbol{X}_i) / q^2(\boldsymbol{X}_i)},
\end{align*}

and define the normalized-ESS as $\mathrm{ESS}/N$:

\begin{align*}
    \frac{\mathrm{ESS}}{N} &= \frac{(\tilde{\mu}^N)^2}
                                   {\frac{1}{N}\sum_i \ind_{\DT}(f(\boldsymbol{X}_i))p^2(\boldsymbol{X}_i) / q^2(\boldsymbol{X}_i)}
\end{align*}

Recall from Equation (4) in~\cite{wahal2019bimc} 
that the relative Root Mean Square Error is defined as:

\begin{align*}
    e_{\mathrm{RMS}} = \sqrt{\frac{\mathbb{V}_q\left[\ind_{\DT}(f(\uq)p(\uq)/q(\uq))\right]}{{\mu}^2N}}
\end{align*}

Approximating $\mathbb{V}_q\left[\ind_{\DT}(f(\uq)p(\uq)/q(\uq))\right]$ 
and ${\mu}$ via samples 
leads to the following estimate for the relative RMSE:

\begin{align*}
    \tilde{e}_{\mathrm{RMS}}^N &= \sqrt{\frac{1}{N}\left(\frac{\sum_i
                        \ind_{\DT}(f(\boldsymbol{X}_i))p(\boldsymbol{X}_i)^2/q(\boldsymbol{X}_i)^2}{N(\tilde{\mu}^N)^2} - 1\right)}\\
                     &= \sqrt{\frac{1}{N}\left(\frac{N}{\mathrm{ESS}} - 1\right)}
\end{align*}

Alternatively,

\begin{align}
    \frac{\mathrm{ESS}}{N} = \frac{1}{N (\tilde{e}_{\mathrm{RMS}}^N)^2 + 1}
\end{align}

The reason why ESS and $e_{\mathrm{RMS}}^N$ are directly related is that
they're both dependent on the $\chi^2$-divergence between $q^*$ and $Q$. The
$\chi^2$-divergence between two probability distributions, like the
Kullback-Leibler divergence, is a measure of the distance between two
probability distributions. For instance, the $\chi^2$-divergence between $q^*$
and $Q$ is defined as:

\begin{align*}
    D_{\mathrm{\chi^2}} = \int \left(\frac{q^*(\uq)}{Q(\uq)} - 1\right)^2 Q(\uq) \mathrm{d}\uq
\end{align*}

And once again, a sample estimate of $D_{\mathrm{\chi^2}}$, denoted
$\tilde{D}_{\mathrm{\chi^2}}$ can be computed as:

\begin{align*}
    \tilde{D}_{\mathrm{\chi^2}} = \frac{1}{(\tilde{\mu}^N)^2}
                                \left(\frac{1}{N}\sum_i\ind_{\DT}(f(\boldsymbol{X}_i)) 
                                      p(\boldsymbol{X}_i)^2/q(\boldsymbol{X}_i)^2 - 1\right)
\end{align*}

Straightforward manipulations lead to:

\begin{align*}
\tilde{e}_{\mathrm{RMS}}^N &= \sqrt{\frac{\tilde{D}_{\chi^2}}{N}}\\
\frac{\mathrm{ESS}}{N}     &= \frac{1}{\tilde{D}_{\chi^2} + 1}
\end{align*}

Hence, $e^N_{\mathrm{RMS}}$ and the normalized-ESS are just
different ways of expressing the mismatch between $q^*$ and $Q$, as measured by
the $\chi^2$-divergence. As a result, one does not contain more information
over the other. Note that the existence of such a relationship isn't totally
unexpected, and has in fact been established elsewhere for other settings in
which importance sampling is employed. In~\cite{agapiou2017importance}, 
a similar relationship
is derived, but for the autonormalized IS estimator. Again 
in~\cite{sanz-alonso2018importance}, a
very similar relationship is derived, but under the assumption that
importance sampling is being employed to approximate some intractable target
distribution. As a result, their derivation is agnostic of the function being
integrated. Ignoring the integrand, which would translate to the indicator
function $\ind_{\DT}(f(\uq))$, in the rare-event setting will not yield
meaningful relationships. Indeed, it is due to the highly non-linear nature of
the indicator function that rare-event probability estimation is so notoriously
difficult. Here, we've shown that this relationship 
 exists in the the rare-event setting as well, but
between the relative RMSE, and the rare-event (or function-) specific ESS
defined in~\cite{mcbook}.

\section{Additional Results}
\label{supplement:additional_results}

In the main text, only the relative RMSE $\tilde{e}_{\mathrm{RMS}}^N$ are
reported. Here, we report the actual IS estimate of the rare-event
probabilities.

\subsection{Toy problem}

\Cref{supplement_table_taylor_green_prob} reports the rare-event
probabilities for experiments in Section 4.1. 

\begin{table}[H]
\scriptsize
    \centering
    \caption{Toy problem}
    \pgfplotstabletypeset[
    every head row/.style={before row=\toprule,after row=\midrule},
    every last row/.style={after row=\bottomrule},
    columns/0/.style={column name=$N$, sci e, std=-10:0},
    columns/1/.style={column name=BIMC, 
        int detect, sci e, precision=3},
    columns/2/.style={column name=Stage-1, 
        int detect, sci e, precision=3},
    columns/3/.style={column name=Stage-2,
        sci e, precision=3, zerofill},
    columns/4/.style={column name=MC,
        sci e, int detect},
    col sep=comma
]{./tables/taylor_green__prob_consolidated.out}
\label{supplement_table_taylor_green_prob}
\end{table}

\subsection{Effect of dimensionality}
\label{supplement:dimensionality_results}

In this subsection, we report the rare-event probabilities for experiments
described in Section 4.2.

\begin{table}[htbp]
    \scriptsize
        \centering
        \caption{Quadratic, $m = 16$}
        \begin{subtable}[h]{0.45\textwidth}
        \pgfplotstabletypeset[
        every head row/.style={before row=\toprule,after row=\midrule},
        every last row/.style={after row=\bottomrule},
        columns/0/.style={column name=$N$, sci e, std=-10:0},
        columns/1/.style={column name=Stage-1, 
            int detect, sci e, precision=3},
        columns/2/.style={column name=Stage-2,
            sci e, precision=3, zerofill},
        columns/3/.style={column name=MC,
            sci e, int detect},
        col sep=comma
    ]{./tables/quadratic_dim_conv_prob_level_1e-03_int_4_amb_16_prob_consolidated.out}
    \caption{$m_{\mathrm{int}} = 4$}
    \end{subtable}
    \begin{subtable}[h]{0.45\textwidth}
    \pgfplotstabletypeset[
        every head row/.style={before row=\toprule,after row=\midrule},
        every last row/.style={after row=\bottomrule},
        columns/0/.style={column name=$N$, sci e, std=-10:0},
        columns/1/.style={column name=Stage-1, 
            int detect, sci e, precision=3},
        columns/2/.style={column name=Stage-2,
            sci e, precision=3, zerofill},
        columns/3/.style={column name=MC,
            sci e, int detect},
        col sep=comma
    ]{./tables/quadratic_dim_conv_prob_level_1e-03_int_8_amb_16_prob_consolidated.out}
    \caption{$m_{\mathrm{int}} = 8$}
    \end{subtable}
    \begin{subtable}[h]{0.45\textwidth}
    \pgfplotstabletypeset[
        every head row/.style={before row=\toprule,after row=\midrule},
        every last row/.style={after row=\bottomrule},
        columns/0/.style={column name=$N$, sci e, std=-10:0},
        columns/1/.style={column name=Stage-1, 
            int detect, sci e, precision=3},
        columns/2/.style={column name=Stage-2,
            sci e, precision=3, zerofill},
        columns/3/.style={column name=MC,
            sci e, int detect},
        col sep=comma
    ]{./tables/quadratic_dim_conv_prob_level_1e-03_int_16_amb_16_prob_consolidated.out}
    \caption{$m_{\mathrm{int}} = 16$}
    \end{subtable}
\end{table}

\begin{table}[htbp]
\scriptsize
    \centering
    \caption{Cubic, $m = 16$}
\begin{subtable}[h]{0.45\textwidth}
    \pgfplotstabletypeset[
    every head row/.style={before row=\toprule,after row=\midrule},
    every last row/.style={after row=\bottomrule},
    columns/0/.style={column name=$N$, sci e, std=-10:0},
    columns/1/.style={column name=Stage-1, 
        int detect, sci e, precision=3},
    columns/2/.style={column name=Stage-2,
        sci e, precision=3, zerofill},
    columns/3/.style={column name=MC,
        sci e, int detect},
    col sep=comma
]{./tables/cubic_dim_conv_prob_level_1e-03_int_4_amb_16_prob_consolidated.out}
\caption{$m_{\mathrm{int}} = 4$}
\end{subtable}
\begin{subtable}[h]{0.45\textwidth}
    \pgfplotstabletypeset[
    every head row/.style={before row=\toprule,after row=\midrule},
    every last row/.style={after row=\bottomrule},
    columns/0/.style={column name=$N$, sci e, std=-10:0},
    columns/1/.style={column name=Stage-1, 
        int detect, sci e, precision=3},
    columns/2/.style={column name=Stage-2,
        sci e, precision=3, zerofill},
    columns/3/.style={column name=MC,
        sci e, int detect},
    col sep=comma
]{./tables/cubic_dim_conv_prob_level_1e-03_int_8_amb_16_prob_consolidated.out}
\caption{$m_{\mathrm{int}} = 8$}
\end{subtable}
\begin{subtable}[h]{0.45\textwidth}
    \pgfplotstabletypeset[
    every head row/.style={before row=\toprule,after row=\midrule},
    every last row/.style={after row=\bottomrule},
    columns/0/.style={column name=$N$, sci e, std=-10:0},
    columns/1/.style={column name=Stage-1, 
        int detect, sci e, precision=3},
    columns/2/.style={column name=Stage-2,
        sci e, precision=3, zerofill},
    columns/3/.style={column name=MC,
        sci e, int detect},
    col sep=comma
]{./tables/cubic_dim_conv_prob_level_1e-03_int_16_amb_16_prob_consolidated.out}
\caption{$m_{\mathrm{int}} = 16$}
\end{subtable}
\end{table}

\begin{table}[htbp]
\scriptsize
    \centering
    \caption{Quadratic, $m = 32$}
\begin{subtable}[h]{0.45\textwidth}
    \pgfplotstabletypeset[
    every head row/.style={before row=\toprule,after row=\midrule},
    every last row/.style={after row=\bottomrule},
    columns/0/.style={column name=$N$, sci e, std=-10:0},
    columns/1/.style={column name=Stage-1, 
        int detect, sci e, precision=3},
    columns/2/.style={column name=Stage-2,
        sci e, precision=3, zerofill},
    columns/3/.style={column name=MC,
        sci e, int detect},
    col sep=comma
]{./tables/quadratic_dim_conv_prob_level_1e-03_int_4_amb_32_prob_consolidated.out}
\caption{$m_{\mathrm{int}} = 4$}
\end{subtable}
\begin{subtable}[h]{0.45\textwidth}
    \pgfplotstabletypeset[
    every head row/.style={before row=\toprule,after row=\midrule},
    every last row/.style={after row=\bottomrule},
    columns/0/.style={column name=$N$, sci e, std=-10:0},
    columns/1/.style={column name=Stage-1, 
        int detect, sci e, precision=3},
    columns/2/.style={column name=Stage-2,
        sci e, precision=3, zerofill},
    columns/3/.style={column name=MC,
        sci e, int detect},
    col sep=comma
]{./tables/quadratic_dim_conv_prob_level_1e-03_int_8_amb_32_prob_consolidated.out}
\caption{$m_{\mathrm{int}} = 8$}
\end{subtable}
\begin{subtable}[h]{0.45\textwidth}
    \pgfplotstabletypeset[
    every head row/.style={before row=\toprule,after row=\midrule},
    every last row/.style={after row=\bottomrule},
    columns/0/.style={column name=$N$, sci e, std=-10:0},
    columns/1/.style={column name=Stage-1, 
        int detect, sci e, precision=3},
    columns/2/.style={column name=Stage-2,
        sci e, precision=3, zerofill},
    columns/3/.style={column name=MC,
        sci e, int detect},
    col sep=comma
]{./tables/quadratic_dim_conv_prob_level_1e-03_int_16_amb_32_prob_consolidated.out}
\caption{$m_{\mathrm{int}} = 16$}
\end{subtable}
\begin{subtable}[h]{0.45\textwidth}
    \pgfplotstabletypeset[
    every head row/.style={before row=\toprule,after row=\midrule},
    every last row/.style={after row=\bottomrule},
    columns/0/.style={column name=$N$, sci e, std=-10:0},
    columns/1/.style={column name=Stage-1, 
        int detect, sci e, precision=3},
    columns/2/.style={column name=Stage-2,
        sci e, precision=3, zerofill},
    columns/3/.style={column name=MC,
        sci e, int detect},
    col sep=comma
]{./tables/quadratic_dim_conv_prob_level_1e-03_int_32_amb_32_prob_consolidated.out}
\caption{$m_{\mathrm{int}} = 32$}
\end{subtable}
\end{table}
\begin{table}[htbp]
\scriptsize
    \centering
    \caption{Cubic, $m = 32$}
\begin{subtable}[h]{0.45\textwidth}
    \pgfplotstabletypeset[
    every head row/.style={before row=\toprule,after row=\midrule},
    every last row/.style={after row=\bottomrule},
    columns/0/.style={column name=$N$, sci e, std=-10:0},
    columns/1/.style={column name=Stage-1, 
        int detect, sci e, precision=3},
    columns/2/.style={column name=Stage-2,
        sci e, precision=3, zerofill},
    columns/3/.style={column name=MC,
        sci e, int detect},
    col sep=comma
]{./tables/cubic_dim_conv_prob_level_1e-03_int_4_amb_32_prob_consolidated.out}
\caption{$m_{\mathrm{int}} = 4$}
\end{subtable}
\begin{subtable}[h]{0.45\textwidth}
    \pgfplotstabletypeset[
    every head row/.style={before row=\toprule,after row=\midrule},
    every last row/.style={after row=\bottomrule},
    columns/0/.style={column name=$N$, sci e, std=-10:0},
    columns/1/.style={column name=Stage-1, 
        int detect, sci e, precision=3},
    columns/2/.style={column name=Stage-2,
        sci e, precision=3, zerofill},
    columns/3/.style={column name=MC,
        sci e, int detect},
    col sep=comma
]{./tables/cubic_dim_conv_prob_level_1e-03_int_8_amb_32_prob_consolidated.out}
\caption{$m_{\mathrm{int}} = 8$}
\end{subtable}
\begin{subtable}[h]{0.45\textwidth}
    \pgfplotstabletypeset[
    every head row/.style={before row=\toprule,after row=\midrule},
    every last row/.style={after row=\bottomrule},
    columns/0/.style={column name=$N$, sci e, std=-10:0},
    columns/1/.style={column name=Stage-1, 
        int detect, sci e, precision=3},
    columns/2/.style={column name=Stage-2,
        sci e, precision=3, zerofill},
    columns/3/.style={column name=MC,
        sci e, int detect},
    col sep=comma
]{./tables/cubic_dim_conv_prob_level_1e-03_int_16_amb_32_prob_consolidated.out}
\caption{$m_{\mathrm{int}} = 16$}
\end{subtable}
\begin{subtable}[h]{0.45\textwidth}
    \pgfplotstabletypeset[
    every head row/.style={before row=\toprule,after row=\midrule},
    every last row/.style={after row=\bottomrule},
    columns/0/.style={column name=$N$, sci e, std=-10:0},
    columns/1/.style={column name=Stage-1, 
        int detect, sci e, precision=3},
    columns/2/.style={column name=Stage-2,
        sci e, precision=3, zerofill},
    columns/3/.style={column name=MC,
        sci e, int detect},
    col sep=comma
]{./tables/cubic_dim_conv_prob_level_1e-03_int_32_amb_32_prob_consolidated.out}
\caption{$m_{\mathrm{int}} = 32$}
\end{subtable}
\end{table}

\begin{table}[htbp]
\scriptsize
    \centering
    \caption{Quadratic, $m = 64$}
\begin{subtable}[h]{0.45\textwidth}
    \pgfplotstabletypeset[
    every head row/.style={before row=\toprule,after row=\midrule},
    every last row/.style={after row=\bottomrule},
    columns/0/.style={column name=$N$, sci e, std=-10:0},
    columns/1/.style={column name=Stage-1, 
        int detect, sci e, precision=3},
    columns/2/.style={column name=Stage-2,
        sci e, precision=3, zerofill},
    columns/3/.style={column name=MC,
        sci e, int detect},
    col sep=comma
]{./tables/quadratic_dim_conv_prob_level_1e-03_int_4_amb_64_prob_consolidated.out}
\caption{$m_{\mathrm{int}} = 4$}
\end{subtable}
\begin{subtable}[h]{0.45\textwidth}
    \pgfplotstabletypeset[
    every head row/.style={before row=\toprule,after row=\midrule},
    every last row/.style={after row=\bottomrule},
    columns/0/.style={column name=$N$, sci e, std=-10:0},
    columns/1/.style={column name=Stage-1, 
        int detect, sci e, precision=3},
    columns/2/.style={column name=Stage-2,
        sci e, precision=3, zerofill},
    columns/3/.style={column name=MC,
        sci e, int detect},
    col sep=comma
]{./tables/quadratic_dim_conv_prob_level_1e-03_int_8_amb_64_prob_consolidated.out}
\caption{$m_{\mathrm{int}} = 8$}
\end{subtable}
\begin{subtable}[h]{0.45\textwidth}
    \pgfplotstabletypeset[
    every head row/.style={before row=\toprule,after row=\midrule},
    every last row/.style={after row=\bottomrule},
    columns/0/.style={column name=$N$, sci e, std=-10:0},
    columns/1/.style={column name=Stage-1, 
        int detect, sci e, precision=3},
    columns/2/.style={column name=Stage-2,
        sci e, precision=3, zerofill},
    columns/3/.style={column name=MC,
        sci e, int detect},
    col sep=comma
]{./tables/quadratic_dim_conv_prob_level_1e-03_int_16_amb_64_prob_consolidated.out}
\caption{$m_{\mathrm{int}} = 16$}
\end{subtable}
\begin{subtable}[h]{0.45\textwidth}
    \pgfplotstabletypeset[
    every head row/.style={before row=\toprule,after row=\midrule},
    every last row/.style={after row=\bottomrule},
    columns/0/.style={column name=$N$, sci e, std=-10:0},
    columns/1/.style={column name=Stage-1, 
        int detect, sci e, precision=3},
    columns/2/.style={column name=Stage-2,
        sci e, precision=3, zerofill},
    columns/3/.style={column name=MC,
        sci e, int detect},
    col sep=comma
]{./tables/quadratic_dim_conv_prob_level_1e-03_int_32_amb_64_prob_consolidated.out}
\caption{$m_{\mathrm{int}} = 32$}
\end{subtable}
\begin{subtable}[h]{0.45\textwidth}
    \pgfplotstabletypeset[
    every head row/.style={before row=\toprule,after row=\midrule},
    every last row/.style={after row=\bottomrule},
    columns/0/.style={column name=$N$, sci e, std=-10:0},
    columns/1/.style={column name=Stage-1, 
        int detect, sci e, precision=3},
    columns/2/.style={column name=Stage-2,
        sci e, precision=3, zerofill},
    columns/3/.style={column name=MC,
        sci e, int detect},
    col sep=comma
]{./tables/quadratic_dim_conv_prob_level_1e-03_int_64_amb_64_prob_consolidated.out}
\caption{$m_{\mathrm{int}} = 64$}
\end{subtable}
\end{table}

\begin{table}[htbp]
\scriptsize
    \centering
    \caption{Cubic, $m = 64$}
\begin{subtable}[h]{0.45\textwidth}
    \pgfplotstabletypeset[
    every head row/.style={before row=\toprule,after row=\midrule},
    every last row/.style={after row=\bottomrule},
    columns/0/.style={column name=$N$, sci e, std=-10:0},
    columns/1/.style={column name=Stage-1, 
        int detect, sci e, precision=3},
    columns/2/.style={column name=Stage-2,
        sci e, precision=3, zerofill},
    columns/3/.style={column name=MC,
        sci e, int detect},
    col sep=comma
]{./tables/cubic_dim_conv_prob_level_1e-03_int_4_amb_64_prob_consolidated.out}
\caption{$m_{\mathrm{int}} = 4$}
\end{subtable}
\begin{subtable}[h]{0.45\textwidth}
    \pgfplotstabletypeset[
    every head row/.style={before row=\toprule,after row=\midrule},
    every last row/.style={after row=\bottomrule},
    columns/0/.style={column name=$N$, sci e, std=-10:0},
    columns/1/.style={column name=Stage-1, 
        int detect, sci e, precision=3},
    columns/2/.style={column name=Stage-2,
        sci e, precision=3, zerofill},
    columns/3/.style={column name=MC,
        sci e, int detect},
    col sep=comma
]{./tables/cubic_dim_conv_prob_level_1e-03_int_8_amb_64_prob_consolidated.out}
\caption{$m_{\mathrm{int}} = 8$}
\end{subtable}
\begin{subtable}[h]{0.45\textwidth}
    \pgfplotstabletypeset[
    every head row/.style={before row=\toprule,after row=\midrule},
    every last row/.style={after row=\bottomrule},
    columns/0/.style={column name=$N$, sci e, std=-10:0},
    columns/1/.style={column name=Stage-1, 
        int detect, sci e, precision=3},
    columns/2/.style={column name=Stage-2,
        sci e, precision=3, zerofill},
    columns/3/.style={column name=MC,
        sci e, int detect},
    col sep=comma
]{./tables/cubic_dim_conv_prob_level_1e-03_int_16_amb_64_prob_consolidated.out}
\caption{$m_{\mathrm{int}} = 16$}
\end{subtable}
\begin{subtable}[h]{0.45\textwidth}
    \pgfplotstabletypeset[
    every head row/.style={before row=\toprule,after row=\midrule},
    every last row/.style={after row=\bottomrule},
    columns/0/.style={column name=$N$, sci e, std=-10:0},
    columns/1/.style={column name=Stage-1, 
        int detect, sci e, precision=3},
    columns/2/.style={column name=Stage-2,
        sci e, precision=3, zerofill},
    columns/3/.style={column name=MC,
        sci e, int detect},
    col sep=comma
]{./tables/cubic_dim_conv_prob_level_1e-03_int_32_amb_64_prob_consolidated.out}
\caption{$m_{\mathrm{int}} = 32$}
\end{subtable}
\begin{subtable}[h]{0.45\textwidth}
    \pgfplotstabletypeset[
    every head row/.style={before row=\toprule,after row=\midrule},
    every last row/.style={after row=\bottomrule},
    columns/0/.style={column name=$N$, sci e, std=-10:0},
    columns/1/.style={column name=Stage-1, 
        int detect, sci e, precision=3, sci zerofill},
    columns/2/.style={column name=Stage-2,
        sci e, precision=3, zerofill},
    columns/3/.style={column name=MC,
        sci e, int detect},
    col sep=comma
]{./tables/cubic_dim_conv_prob_level_1e-03_int_64_amb_64_prob_consolidated.out}
\caption{$m_{\mathrm{int}} = 64$}
\end{subtable}
\end{table}

\clearpage

\subsection{Effect of rarity}
\label{supplement:rarity_results}
This subsection reports the rare-event probabilities for experiments
described in Section 4.3.
\begin{table}[H]
\scriptsize
    \centering
    \caption{Quadratic, $m = 16$, $m_{\mathrm{int}} = 8$}
\begin{subtable}[h]{0.45\textwidth}
    \pgfplotstabletypeset[
    every head row/.style={before row=\toprule,after row=\midrule},
    every last row/.style={after row=\bottomrule},
    columns/0/.style={column name=$N$, sci e, std=-10:0},
    columns/1/.style={column name=Stage-1, 
        int detect, sci e, precision=3},
    columns/2/.style={column name=Stage-2,
        sci e, precision=3, zerofill},
    columns/3/.style={column name=MC,
        sci e, int detect},
    col sep=comma
]{./tables/quadratic_dim_conv_prob_level_1e-04_int_8_amb_16_prob_consolidated.out}
\caption{$\mathcal{O}(10^{-4})$}
\end{subtable}
\begin{subtable}[h]{0.45\textwidth}
    \pgfplotstabletypeset[
    every head row/.style={before row=\toprule,after row=\midrule},
    every last row/.style={after row=\bottomrule},
    columns/0/.style={column name=$N$, sci e, std=-10:0},
    columns/1/.style={column name=Stage-1, 
        int detect, sci e, precision=3, sci zerofill},
    columns/2/.style={column name=Stage-2,
        sci e, precision=3, zerofill},
    columns/3/.style={column name=MC,
        sci e, int detect},
    col sep=comma
]{./tables/quadratic_dim_conv_prob_level_1e-05_int_8_amb_16_prob_consolidated.out}
\caption{$\mathcal{O}(10^{-5})$}
\end{subtable}
\begin{subtable}[h]{0.45\textwidth}
    \pgfplotstabletypeset[
    every head row/.style={before row=\toprule,after row=\midrule},
    every last row/.style={after row=\bottomrule},
    columns/0/.style={column name=$N$, sci e, std=-10:0},
    columns/1/.style={column name=Stage-1, 
        int detect, sci e, precision=3},
    columns/2/.style={column name=Stage-2,
        sci e, precision=3, zerofill},
    columns/3/.style={column name=MC,
        sci e, int detect},
    col sep=comma
]{./tables/quadratic_dim_conv_prob_level_1e-06_int_8_amb_16_prob_consolidated.out}
\caption{$\mathcal{O}(10^{-6})$}
\end{subtable}
\end{table}

\begin{table}[H]
\scriptsize
    \centering
    \caption{Cubic, $m = 16$, $m_{\mathrm{int}} = 8$}
\begin{subtable}[h]{0.45\textwidth}
    \pgfplotstabletypeset[
    every head row/.style={before row=\toprule,after row=\midrule},
    every last row/.style={after row=\bottomrule},
    columns/0/.style={column name=$N$, sci e, std=-10:0},
    columns/1/.style={column name=Stage-1, 
        int detect, sci e, precision=3},
    columns/2/.style={column name=Stage-2,
        sci e, precision=3, zerofill},
    columns/3/.style={column name=MC,
        sci e, int detect},
    col sep=comma
]{./tables/cubic_dim_conv_prob_level_1e-04_int_8_amb_16_prob_consolidated.out}
\caption{$\mathcal{O}(10^{-4})$}
\end{subtable}
\begin{subtable}[h]{0.45\textwidth}
    \pgfplotstabletypeset[
    every head row/.style={before row=\toprule,after row=\midrule},
    every last row/.style={after row=\bottomrule},
    columns/0/.style={column name=$N$, sci e, std=-10:0},
    columns/1/.style={column name=Stage-1, 
        int detect, sci e, precision=3},
    columns/2/.style={column name=Stage-2,
        sci e, precision=3, zerofill},
    columns/3/.style={column name=MC,
        sci e, int detect},
    col sep=comma
]{./tables/cubic_dim_conv_prob_level_1e-05_int_8_amb_16_prob_consolidated.out}
\caption{$\mathcal{O}(10^{-5})$}
\end{subtable}
\begin{subtable}[h]{0.45\textwidth}
    \pgfplotstabletypeset[
    every head row/.style={before row=\toprule,after row=\midrule},
    every last row/.style={after row=\bottomrule},
    columns/0/.style={column name=$N$, sci e, std=-10:0},
    columns/1/.style={column name=Stage-1, 
        int detect, sci e, precision=3},
    columns/2/.style={column name=Stage-2,
        sci e, precision=3, zerofill},
    columns/3/.style={column name=MC,
        sci e, int detect},
    col sep=comma
]{./tables/cubic_dim_conv_prob_level_1e-06_int_8_amb_16_prob_consolidated.out}
\caption{$\mathcal{O}(10^{-6})$}
\end{subtable}
\end{table}

\subsection{Failure}
\label{supplement:failure_results}
Here, rare-event probabilities for experiments performed in Section 5.
\begin{table}[H]
\scriptsize
    \centering
    \caption{Failure case F1}
    \begin{filecontents*}{./tables/f1_prob_consolidated.dat}
10,0.013096,0.0011028
100,0.0006578,0.0011887
1000,0.0013088,0.0012069
10000,0.0011911,0.0011647
1e+05,0.0011862,0.0011733
\end{filecontents*}
\pgfplotstabletypeset[
    every head row/.style={before row=\toprule,after row=\midrule},
    every last row/.style={after row=\bottomrule},
    columns/0/.style={column name=$N$, sci e, std=-10:0},
    columns/1/.style={column name={$\epsilon_{\mathrm{abs}} = 1 - 10^{-3}$}, 
        int detect, sci e, precision=3},
    columns/2/.style={column name={$\epsilon_{\mathrm{abs}} = 1 - 10^{-4}$},
        sci e, precision=3, zerofill},
    col sep=comma
]{./tables/f1_prob_consolidated.dat}
\end{table}

\begin{table}[H]
\scriptsize
    \centering
    \caption{Failure case F2}
    \begin{filecontents*}{./tables/f2_prob_consolidated.dat}
10,3.6394e-05,0.0003134
100,3.6035e-07,0.001064
1000,0.00010948,0.00085886
10000,0.00013903,0.00086958
1e+05,0.00069575,0.00088182
\end{filecontents*}
\pgfplotstabletypeset[
    every head row/.style={before row=\toprule,after row=\midrule},
    every last row/.style={after row=\bottomrule},
    columns/0/.style={column name=$N$, sci e, std=-10:0},
    columns/1/.style={column name={$N_{\mathrm{MPMC}} = 10^{6}$}, 
        int detect, sci e, precision=3},
    columns/2/.style={column name={$N_{\mathrm{MPMC}} = 10^{7}$},
        sci e, precision=3, zerofill},
    col sep=comma
]{./tables/f2_prob_consolidated.dat}
\end{table}

\bibliographystyle{siamplain}
\bibliography{refs}